\documentclass[3p,authoryear,a4paper]{elsarticle}
\usepackage{graphicx} % Required for inserting images
\usepackage{rotating}
\usepackage{makecell} % \makecell for wrapped headers
\usepackage{adjustbox} % flexible scaling without shrinking unnecessarily
\usepackage{subcaption}
\usepackage{parskip} % paragraph formatting
\usepackage{xcolor} % for drafting to highlight areas that need work
\usepackage{booktabs} % tables
\usepackage{varioref}
\usepackage{amsthm}
\newtheorem{remark}{Remark}[section]

\usepackage{amsmath}
\numberwithin{equation}{section}

\usepackage{chngcntr}
\counterwithin{figure}{section}
\counterwithin{table}{section}

\usepackage[section]{placeins} % forces figures to appear in the section they were defined

\usepackage{fontspec}  
\usepackage{natbib} % recommended bibliography package
\usepackage{url} % for formatting URLs in bibliography

\usepackage{array,booktabs}
\newcolumntype{m}{>{$}c<{$}} % math-mode centered column

\usepackage{hyperref} % link table of contents to sections of the report
\hypersetup{
    colorlinks=false,
    linktoc=all, 
    hidelinks
}

\usepackage{booktabs}
\usepackage{tabularx}     % for X columns that auto-wrap
\usepackage{threeparttable}

\usepackage{tikz}
\usepackage{pgffor}
\usetikzlibrary{arrows.meta}
\usetikzlibrary{positioning}
\usetikzlibrary {shapes.geometric}

\usepackage{amsmath} % for aligned
\usepackage{amssymb}
\usepackage{listofitems} % for \readlist to create arrays
\usetikzlibrary{arrows.meta} % for arrow size
\usepackage[outline]{contour} % glow around text
\contourlength{1.4pt}
\colorlet{myred}{red!80!black}
\colorlet{myblue}{blue!80!black}
\colorlet{mygreen}{green!60!black}
\colorlet{myorange}{orange!70!red!60!black}
\colorlet{mydarkred}{red!30!black}
\colorlet{mydarkblue}{blue!40!black}
\colorlet{mydarkgreen}{green!30!black}

\usepackage[newfloat]{minted}   % for \listoflistings and the listing counter

\setminted{
  breaklines,
  breakanywhere,
  autogobble,
  linenos,
  fontsize=\footnotesize,
  tabsize=2
}

\tikzset{
  >=latex, % for default LaTeX arrow head
  node/.style={thick,circle,draw=myblue,minimum size=22,inner sep=0.5,outer sep=0.6},
  node in/.style={node,green!20!black,draw=mygreen!30!black,fill=mygreen!25},
  node hidden/.style={node,blue!20!black,draw=myblue!30!black,fill=myblue!20},
  node convol/.style={node,orange!20!black,draw=myorange!30!black,fill=myorange!20},
  node out/.style={node,red!20!black,draw=myred!30!black,fill=myred!20},
  connect/.style={thick,mydarkblue}, %,line cap=round
  connect arrow/.style={-{Latex[length=4,width=3.5]},thick,mydarkblue,shorten <=0.5,shorten >=1},
  node 1/.style={node in}, % node styles, numbered for easy mapping with \nstyle
  node 2/.style={node hidden},
  node 3/.style={node out}
}
\begin{document}

\begin{frontmatter}

\title{Reinforcement Learning for Micro-Level Claims Reserving}

\cortext[cor]{Corresponding author.}
\address[UMelb]{Centre for Actuarial Studies, Department of Economics, The University of Melbourne}
\address[insureAI]{insureAI and University of the Witwatersrand}
\address[UNSW]{School of Risk and Actuarial Studies, UNSW Australia Business School, UNSW Sydney}
\address[ETH]{Department of Mathematics, ETH Zurich}

\author[UMelb]{Benjamin Avanzi}
\ead{b.avanzi@unimelb.edu.au}

\author[insureAI]{Ronald Richman}
\ead{ronaldrichman@gmail.com}

\author[UNSW]{Bernard Wong}
\ead{bernard.wong@unsw.edu.au}

\author[ETH]{Mario W{\"u}thrich}
\ead{mario.wuethrich@math.ethz.ch}

\author[UMelb]{Yagebu Xie\corref{cor}}
\ead{yagebuxie@gmail.com}

\begin{abstract}

Outstanding claim liabilities are revised repeatedly as claims develop, yet most modern reserving models are trained as one-shot predictors and typically learn only from settled claims. We formulate individual claims reserving as a claim-level Markov decision process in which an agent sequentially updates outstanding claim liability (OCL) estimates over development, using continuous actions and a reward design that balances accuracy with stable reserve revisions. A key advantage of this reinforcement learning (RL) approach is that it can learn from all observed claim trajectories, including claims that remain open at valuation, thereby avoiding the reduced sample size and selection effects inherent in supervised methods trained on ultimate outcomes only. We also introduce practical components needed for actuarial use—initialisation of new claims, temporally consistent tuning via a rolling-settlement scheme, and an importance-weighting mechanism to mitigate portfolio-level underestimation driven by the rarity of large claims. On CAS and SPLICE synthetic property–casualty datasets, the proposed Soft Actor–Critic implementation delivers competitive claim-level accuracy and strong aggregate OCL performance, particularly for the immature claim segments that drive most of the liability.

\end{abstract}

\begin{keyword} Reserving \sep Neural networks \sep Reinforcement learning \sep Outstanding Claim Liability 

JEL codes:  %https://www.aeaweb.org/econlit/jelCodes.php?view=jel#L
G22	\sep %Insurance • Insurance Companies • Actuarial Studies
C45	\sep %Neural Networks and Related Topics
C53	%Forecasting and Prediction Methods • Simulation Methods

MSC classes:
%91B06 \sep %Decision theory
% 91B16: Utility theory
%60G51 \sep % Processes with independent increments
%93E20 \sep % Optimal stochastic control
91G70 \sep 	%Statistical methods; risk measures [See also 62P05, 62P20] in Actuarial science and mathematical finance
91G60 \sep 	%Numerical methods (including Monte Carlo methods) in Actuarial science and mathematical finance
62P05 \sep 	%Applications of statistics to actuarial sciences and financial mathematics
%62H12 %\sep 	%Estimation in multivariate analysis
91B30 %\sep % Risk theory, insurance

\end{keyword}
\end{frontmatter}

\section{Introduction}

\subsection{Background} \label{Background}

The prediction of outstanding claims liabilities (OCL) is a core task that underpins solvency and capital management for all insurers. In non-life insurance, policyholders pay premium(s) to receive coverage for a non-life risk. In the event of an accident, a claim may be incurred, and the insurer is liable to cover some or all of the loss.

Early chain ladder models \citep*[e.g., ][]{Mack1993} were limited by low computational power, and hence relied on strong homogeneity assumptions and followed macro-level reserving formulations, which deal with aggregate claims data. Limitations to storage capacity meant that data was stored at the aggregate level as well with no easy way to extract individual claims data. With improvements to computational power came progressively more complex reserving models. These improvements came in two major forms. Firstly, the introduction of micro-level reserving formulations (individual claims level reserving) by \citet{Arjas1989} and \citet{Norberg1993, Norberg1999} proposed initial ideas to work with individual level claims data. Aggregate data are not necessarily sufficient statistics for the estimation of model parameters, and hence the additional granularity and incorporation of claim-level covariates within a micro-level reserving model can yield improvements in predictive power. \citet{Antonio2012} implemented the framework of \citet{Norberg1993, Norberg1999} and showed micro-level results outperformed traditional macro-level methods in their case study. \citet{Avanzi2015} extends this by relaxing the impractical Poisson distribution assumption. Secondly, machine learning has become an active area of research in developing state-of-the-art reserving models for both macro- and micro-level formulations. With increasing computational power, the reserving literature commonly explores the techniques of GLMs \citep{Taylor2016, England2002, Wuthrich2008}, decision trees \citep{Baudry2019, Lopez2019, Duval2019}, and neural networks \citep{Gabrielli2018, Kuo2019, Delong2020, Gabrielli2021}.

An insurer, in any given calendar period, must make a decision on how much to hold in reserve to meet its outstanding claims liabilities (OCL). Regulators (APRA in Australia) require insurers to produce central estimates of the OCL. This task should be performed using the latest available claims information, and the decision is made sequentially every calendar period. This motivates the natural representation of the reserving problem as a Markov Decision Process (MDP) \citep{Howard1960}, which in turn, motivates the use of RL techniques. Note that in most jurisdictions, calculation of a risk margin is also required to reflect aleatoric and epistemic uncertainty; this isn't considered here.

The application of RL in the actuarial literature is still nascent, and particularly so for reserving. \citet{Dong2025} propose a discrete action space framework for aggregate reserving using proximal policy optimisation (PPO), targeting a conditional value-at-risk (CVaR) objective. Their model is trained on two CAS datasets and tested on simulated data obtained through a random draw with a random Gaussian macroeconomic shock applied. They report a boost in performance across all metrics compared to the chain ladder, Bornhuetter-Ferguson, and a stochastic bootstrap method. To the best of our knowledge, this is the only work that applies RL to reserving in the literature. Our contribution is distinct in that we propose a reserving approach working at the transactional level, targeting a central estimate of the OCL. Moreover, we allow smooth adjustments to OCL estimates over developments rather than discrete aggregate reserve adjustments. Finally, we detail implementational mechanisms to address data splitting, hyperparameter tuning, and downward bias correction relevant to micro-level reserving. Outside reserving, \citet{Krasheninnikova2019} applied RL to find an insurance renewal pricing strategy in discrete state space and action space. \citet{Palmborg2023} extends this to a more general premium control formulation and generalises to a non-finite state space with a tabular learning approach called SARSA. \citet{Chong2023} apply deep RL to hedge variable annuities, while \citet{Wekwete2023} apply deep RL to reduce the need for human judgment in asset liability management. It is worthwhile to mention that the field of RL originated largely from optimal control theory, which is not new in the actuarial literature \citep[e.g.,][and the references therein]{Gerber1979, Asmussen1997, Martin-Lf1983, Martin-Lf1994}. With greater computational power, RL can work with far more complex dynamic systems than in traditional optimal control theory, and particularly when there is no model of the underlying environment dynamics. \citet{Palmborg2023} in particular take heavy inspiration from \citeauthor{Martin-Lf1983}'s (\citeyear{Martin-Lf1983, Martin-Lf1994}) formulation of an optimal premium rule in simple settings. Applications of RL have been explored more so in other non-actuarial literature, such as algorithmic trading \citep{Moody2001, Deng2017}, finance \citep{Buehler2019, Hambly2023}, inventory management \citep{Boute2022, Moor2022}, and even particle physics \citep{Meier2012}.

\subsection{Motivations for using RL}

As we argued above, the recursive nature of RL makes it a natural solution to the reserving problem. There are several potential advantages (some purely theoretical) to using RL for reserving, which are elaborated in this section.

    While standard/vanilla RL algorithms are not theoretically robust to non-stationarity \citep{Sutton2018}, an educated choice of design and features for the RL implementation may give it an edge compared to supervised methods when dealing with gradual shifts (rather than abrupt regime changes). RL is a sequential approach, and its implementation will likely be such that the model gets data input (approximately) chronologically. This is particularly so in the soft actor-critic (SAC) algorithm that we use (see also Section \ref{RL lit review} below). The SAC resamples recent past experiences, so that more emphasis is placed on recent experiences rather than distant past experiences. 

Importantly, an RL approach allows learning from open claims. In contrast,  supervised learning approaches require claims to be settled before we can learn from them, since their ``supervision'' requires a target -- here the actual ultimate loss (UL) or equivalently outstanding claims liability (OCL). Some supervised methods have attempted to circumvent this constraint by predicting a series of sequential payments rather than the UL or OCL \citep[e.g.,][]{Gabrielli2021, Schwab2024}, but this approach can be controversial as errors may compound. Approaches that do target the UL/OCL use exclusively settled claims for training, which does not make full use of the available data, and leads to biases (mainly towards quick-to-settle claims; see also \ref{downward bias appendix}). RL avoids this issue altogether as it is not a supervised learning algorithm, and hence does not require targets to train on.

Additionally, RL conceptually allows for a continuously learning model without the need to retrain every time new data is received. 
In contrast, supervised learning methods require periodic refitting with an updated dataset in order to mitigate concept drift. An attractive idea with RL is that it can (theoretically) continue to integrate new experiences over time through gradient updates, without the need to retrain the model. This is very much an adjacent idea to the first advantage pertaining to non-stationarity. However, this idealistic, continuously learning model is still likely a pipe dream due to a problem known as ``catastrophic forgetting", where neural networks can tend to forget quickly what it has learned in the past upon learning something new \citep{French1999}. Although not explored in this paper, this idea endows RL with potential that can be tested in future work.

A further motivation for reinforcement learning, beyond its sequential structure, is that it provides a principled framework for \emph{temporal credit assignment}. In reserving, the quality of an estimate at an early development period can only be fully assessed once the claim eventually settles. However, the ability to design rewards in an RL approach allows intermediate feedback to the model before a claim is ever settled. Through well-designed rewards, reinforcement learning naturally propagates information about accuracy backward through time via ``temporal-difference" updates. This allows early-period estimation behaviour to be shaped by long-horizon outcomes without requiring intermediate labels. From this perspective, reinforcement learning can be viewed as a generalisation of classical reserving techniques that update estimates recursively, but with function approximation and optimisation driven directly by long-run estimation performance.

Relatedly, the ability to design the reward structure with which the model learns can be very powerful. One can incorporate expert knowledge and judgment, as well as potentially tailor a model to specific characteristics of a portfolio. While the design requires some creativity and experimentation, the ability to directly influence the behaviour of the model is essentially unmatched in supervised learning techniques.

In this paper, we develop an implementation of RL for micro-level reserving with detailed benchmarking to demonstrate its feasibility. While the performance is already competitive with benchmarks, it opens the door to many exciting possible improvements to the approach by bringing in more advanced techniques within the RL literature. Our contributions are detailed in the next section.

\subsection{Contributions}

In this paper, we develop a broad conceptual framework to apply RL on the micro-level reserving problem, including core definitions for its Markov Decision Process (MDP) representation. As is usual in the micro-level reserving literature, we focus on forecasting the OCL for open claims (RBNS), but as explained above, we do make use of open claims data. 
In our framework, the RL algorithm learns over time from a sequence of regularly time-spaced observations (e.g., quarters, but this could be arbitrarily adjusted).

We benchmark the performance of our RL approach against that of the chain ladder and a vanilla feed-forward neural network (FNN). We first compare all three models on a property casualty synthetic dataset from the Casualty Actuarial Society \citep{CAS2025}, which is designed to ``replicate selected modelling problems that are commonly faced by members of the CAS''. To properly study the properties of the models, we then use simulated data from the SPLICE simulator \citep{Avanzi2021,Avanzi2023,R-SPLICE}. Because data are simulated, we can benchmark those models using ``actuals'' (simulated beyond the valuation date), and importantly, results can be averaged over multiple simulations. With simulated data of low complexity (agreeing with chain ladder assumptions), we show that the RL performs decently relative to chain ladder. We then run all three models on high complexity data (also from SPLICE), and show that RL is able to remain competitive (if not better) relative to the FNN model. 

The implementation of this framework is far from trivial, and required us to address a number of critical issues, including (but not limited to) the following ones:
\begin{itemize}
    \item Design decisions made in the MDP are critical in determining how performant the RL model will be. The most important component here is the design of reward signals from which the RL model learns. This involved developing a reward function that encourages accurate predictions while also developing additional reward components that encourage ideal behaviour of stable predictions over the lifetime of a claim. 
    \item When a claim is notified and first ``passed'' through the RL algorithm, it needs to be initialised. We discuss how to do so in depth. A good initialisation is particularly material for shorter claims, because the model will have only a few opportunities to adjust a poor starting estimate. We propose a credibility-like approach (balancing period of origin data with whole past data, depending on the maturity of that period of origin), with explicit adjustment for any potential changing claims mix over the training period.
    \item We introduce a ``rolling settlement'' tuning procedure, whereby temporal splitting of data is carefully performed to prevent data leakage during training and validation. Traditionally, RL does not have a need for data splitting to facilitate validation and testing, as data is typically plentiful. However, there is such a need in the reserving context for hyperparameter tuning and for comparisons with supervised methods.
    \item Optimising average claim-level accuracy alone can lead to systematic underestimation of aggregate outstanding liabilities due to the rarity and materiality of large claims; we explicitly address this issue through an `OCL' importance-weighting mechanism.
    The same adjustment is applied to FNN, and hence does not affect the comparability of these two methods.
\end{itemize}

\subsection{Outline of Paper}

The paper is structured as follows. Section \ref{RL lit review} provides a relatively non-technical introduction to RL before we formulate the framework for applying RL to micro-level reserving at a high level in Section \ref{Micro-Formulation}. While Section \ref{Micro-Formulation} discusses only the crucial elements of the formulation, there are many implementational decisions necessary for a performant RL model that we develop in Section \ref{Implementational Choices}. We then develop a novel data splitting approach for the purposes of validation and evaluation in Section \ref{train_test_split}. In the initial evaluation of our models, we encountered a significant downward bias in the portfolio-level predictions, which we discuss and propose solutions for in Section \ref{Downward Bias Adjustment}. In Section \ref{Evaluation}, we discuss the evaluation procedure of the RL model against benchmark models, with performance on simulated property--casualty datasets from CAS \citep{CAS2025} in Section \ref{Results_CAS} and SPLICE \citep{Avanzi2023} data presented in Section \ref{Results}. We conclude by summarising and discussing the main results in Section \ref{S_Conclusion}.

\section{Introduction to Reinforcement Learning} \label{RL lit review}

Reinforcement learning (RL) is a field of study that concerns optimal sequential decision-making \citep{Silver2015}. The RL approach entails a learning agent that interacts with an environment and learns through reward/penalty signals. Through the objective of wanting to maximise rewards (or minimise penalties), the agent improves its behaviour towards some optimal policy of behaviour. Famous examples of RL results include models that can play Atari games \citep{Mnih2013} and beat the best human players at Go and Chess \citep{Silver2017}. Below, we give a brief introduction to the general ideas of RL, but for a more comprehensive treatment, we refer the reader to \citet*{Sutton1988}, \citet{Silver2015}, and \citet{Wuthrich2025}.

\subsection{Overview}

In a very simple (and classic) illustration to capture the essence of this concept \citep[adapted from lecture 1 of][]{Silver2015}, a rat (the agent) learns from trial and error in an experiment. The environment in this scenario might be a researcher who provides stimuli, such as ringing a bell. If the rat pulls the lever after the sound of the bell ringing, it is rewarded with some cheese, while if it pulls the lever before the bell rings, it is given a shock. Without any training, the rat is unaware of these reward signals and environment dynamics. However, over time, it will learn that the bell ringing is an important environmental signal, and that pulling the lever after the bell maximises its reward (optimal policy). For simplicity, let us assume this example world operates in discrete time steps, such that if the bell rings, the rat can, at the earliest, pull the lever one time step later. This is so that the speed at which the rat reacts is not a consideration.

RL concerns ``sequential decision making", which requires that the problem at hand be formulated as a Markov Decision Process (MDP) $\langle \mathcal S, \mathcal A, \mathcal P, \mathcal R, \gamma\rangle$ \citep{Howard1960}. The discrete-time RL formulation is such that when the agent interacts with the environment, it observes a state $s_t\in \mathcal S$, decides to take an action $a_t\in \mathcal A$ as per its policy, and receives a reward signal $r_t\in \mathcal R$. Subsequently, the environment evolves according to some unknown underlying dynamics $\mathcal P$ (denoting transition probabilities) due (at least partly) to the agent's action $a_t$, and the agent will then observe the next state $s_{t+1}$, repeating the cycle. Lastly, $\gamma\in (0, 1)$ is a discount factor that controls how much less the agent should value tomorrow's reward compared to today's.

While this interaction with the environment occurs, the agent's behaviour is called the policy $\pi(a\mid s):=\mathbb P(A_t=a\mid S_t=s)$. The core objective is for the agent to reach some deterministic optimal policy $\pi^*$ that maximises rewards. To achieve this, the RL model either directly optimises a parameterised policy (policy-based), or optimises estimates of ``value functions" (value-based):

\begin{itemize}
    \item The (state)-value function $v_\pi(s)$ is the expected present value of total future rewards given we are currently in state $s$ and continue to follow the policy $\pi$ hereinafter. That is,
    \begin{equation} \label{state-value func}
    v_\pi(s)=\mathbb E_\pi[G_t\mid S_t=s], \quad G_t=\sum_{k=0}^\infty \gamma^k r_{t+k+1}.
    \end{equation}
    where $G_t$ denotes the ``goal" (i.e., objective) the agent wants to maximise.
    \item The action-value function $q_\pi(s,a)$ is the expected present value (EPV) of total future rewards given we are currently in state $s$, take action $a$, and then continue to follow the policy $\pi$ hereinafter. In this case, we have then
    \begin{equation}
    q_\pi(s,a)=\mathbb E_\pi[G_t\mid S_t=s, A_t=a].
    \end{equation}
\end{itemize}

The intuition here is that value-based methods preference actions based on some underlying computed expected future value that the action will yield, whereas policy-based methods simply increase the probability of an action that yielded good returns. In our example, if the rat is policy-based, it would increase its probability of pulling the lever after the bell upon getting cheese. If it is value-based, then it would intrinsically compute some underlying value associated with pulling the lever at any point in time. Modern methods often combine these approaches in the actor–critic framework \citep{Sutton2018, Konda1999}, which separately implements both policy improvement (actor) and value estimation (critic). In ``deep RL'', these are typically represented by two separate neural networks. By having both a policy and a value function approximation, we get the best of both worlds. In practice, this yields faster and more stable convergence due to variance reduction and greater sample efficiency \citep{Konda1999, Xu2021}.

At every time step, the agent faces two choices: either exploit its policy's current best known action to get a known maximum EPV, or take a risk and explore some action whose consequences are unknown to the model in the hopes of finding a better policy. This is known as the exploration vs exploitation trade-off in RL; see also Remark \ref{A comment on the Exploration vs Exploitation Trade-off} for additional commentary specific to our reserving context. Using the rat example, suppose now that there are multiple stimuli, such as a flash of light in addition to the bell. Further, suppose that if the rat pulls the lever after a flash of light, it has a 75\% chance of getting some cheese, and 25\% chance of being shocked, whereas if it pulls the lever after the bell, it will get some cheese with 100\% chance. If the rat discovers the pattern behind the flash of light first, then it has the option to continue exploiting this newfound pattern, or to explore something new, like waiting until after the bell rings. Exploration could backfire, but in this case, it would lead to the discovery of a new optimum. In the context of micro-level reserving, suppose that there is a structural break at some point in time that increases settlement speed of claims. Prior to this structural break, the agent may tend to adjust the OCL estimate downward at a slow pace on average. After the structural break, if the agent continues to exploit its pre-break policy, it will still be rewarded, but if the agent undertakes exploration, it could, by chance, discover that a faster rate of OCL estimate reduction is better.

Of course, because the value functions pertain to the expected (discounted) future gains, they must in practice be estimated since the underlying environment dynamics are not known. There are two broad methods to do so – Monte Carlo (MC) and temporal differencing (TD), with TD being more widely adopted. We refer to the aforementioned sources for a more detailed discussion of these methods. In short, TD gives us the ability to perform updates before a learning episode terminates (or in cases where there is no termination). An episode is simply one run of interaction between the agent and the environment, e.g., a single claim's development until settlement is one episode. Using the example of a claim, TD performs updates at each time step before the claim is ever settled, whereas MC would do all the updates upon settlement. There is a bias-variance trade-off here: MC estimation is unbiased, but has high variance, whereas TD estimation has low variance, but introduces bias.

The state and action spaces ($\mathcal S, \mathcal A$) simply pertain to the range of possible values for the state and actions, and these can be either discrete or continuous. At its simplest, for discrete, finite state and action spaces (e.g., the rat example), combinations of states and actions mapping to a value function can be stored in a table (called tabular learning). However, as the size of these spaces grow (or become infinite), this tabular approach quickly becomes untenable. For a claim, its outstanding claims liabilities can, in theory, be any dollar amount in $\mathbb R^+$ (up to some maximum). We cannot represent this continuous range of possible OCL in a table. Hence, a far more practical approach is to approximate this mapping via some function, known as functional approximation. In deep RL, this functional approximation uses neural networks \citep{Mnih2013}. For a review on neural networks, we refer to \citet{Wuthrich2025} and \citet{Goodfellow2017}. 

After specifying an MDP for the problem, we must use an RL algorithm to train the model. RL algorithms specify

\begin{itemize}
    \item a representation for the policy and value functions (typically neural networks); 
    \item whether the learning process is on-policy or off-policy. On-policy algorithms are such that the models learns policy $\pi$ while also following the same policy $\pi$, whereas off-policy algorithms learn a policy $\pi$ by sampling the experiences of another different policy $\mu$;
    \item how the agent conducts exploration.
\end{itemize}

\subsection{RL algorithms}

In this section, we briefly cover two popular RL algorithms: proximal policy optimisation \citep[PPO][]{schulman2017b} and soft actor-critic \citep[SAC][]{Haarnoja2018}. Before that, we note that the most significant difference between these two algorithms is that fact that PPO is an on-policy algorithm, while SAC is an off-policy algorithm. Because off-policy algorithms can learn from a different policy, they are able to employ experience replay \citep{Lin1992}, whereby past experiences are stored in a replay buffer (essentially a list) to learn from later. This technique boosts sample efficiency by re-using past experiences and breaks temporal correlation to potentially help improve training stability.

\subsubsection{PPO}

Proximal Policy Optimisation \citep[PPO,][]{schulman2017b} and its predecessor, Trust Region Policy Optimisation \citep[TRPO,][]{schulman2017a} are on-policy methods. \citet{schulman2017a} espouse that policy optimisation can be classified into three broad categories: policy iteration, policy gradient, and gradient-free methods. Policy iteration is simply the value-based method of optimisation described previously. For policy-based methods, it can be further broken down into those that optimise a differentiable objective using the gradient (policy gradient methods) vs those that do not use the gradient (gradient-free methods). In particular, prior to TRPO, gradient-based methods were not able to consistently beat out gradient-free methods despite having better sample efficiency. In practice, TRPO and PPO are popular because they are ``effective for optimising large nonlinear policies such as neural networks” \citep{schulman2017a}.

PPO is generally considered an improvement to TRPO because it retains many benefits of TRPO, yet is much simpler to implement, and enjoys better sample efficiency empirically. The innovation of PPO (and TRPO) is that it only allows small changes to the policy within a small and safe range at each step of gradient update. While TRPO has stronger theoretical guarantees, PPO enjoys more stable convergence and better performance empirically thanks to its more parsimonious approach, while retaining the core idea of limiting the size of policy updates.

\subsubsection{SAC}

Soft Actor-Critic (SAC) is an off-policy method that aims to maximise the expected gain plus an entropy term, where higher entropy means more exploration. In the words of \citet{Haarnoja2018}, SAC aims to ``succeed at the task while acting as randomly as possible". Prior deep RL methods based on the ``maximum entropy" approach were restricted to discrete actions, while SAC is able to handle continuous actions. SAC was able to outperform many previous on-policy and off-policy methods, with the additional empirical benefit that it is stable in the sense that it achieves similar performance across different random seeds.

The general maximum entropy objective that the agent seeks to maximise \citep{Ziebart2010} is
\begin{equation}
J(\pi) = \mathbb E\left[ \sum_{t=0}^T \gamma^t \{ r(s_t,a_t) + \alpha \mathcal H(\pi(\cdot\mid s_t)) \} \right],
\end{equation}
where $\gamma$ is the discount factor, $r(s_t,a_t)$ is the reward given action $a_t$ taken in state $s_t$, $\mathcal H(\pi(\cdot \mid s))$ is the entropy term, and $\alpha>0$ is the ``temperature" coefficient that controls how much the agent is encouraged to explore. An agent that explores more will have higher entropy $\mathcal H$, and thus be rewarded for it since it would increase the objective function. We show this objective to demonstrate that the crux of the idea is that a maximum entropy objective is designed to prefer policies that achieve high reward while also being as stochastic as possible; the agent must maximise the cumulative discounted sum of the rewards of its actions, $r(s_t,a_t)$, and the amount of exploration it undertook ($\mathcal H$). Note that entropy originates from information theory as a measure of average uncertainty, where high entropy denotes high uncertainty (stochasticity) \citep{Cover2006}. 

\section{Micro-reserving RL Formulation} \label{Micro-Formulation}

\subsection{Overarching RL Framework} \label{overarching RL framework}

Let us define the following notation: $t=1,...$ denotes the calendar period; $i=1,...,I$ denotes the accident period; $j=1,...,J$ denotes the development period \textit{since accident}; and $\tau=1,...$ denote the periods since notification of the claim. Hence, $t=i+j-1<I+J$. This convention is adopted throughout the paper unless stated otherwise.

We address the application of reinforcement learning to micro-level reserving through a modular and extensible framework. The reserving problem is first cast as a Markov decision process (MDP) $\langle \mathcal S, \mathcal A, \mathcal P, \mathcal R, \gamma\rangle$, which formalises claim development as a sequential estimation problem with delayed feedback. Within this representation, an RL agent learns to update outstanding claim liability estimates over time by interacting with transactional claim data. We then specify an appropriate reinforcement learning algorithm for this setting and discuss how its design choices align with actuarial reserving objectives. Finally, the proposed framework should be evaluated empirically to assess its ability to produce accurate and stable claim-level and portfolio-level reserve estimates relative to established benchmarks, and any results fed back into the design of the framework.

\begin{figure}[tb]
    \centering
    \includegraphics[width=1\linewidth]{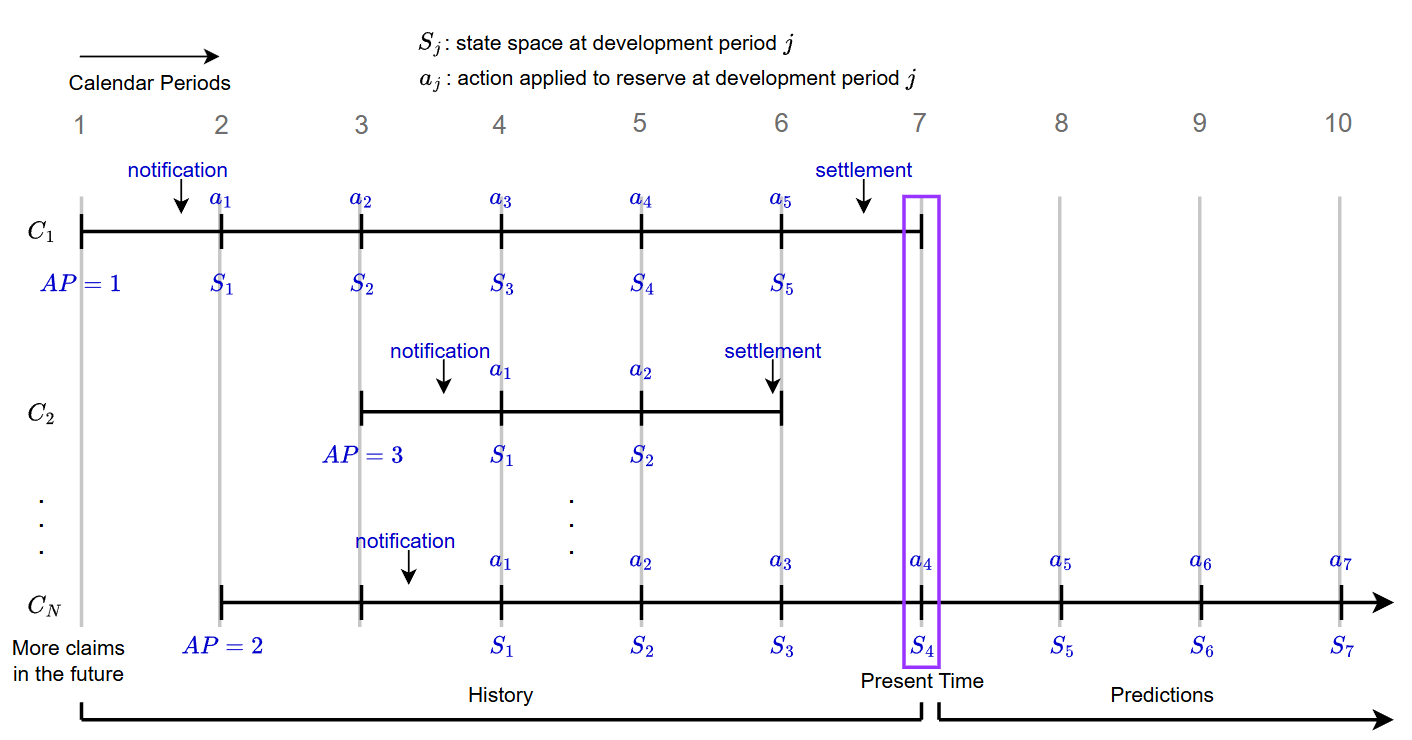}
    \caption{Illustration of the micro-level reserving formulation in action}
    \label{fig:micro_formulation}
\end{figure}

\subsection{Micro-level Framework Design and MDP Formulation} \label{MDP formulation}

A micro-reserving framework approaches the formulation on the basis of claim-level transactional data. Consequently, this works for reported but not settled (RBNS) claims only. In practice, a settled/closed claim can also reopen. However, we do not consider this possibility in our paper.

To prime our micro-level reserving framework, it is useful to consider an analogy of a chessbot learning from many concurrently occurring games: every move made on every board up to the current time will have contributed to updating the bot's strategy. Each board would be a realisation of the environment. For reserving, this is very much a similar idea: we can consider each claim to be a game where the objective is to guess the outstanding claims liabilities (OCL) as accurately as possible at each development for claims developing concurrently. There is the added complication that these reserving ``games" do not start or finish at the same time. We formalise this in Figure \ref{fig:micro_formulation}, which illustrates an example of an insurer with many developing claims in calendar time, with the present time (i.e., valuation period) being the end of calendar period 7. The agent then would have seen the developments of all notified claims up to this time, and can learn from these to predict the OCL for each currently open claim.

A subtle but important point to make clear is that our predictions target the OCL in settlement year dollars (so that base and superimposed inflation are accounted for). To use the OCL was a deliberate choice over targeting the ultimate loss (UL) for technical reasons. In the implementation, we need to specify the bounds for the target. It is easy and clean to specify zero as the lower bound for the OCL, whereas there is no easy way to specify a lower bound for targeting the UL, since this lower bound would naturally be in line with cumulative payments, and those can change at every time step. 

A natural question is where ``control'' arises in our formulation, since the insurer’s reserve estimates do not influence the physical evolution of claims. In our framework, actions do not control claim outcomes; instead, they control the \emph{sequence of estimates} produced by the model.
The action at each development period determines how the previous OCL estimate is updated, and this updated estimate becomes part of the next state observed by the agent. The environment transition is therefore deterministic with respect to the agent’s action, even though claim evolution itself is exogenous. This reframes the problem as \emph{sequential prediction with delayed feedback}, where reinforcement learning is used as a temporal-difference learning mechanism to optimise a sequence of interdependent predictions rather than a physical control system. Such ``prediction-as-control'' formulations are common in reinforcement learning when the objective is to optimise long-horizon forecasting accuracy under partial and delayed information rather than to intervene in the environment directly. In a more complex setting, one could argue that reserving decisions may have an indirect impact on the state space, but we are not considering this here.

\subsection{Markov Decision Process (MDP)}

We now develop the environment and reward formulations as per the framework described above. Formally, every claim follows an MDP $\langle \mathcal S, \mathcal A, \mathcal P, \mathcal R, \gamma\rangle$ \citep{Howard1960}, which we develop below. In this section, we focus only on the general, theoretical formulation, and leave specific choices made in the implementation of this paper to Section \ref{choices in MDP formulation}.

\textbf{Note:} Elements of the MDP are indexed by time (since they evolve over time). It is sometimes more convenient to index by the time since notification $\tau$ (such as in Figure \ref{fig:micro_formulation}), and other times more convenient to index by the development period $j$ (such as in the formulation below). It will be clear which we are using, but it is useful to keep in mind that the indexing is merely anchoring the MDP to a particular calendar period in time.

\begin{remark}
    $\mathcal P$ denotes the underlying environment dynamics (transition probabilities of the Markov Chain). We assume this to be unknown, and do not explicitly estimate them (as is generally standard in RL), so they are not further discussed.
\end{remark}

\begin{remark} \label{continuous MDP remark}
    This paper formulates the reserving problem in a discrete time, continuous state and continuous action space setting. A discrete action space will also work \citep[as in][]{Dong2025, Moody2001}, and is a simpler problem to solve at the cost of less fine-grained control. If fine-grain control is not necessary, it could make more sense to reduce the complexity of the problem. Although not explored in this paper, this adaptation should be rather simple as one only needs to restrict the action space to a set of discrete values, e.g., $\mathcal A=\{ -0.10, -0.066, -0.033, 0, 0.033, 0.066, 0.10 \}$ in \citet{Dong2025}.
\end{remark}

\subsubsection{State space $\mathcal S$}
The state space $\mathcal S$ contains information about the environment for the agent to observe. There are three things that must be included: the accident period (AP), development period (DP), and the previous OCL estimate. Therefore, the bare minimum state space at development period $j$ is 
\begin{equation}\label{E_minSS}
        \{ AP=i, DP=j, \hat{\text{OCL}}_{j-1}\},
\end{equation}
where $\hat{\text{OCL}}_j$ is the OCL predicted at development period $j$ with $\hat{\text{OCL}}_0$ initialisation. Note that the initialisation is quite important, because we can see that at development period $j=1$, our state space starts with the initialisation value $\hat{\text{OCL}}_0$. This means that the sequential predictions for RL begin with how we choose to initialise it. A naive zero or constant initialisation produces poor results because it is not suitable for claims of different sizes. One can imagine an extreme example in which if a claim with OCL \$1000 is initialised at \$1 million, then it would be difficult to adjust quickly (because the action space is restricted; refer to the next section). As such, we consider initialisation at length in Section \ref{Initialisation}.

As aforementioned, we must include the previously predicted OCL in the state space so that the environment is designed such that the action taken is an adjustment to the previous prediction. This is to satisfy the requirement of the MDP formulation, whereby actions taken should impact the state; see also the last paragraph of Section \ref{MDP formulation}. The inclusion of the accident and development periods is standard and necessary in all reserving models because claim sizes vary between accident periods, and the OCL is, of course, generally decreasing as the claim develops. 

It goes without saying that simply using the bare minimum state space in \eqref{E_minSS} will not yield very good results because it is too simplistic. We discuss our choices for the full state space in Section \ref{choices in MDP formulation}.

\subsubsection{Action space $\mathcal A$}

As we discussed above, we want to design the system such that the action taken by the agent is an adjustment to the previous OCL estimate. We choose to define 
\begin{equation}
    \hat{\text{OCL}}_{j}=\hat{\text{OCL}}_{j-1}\cdot \exp(a_j),
\end{equation}
where $a_j\in \mathcal A$ is a scaling factor applied to the previous OCL estimate and the initialisation of $\hat{\text{OCL}}_0$ is discussed in Section \ref{Initialisation}. 

We define the action space to be $\mathcal A=[-\ln(K), +\ln(K)], \; K\in\mathbb R$ to (arbitrarily) impose the constraint that in any single development period, the OCL estimate cannot increase or decrease by more than a factor of $K$. A constraint of $K$ is imposed for stability purposes so that the RL agent doesn't get stuck at low estimates or suddenly explode with a huge OCL estimate when the agent is undertaking exploration.

Treating the action as a scaling factor is an existing approach \citep[e.g.][]{Xu2019} because it is straightforward and is percentage-change based rather than being based on absolute-dollar-change. 
Furthermore, we work in the log-space and then exponentiate so that the action space is symmetrical about zero, with the intention that exploration should not be unbalanced in one direction or the other simply because of the action space definition. A multiplicative action is thus required.
\begin{remark}
    $K$ is a tunable hyperparameter for which there is a tradeoff: for larger $K$, there is greater model flexibility at the cost of a larger action space to explore, while for smaller $K$, there is a smaller action space to explore, but more constraints in how much the model can adjust estimates per development period. We observed that $K=2$ often led to better performance than larger values of $K$, which introduces a high degree of restrictiveness. However, a larger action space may lead to more instability, and the state space is not rich enough and/or the volume of data is not large enough for the model to learn as well.

    To assess the stability of the learned policy, we examined the empirical distribution of actions taken by the trained agent across all claims and development periods; see Figures \ref{fig:actions_hist} and \ref{fig:actions_hist_C5} in the Appendix. We find that actions are tightly concentrated around zero (exponentiation of the action concentrated around 1), indicating that the policy predominantly makes small incremental adjustments to OCL estimates. Clipping at the action bounds occurs infrequently during testing (where exploration is turned off). In fact, large adjustments are largely confined to early development periods, where initial uncertainty is highest (see Figures \ref{fig:actions_hist_by_qns} and \ref{fig:action_hist_qsn_C5}). In later development periods, large adjustments are rare, suggesting that the constrained action space primarily acts as a stabilisation device during learning rather than a binding restriction at convergence. These diagnostics support the interpretation that the learned policy operates well within the admissible action range and that the choice of action bounds does not materially distort the learned reserving behaviour.

\end{remark}

\begin{remark}
    The current action space formulation, where actions are multiplicative adjustments, assumes that the OCL is strictly positive. For circumstances where negative OCLs need to be considered, there are no theoretical reasons why it could not be accommodated. However, an alternative definition for how the action adjusts the previous OCL would be required. This is not further considered in our paper.
   
\end{remark}

\subsubsection{Reward signal $\mathcal R$}

 $\mathcal R$ is the reward function that the agent seeks to maximise. This is a critical design choice, as it drives how the agent learns. It can be difficult to develop in practice, and will require some trial and error even beyond theoretical grounding. Our reward signal $\mathcal R$ will be a combination of multiple components. 

Evidently, there must be a signal for the accuracy of the model's predictions upon settlement so that the model knows how well it performed for the given claim. However, as the accuracy signal will be very sparse (only available upon settlement of the claim), the model will find it difficult to converge to an optimal policy. Furthermore, we encourage stable evolutions in the model's prediction of ultimate loss (UL) over time (calculated as the sum of the predicted OCL and cumulative paid) by rewarding small adjustments over large adjustments. This imposes a cost to large adjustments to the implicit UL, so the agent would not do this unless it has high ``confidence". This is a desirable property for a reserving model from a financial management perspective, but also, it allows for denser reward signals during the development of the claim, which helps learning.

While the specific reward formulas are given in Section \ref{reward formulas}, we will use the following analogy to tease out our main ideas. Suppose we are training an RL agent to play chess. Just like how we want to correctly predict a claim, we want the chess bot to win games. Therefore, upon the end of a game, we can assign a reward of $+1$ for a win, $-1$ for a loss, and $0$ for a draw. However, with only this reward signal at the end of the game, it can be difficult for the agent to ``understand" how it won. That is, what moves were good and what moves were bad. In this sense, we may employ some human knowledge about the underlying theory of the game of chess to create intermediary rewards during the course of a game to reward ``good" moves and penalise ``bad" moves with, say, $+/-0.5$ respectively. In the same way, we reward stability of estimates for our reserving model.

There are many possible reward signals that would be available in practice for a reserving model. Inclusion of covariates into the model could allow for many informative reward signals. 
Additionally, case estimates make for useful proxies of the true unknown OCL, with which we can build denser reward signals before the claim is settled.

\subsubsection{Discount $\gamma$} \label{S_gamma1}

The discount factor $\gamma$ can be chosen or tuned. In practice, this may be determined by risk appetite rather than tuning, as it reflects how much we value correct predictions early compared to later. In our context, correct predictions are valuable at all times, but more so early when there is more OCL at stake. This translates to a large but less than 1 value for $\gamma$.

\subsection{Description of one iteration}

We describe here one iteration of learning from calendar period $t$ to $t+1$ for a single claim. This is a walkthrough of the learning process for a single claim in Figure \ref{fig:micro_formulation}. We provide a concrete numerical example of this process in \ref{A_example}, based on our choices outlined in Section \ref{choices in MDP formulation}.

Assume we are currently in calendar period $t$. Let $C_n(t, i_n)$ denote claim $n$ with accident period $i_n$ at development period $j_n:=t+1-i_n$. Steps are then as follows:
\begin{enumerate}
    \item The agent observes the current state space $\{AP=i_n, DP=j_n, \hat{\text{OCL}}_{j_n-1}, ... \}$;

    \item The agent takes an action $a_{j_n}$ based on the state space observed, that is, the predicted OCL at development period $j_n$ becomes $\hat{\text{OCL}}_{j_{n}}=\hat{\text{OCL}}_{j_n-1} \cdot \exp(a_{j_n})$;

    \item The agent then receives a reward signal $r_{j_n}$ for its prediction of $\hat{\text{OCL}}_{j_{n}}$;
    
    \item Then, at the next development period $j_{n+1}$, the agent observes the state space $\{AP=i_{n+1}, DP=j_{n+1}, \hat{\text{OCL}}_{j_{n}}, ... \}$ and the cycle repeats.
\end{enumerate}

\begin{remark} 
\label{A comment on the Exploration vs Exploitation Trade-off}

Given there are adverse financial consequences to inaccurate estimates, once the insurer has a reasonably trained model $M_1$ at present calendar period $t$ to deploy, they may choose to exploit model $M_1$ deterministically (as opposed to allowing exploration) at calendar period $t+1$ and beyond. However, this would mean $M_1$ lose the ability to learn from the new incoming data since we are no longer exploring. We briefly describe some possible ways to get around this.

First, the simplest workaround is to run 2 parallel agents/models. When a model $M_1$ is to be frozen for deployment, duplicate it, and the clone $M_2$ will continue to be able to explore and learn from the incoming data. The agent and its clone will then be periodically evaluated, and the production agent can be replaced if the clone starts to perform better. The drawback of this approach is that subtle shifts in underlying dynamics may take too long to manifest as a noticeable difference in performance between the production agent and its clone. There will need to be a certain level of judgment as to when the clone should replace its original.

Alternatively, one could define a tight set of constraints to allow exploration in some tolerably safe manner to address the issue of timeliness, at the cost of taking on some risk of exploration.

Though not explored further in this paper, these are worthwhile to investigate in future work.

\end{remark}

\section{Implementational Choices} \label{Implementational Choices}

In this section, we discuss important and nuanced topics regarding the choices involved in the MDP formulation, how to initialise the model at the start of a claim, and which RL algorithm to use.

\subsection{Choices in the MDP Formulation} \label{choices in MDP formulation}

This section continues on from the development of the MDP in Section \ref{MDP formulation} and develops the key choices involved in the MDP formulation. 

\subsubsection{State space $\mathcal S$}

We augment the bare minimum state space of accident period, development period, and previous OCL prediction with the following additions to the state space are derived from the information available to us at each point in time for a claim's development; see \ref{Data Description} for a description of the data we used. Firstly, for the CAS data:

\begin{itemize}
    \item $P_j$ is the cumulative amount paid by development period $j$
    \item $\text{repdel}=\lceil \text{notification time}\rceil - AP$ is the claim's reporting delay. This is potentially an important predictor because it may be the case that claims with longer reporting delays can have larger claim sizes.
    \item $\text{past predictions}_j$ is the (up to) past $n$ predictions made by the model (where $n$ is tunable), with zero-padding when there are less than $n$ previous predictions. We remark that the $n$ past predictions are not necessary, but rather an extension we tested and found to slightly improve performance. This is easily implemented for RL, but not implemented for FNN, which is another minor technical advantage of RL being a sequential approach.
\end{itemize}

Therefore, the state space at development period $j$ is 
\begin{equation}
    \{ AP=i, DP=j, \hat{\text{OCL}}_{j-1}, P_j, \text{repdel}, \text{past predictions}_j \}
\end{equation}

The SPLICE data allows for a richer state space, as it contains case estimates, and allows us to set the time unit to quarters rather than years. It has the following \textit{additions} to the state space

\begin{itemize}
    \item $\text{txn\_types}_j$ are the (one-hot-encoded) transaction types for transactions that occurred in development period $j$.
    \item $n_j^\text{pay}$ is the number of payments made so far by development period $j$, with $n_1^\text{pay}=0$.
    \item $\text{AQ} \in \{1,2,3,4\}$ is the quarter of the year of the accident period. The idea of including the quarter of the accident period is that it could capture any cyclical differences in claim mixtures that would not be detectable with just the accident period. Note that we are working with quarters as the time unit in our implementation, but this can be generalised to an arbitrary time unit easily.
    \item $\text{DQ}_j \in \{1,2,3,4\}$ is the quarter of the current development period $j$. The idea here is similar as that of $\text{AQ}$; it can help capture any cyclical differences in claim development patterns.
    \item $\text{case}_j$ is the case estimate of the OCL at development period $j$.
\end{itemize}

Therefore, the state space at development period $j$ for the SPLICE data in general is 
\begin{equation}
    \{ AP=i, DP=j, \hat{\text{OCL}}_{j-1}, \text{txn\_types}_j, P_j, \text{repdel}, n^\text{pay}_j, \text{AQ}, \text{DQ}_j, \text{past predictions}_j, \text{case}_j \}
\end{equation}

However, for complexity 1, the SPLICE-generated data is simple and consistent with chain-ladder assumptions; therefore, such a state space is too complex and leads to subpar results. Instead, we reduce the state space to:

\begin{equation}
    \{ AP=i, DP=j, \hat{\text{OCL}}_{j-1}, P_j \}
\end{equation}

\begin{remark}
While signals like covariates and case estimates are easy to include here in the state space without much thought, there are many possible ways to make use of them in the design of the RL framework, such as inclusion in the state space or the reward space or both, and it is not obvious what the best approach would be. 
We have chosen to leave the inclusion of such additional reward signals to future work so that our formulation remains as foundational as possible, and to keep the paper to a reasonable length. 

Nevertheless, our formulation and implementation without these additional informative reward signals already performs remarkably well on simulated data. It is likely that this performance could be further improved.
\end{remark}

\subsubsection{Reward function $\mathcal R$} \label{reward formulas}
Finally, we discuss the components of the reward function $\mathcal R$ that the agent seeks to maximise. From a reserving perspective, the reward is designed to encourage reserve revisions that are accurate at settlement, while avoiding unnecessary volatility during periods with no new payment information. In particular, reserve changes are penalised unless justified by observed claim activity, reflecting the practical preference for stable reserves that adjust primarily when payments occur.

To construct reward terms, we will use the function $h(y, \hat y)=1-g(y,\hat y)$ ($y$ is the actual value, $\hat y$ is the predicted value), where 
\begin{equation}
    g(y, \hat y) = \frac{\lvert\hat y - y\rvert}{(\lvert y \rvert + \lvert \hat y \rvert)/2}
\end{equation}
is the summand of the sMAPE metric (symmetric mean absolute percentage error), and is used because it penalises under- and over-estimation equally (symmetry) and it is also bounded, which is desirable for RL reward formulations. The function $h$ simply maps $g$ to $[-1,1]$, which is preferable for RL.

The reward signal consists of an accuracy component calculated at settlement time $\tau=T_n$, as well as stability and smoothing components calculated in periods before settlement $\tau=1,...,T_n-1$. The high-level idea is that the accuracy component rewards how good the predictions were at each development period, the stability component rewards the agent the more it keeps its implied ultimate loss estimate stable, and the smoothing component, which progressively penalises large adjustments/actions in later periods to encourage the agent to ``make up its mind" early rather than leaving it till later.
\begin{itemize}
    \item \textit{Accuracy component $r_{\text{acc}}$:} At settlement of the claim $n$ $(\tau=T_n)$, We have
    \begin{equation}\label{E_racc}
    r_{\text{acc}} = C\cdot \frac{\sum_{\tau=1}^{T_n-1} \gamma^{\tau-1}\cdot h\left(\text{OCL}_\tau, \hat{\text{OCL}}_{\tau}\right)}{\sum_{\tau=1}^{T_n-1} \gamma^{\tau-1}},
    \end{equation}    
    where $\text{OCL}_\tau$ is the true OCL, and $\hat{\text{OCL}}_{\tau}$ is the predicted OCL as at $\tau$ periods since notification. The reason we divide by $\sum_{\tau=1}^{T_n-1} \gamma^{\tau-1}$ is to remove the imbalance of signal strength for longer vs shorter claims since notification. Just because a claim has been developing for a long time since notification doesn't mean it should be more important than a claim that hasn't been developing for a long time since notification.

    Note that intermediate true OCL values cannot be observed. As a consequence, $r_{\text{acc}}$ is evaluated at settlement using realised payments and the ultimate claim amount, and interpreted as a terminal reward that attributes estimation accuracy backward to the full prediction path $\hat{\mathrm{OCL}}_{1:T_n-1}$. The formulation above should therefore be understood as a path-based accuracy reward evaluated ex post, rather than as requiring intermediate targets in practice.

    In implementation, we multiply $r_\text{acc}$ by a (tuned) factor of $C$ to increase the size of the accuracy component relative to the subsequent smoothing and payment components.

    \item \textit{Stability component $r_{\text{stab},\tau}$:} We have, for claim $n$,
    \begin{eqnarray}
    r_{\text{stab},\tau} &=& -\alpha_{\text{stab},\tau}\left( \frac{|a_\tau|}{\ln K} \right)^2, \quad \tau=1; \label{E_rstab1}\\
    &=& \alpha_{\text{stab},\tau} \left[ \gamma h\left(\hat{\text{UL}}_{\tau}, \hat{\text{UL}}_{\tau-1}\right) - h\left(\hat{\text{UL}}_{\tau-1}, \hat{\text{UL}}_{\tau-2}\right) \right], \quad \tau=2,...,T_n-2; \label{E_rstab} \\
    &=& \alpha_{\text{stab},\tau} \left[ 0 - h\left(\hat{\text{UL}}_{\tau-1}, \hat{\text{UL}}_{\tau-2}\right) \right],\quad \tau=T_n-1. \label{E_rstablast}
    \end{eqnarray}
    where 
    \begin{equation}
        \alpha_{\text{stab},\tau} =
        \begin{cases}
        1, & \text{if there is no payment in period } \tau;\\
        0, & \text{if there are payments in period } \tau.
        \end{cases}
    \end{equation}
    The first \eqref{E_rstab1} $\in [-1,0]$ discourages overly large movements at the onset (recall that $K$ defined the action space bounds). The expression in square brackets in \eqref{E_rstab} follows the potential-based reward-shaping formulation proposed by \citet{Ng1999}, such that when added to the accuracy component of the reward, it will not change the optimal policy under only the accuracy reward component, since it cancels out as a telescoping sum; see also \eqref{E_rstablast}. As such, this is simply a reward shaping term that helps the agent reach the optimal policy more efficiently. Note that $r_{\text{stab},\tau}$ is formulated in terms of UL rather than OCL, as UL does not change over time, whereas OCL does.

    Operationally, this means that the agent is discouraged from revising reserves in quiet development periods, while being allowed to react more aggressively when payments provide new information.

    \begin{remark}
        
        The stability reward is motivated by potential-based reward shaping \citep{Ng1999}, in which differences of a potential function are added to the reward without altering the optimal policy. In our setting, however, the stability term is multiplied by a gating factor $\alpha_\text{stab}$ that switches off once claim payments occur. This modification breaks the strict theoretical invariance guarantees of potential-based shaping. We therefore use the shaping interpretation heuristically: the stability term is designed to improve empirical learning stability and prevent erratic reserve revisions in early development periods, rather than to preserve policy invariance in a formal sense.
    \end{remark}

    \item \textit{Smoothing component $r_{\text{smooth},\tau}$:} At development periods $\tau=1,2,...,T_n-1$ for claim $n$,
    \begin{equation}
    r_{\text{smooth},\tau}
    = -\alpha_{\text{smooth},\tau} \cdot \left( \frac{|a_\tau|}{\ln K} \right)^2,
    \quad
    \alpha_{\text{smooth},\tau}
    =
    \begin{cases}
    \min(1, \frac{m+1}{M}), & \text{if there is no payment in period } \tau;\\[4pt]
    0, & \text{if there are payments in period } \tau.
    \end{cases}
    \end{equation}
    
    where $m$ is the number of predictions made for the claim since notification, and $M$ is the tuned number of warm-up steps. While the agent tries to maximise the large but delayed lump-sum reward $r_\text{acc}$, $r_\text{smooth}$ encourages more stability in its adjustments by penalising any action, with less penalty on smaller actions through a quadratic reward term. However, we want to remove this penalty whenever there is a payment, so that the agent is free to make adjustments without penalty. Moreover, $\alpha_\text{smooth}$ ramps up from $0$ to $1$ over the first $M$ predictions so that the agent is encouraged to settle near the correct UL estimate early, so that it doesn't need to adjust much later on.

\end{itemize}

Both the stability reward $r_{\text{stab}}$ and the smoothing penalty $r_{\text{smooth}}$ discourage abrupt changes in predicted OCL, but they operate at different levels. The stability term penalises deviations relative to a reference ultimate loss estimate, whereas the smoothing term penalises period-to-period variability and enforces temporal regularity in the prediction path, that is, it increases the resistance to taking action (adjusting OCL) in later claim developments.

\subsubsection{Discount factor $\gamma$}
The discount factor $\gamma=0.99$ is arbitrarily chosen so that we value correct predictions now slightly more than correct predictions in the future, but not by too much; refer to \eqref{state-value func} and Section \ref{S_gamma1}.

\subsubsection{Example}

An example of calculations over the full length of one claim, as well as for an iteration of learning, is provided in \ref{A_example}.

\subsection{Initialisation} \label{Initialisation}

Initialisation of the RL model is an important aspect of the implementation. A poor initialisation can destabilise training and lead to the model failing to learn in time.

In practice, the reserving actuary will either have case estimates available as an initial value, or an estimate derived from the underwriting process (e.g., premium and expected loss ratio). Insurers generally have an a priori estimate of the expected cost of claims associated to a policy when they sell it.

However, as we are working with synthetic data, we don't have access to these priors. They may also be costly to obtain or derive. Hence, we develop an initialisation scheme based on historical data only. We develop a linear-credibility-like approach (similar to \citet{Bornhuetter1972}, but using an overall mean in place of an arbitrary prior) to initialise a claim with accident period $i$ using historical data of all claims settled by valuation calendar period of $T$. The historical data consists of all available \textit{settled} training claims.  

\subsubsection{Credibility mixing}
The idea is that we want to get an accident-period-specific mean $_0\text{UL}_{i}$ by mixing the empirical (individual) mean of the accident period $\bar{\text{UL}}_{i\Sigma}$ (row in a claims triangle), with the overall (collective) mean $\bar{\text{UL}}_{\Sigma\Sigma}$ (whole triangle):
\begin{equation}
    _0\text{UL}_{i} = z_{i} \bar{\text{UL}}_{i\Sigma} + (1-z_{i})\bar{\text{UL}}_{\Sigma\Sigma},
\end{equation}
where 
\begin{equation} \label{E_ULsumsum}
    \bar{\text{UL}}_{\Sigma\Sigma} = \sum_{i=1}^I w_i \bar{\text{UL}}_{i\Sigma}, \quad w_i=\frac{n_i}{n_\text{settled training claims}} = \frac{\text{no. settled training claims from accident period }i}{\text{no. settled training claims}}.
\end{equation}
In practice, the actuary could replace $\bar{\text{UL}}_{\Sigma\Sigma}$ with their a priori estimate of the ultimate claim size; we use \eqref{E_ULsumsum}. The actuary may even use their a priori estimate to initialise $_0\text{UL}_{i}$ directly.

The ``credibility'' weight $z_{i}$ should be large for earlier accident periods when we have more settled claims in the training set (more data, hence more credible), and decrease for later accident periods, where there may only be a few settled claims and so we must rely on the overall mean. Such an objective can be achieved by using the formula from \citet{Taylor2000}
\begin{equation}\label{E_Taylorz}
    z_{i}=\frac{1}{\pi_i},
\end{equation}
where $\pi_i$ is the (chain ladder) age-to-ultimate factor for claims with accident period $i$ at the valuation period $T$ for a cumulative paid triangle. Earlier accident periods will have smaller $\pi_i$ at valuation period $T$, and hence larger credibility $z_i$. 

\subsubsection{Adjustment to claims mix}

The credibility factor \eqref{E_Taylorz} stems from a vanilla chain ladder estimation procedure, which can present issues. A feature of the simulated data we use in this paper (outlined in Section \ref{Results}) is that longer claims progressively become larger in size over time, even in real terms. This is a problem because claims with later accident periods in the \textit{settled} training set will tend to be biased towards claims that settle quickly, and hence may grossly under-represent the potential size of the claim being initialised. This issue of a changing mix of claims leads to worse and worse underestimation for later accident period claims. It should be noted that the issue here is not necessarily that the initialisation is bad, but rather that the initialisation \textit{deteriorates} for more recent accident periods, and we can't expect RL to pick up on that since it's a feature of how the data was split. To address this issue, we adapt the PPCI (payments per claim incurred) triangle of \citet{Taylor2000}:
\begin{enumerate}
    \item Using the training set of settled claims by valuation period $T$, we construct a PPCI triangle of average claim amounts. Rather than the cumulative paid amount in each cell, we have the payments per claim incurred by dividing $C_{i,j}$ by the number of claims incurred to date $N_{i,j}$.
    \item For each AP $i$ in the PPCI triangle, we have an age-to-ultimate factor for the PPCI triangle $\pi'_i$ (where $^\prime$ is to differentiate from the cumulative paid triangle's age-to-ultimate factor from earlier), which ``develops" the PPCI to an ultimate level. We can use this to ``correct" for the difference in mixture of claims. Specifically,
    \begin{itemize}
        \item The AP mean should be adjusted to $\bar{\text{UL}}_{i\Sigma}^\text{adj} = \pi'_i \cdot \bar{\text{UL}}_{i\Sigma}$
        \item The overall mean $\bar{\text{UL}}_{\Sigma\Sigma}$ should also be adjusted to
        \begin{equation}
            \bar{\text{UL}}_{\Sigma\Sigma}^\text{adj} = \frac{1}{n_{\text{training claims}}}\sum_{i=1}^I \pi'_i \cdot \text{UL}_{i\Sigma}.
        \end{equation}   
    \end{itemize}
\end{enumerate}
The initialisation now becomes
\begin{equation}
_0\text{UL}_{i} = z_i\cdot \bar{\text{UL}}_{i\Sigma}^\text{adj} + (1-z_{i}) \bar{\text{UL}}_{\Sigma\Sigma}^\text{adj},
\end{equation}

and $_0\text{OCL}_i$ is simply $_0\text{UL}_i$ minus the cumulative paid at notification. 

If $_0\text{UL}_i$ is less than the cumulative paid, then $_0\text{OCL}_i<0$, which would condemn the RL algorithm to negative predictions forever after. Hence, we pragmatically initialise with a small (but not too small) $\bar{\text{UL}}_{\Sigma\Sigma}^\text{adj} / K^2$, where $K$ is the action space constraint. Essentially, this allows RL, in two periods, to scale the claim to get back to the average adjusted ultimate loss in two periods, or down to something really small.

\begin{remark}
The estimate $_0 \text{UL}_i$ could potentially be used to calculate an IBNR estimate in conjunction with a projection of counts, in a fashion similar to PPCI. This is a very rudimentary idea that is in line with how IBNRs are estimated for micro-level reserving in general, that is, by multiplying an expected severity by (expected) IBNR claim counts \citep[e.g.,][]{Antonio2012, Avanzi2015}.
\end{remark}

\subsection{RL Algorithm Choice} \label{RL Algorithm Choice}

The PPO and SAC algorithms described in \ref{RL lit review} both support a continuous action space.  The most significant difference is that PPO is an on-policy algorithm, while SAC is off-policy. This means that the SAC is able to keep a replay buffer of past experience, which makes it more sample efficient. Given that we are working with potentially limited data for reserving purposes, we decided to use the SAC algorithm.

\section{Data Splitting for Evaluation and Tuning} \label{train_test_split}

In this section, we thoroughly develop the procedure of splitting the data for validation and testing purposes. This results in a novel approach born from the intersection of RL and micro-level reserving. 

Traditionally, RL does not have a concept of train-test-split. There usually is no need to carefully define a training, validation and test set because RL applications often stem from games or other contexts where data scarcity is not a concern. In contrast, reserving data is often scarce. Moreover, the reserving ``game'' faces changing rules (claims characteristics and their payment patterns) over time. Therefore, we needed to develop our own bespoke approach, called ``rolling settlement'', because existing approaches in the reserving literature for splitting the data either do not respect the temporal nature of the data, or do not utilise RL's advantage of being able to learn from open (yet to settle) claims. 

There are really two (related) splitting procedures to discuss in this section. Focusing on model evaluation (on a test set) first, we precisely define and discuss the existing techniques for splitting the overall data into a training and test set that we are aware of. We are unaware of such a comprehensive summary and comparison in the existing literature. Focusing on hyperparameter tuning next, we explain and argue that none of the existing techniques used in reserving are appropriate for RL hyperparameter tuning. Consequently, we develop a new approach, which we have called ``rolling settlement".

\subsection{Train-Test-Split for Model Evaluation}

Let us first define a list of data splitting techniques  for ease of comparison and reference in Table \ref{tab:splitting}.

\begin{table}[htb]
\centering
\begin{threeparttable}
\caption{Splitting Approaches: given $n$ claims in a dataset, possible ways to partition the claims}
\label{tab:splitting}
\renewcommand{\arraystretch}{1.15}

\begin{tabularx}{\linewidth}{l X}
\toprule
Splitting Technique & Description \\
\midrule
Na\"ive Split by Claim (NSC) &
Randomly split the $n$ claims into $k$ groups (e.g., by claim ID).  
This ignores the temporal nature of development and leads to data leakage. For example, if $k=2$ such that we naively split the data into a training and test set in this way, the training set would include information (e.g., the ultimate loss) that in practice would be from the future, and so would be unrealistic. \\

Censored Split by Claim (CSC) &
First split by calendar time into $k$ groups; within each group keep only claims that \textbf{settle} by the end of that group.
Example: if calendar periods $[1,80]$ are split into $[1,40]$ and $[41,80]$, then claims settling within $[1,40]$ form the training set, while the rest form the test set. \\

Temporal Split (TS) &
A temporal split involves splitting by calendar time into $k$ groups, similar to the CSC, but this time, the developments of a claim are assigned to the calendar time that it falls into.  
For example, a claim with accident period $35$ settling in period $50$ contributes its developments up to $40$ to the training set $[1,40]$, while developments in $[41,50]$ belong to the test set. \\
\bottomrule
\end{tabularx}
\end{threeparttable}
\end{table}

The key distinction between CSC and TS is that when targeting the OCL, the former is suitable for supervised methods, while the latter is suitable for RL such that a claim can be partitioned and attributed to multiple splits. TS is natural for RL because it takes advantage of RL's ability to learn from still-developing claims (whereas they are useless for supervised methods as the target is unknown). CSC is suitable for supervised methods targeting the OCL, but not for RL, since it removes open claims and hence does not allow RL to work with the still-open claims. To the best of our knowledge, in the current literature, TS is seen only when targeting a sequence of payments rather than the OCL.

For our approach in this work, we construct the train and test set by splitting the overall data using a TS for RL, and a CSC for the feed-forward neural network. Formally, let $t=1$ denote the accident period of the earliest claim (calendar period 1), and $t=M$ denote the accident period of the latest starting claim in the dataset (calendar period $M$) and $N\geq M$ to be the latest possible calendar period observed (determined by the maximum observation periods after accident), so that all claims start in a calendar period $t\in \{1,2,...,M\}$, and have settled by calendar period $N$. For our dataset, $M=40$ and $N$ is technically unbounded. We decide that the time interval $\{1,2,...,M\}$ will constitute the training set, and $\{M+1,..., N\}$ will constitute the test set. That is, we are pretending that the insurer is currently at calendar period $M=40$, and would like to predict the OCL for all open claims at the current time.

\subsection{Rolling Settlement Approach to Hyperparameter Tuning} \label{validation}

We next consider data splitting for the purposes of hyperparameter tuning. The innovation of this section is to motivate and develop the extension of temporal splitting to incorporate validation, not just train-test-split. We first discuss what is done in the RL literature, then consider several alternatives from the reserving literature and why they are not suitable for us, and finally, we present the rolling settlement approach that we use for our RL reserving model.

\begin{remark} \label{R_validation}
``Validation" in this section pertains to hyperparameter tuning purposes. This is not to be confused with validation for variance reduction purposes to prevent supervised methods from overfitting over too many training epochs. We will discuss early stopping after developing the rolling settlement approach for hyperparameter tuning purposes.
\end{remark}

\subsubsection{Validation in the RL Literature}

Hyperparameter tuning is generally an afterthought in the RL literature, because many problems being tackled (e.g., games and robotics) are such that the cost of interacting with the environment is negligible (or there are simulations of the environment). In these cases, multiple agents with different hyperparameter combinations are trained on different instances of the environment (different seeds), and the highest average performance configuration is chosen \citep{schulman2017b, Haarnoja2018}.

There are also some other novel approaches like population-based training \citep{Jaderberg2017}, and meta-gradient approaches \citep{Xu2018} that could be explored, but we have decided to leave this for potential future work, as this is not the central focus of this paper.

\subsubsection{Validation in the Reserving Literature}
\medskip
\textbf{Cross-Validation (CV)}

Traditionally, for non-time-series data, one performs $k$-fold cross-validation for hyperparameter tuning \citep{Kohavi1995}. This approach would involve a naive split into $k$ groups called folds, iterating through each fold as the validation set, training on the remaining $k-1$ folds, and taking the average loss through all folds as the score for a given hyperparameter configuration. 

However, this uses the NSC approach, which leads to temporal leakage.

Furthermore, cross-validation would split the training set into folds by claim ID or transaction rather than time, which isn't the same problem as what we are tackling in reality. In reality, we want to use \emph{past} claims data to predict the current OCL (\emph{future} payments). This inherently motivates a temporal split. However, cross-validation necessitates a split by training observation (claim ID or transaction time), which means that the problem is actually using some claims data to predict other claims, without respecting the temporal nature of the claims. Because of this mismatch, this would mean the best hyperparameter combination we find is not guaranteed to actually be best for the real problem at hand, even notwithstanding the bigger issue of temporal leakage.

This issue motivates the next approach of censored hold-one-out validation.

\newpage

\textbf{Censored Hold-One-Out Validation (CHOOV)}

To alleviate these issues, one may consider splitting the training set by claim settlement time as in the CSC approach. The entire calendar time of the training set $\{1,2,...,M\}$ is partitioned into two parts by time $t_1$. That is, claims that settle in $\{1, ..., t_1\}$ comprise the training set and $\{t_1+1,M\}$ comprise the validation set. This is essentially applying the CSC technique twice. In this way, no future information is leaked to the past splits. \citet{avanzi2025} adopt this approach.

Specifically, we illustrate how this would look for RL and supervised learning methods to help build up to the rolling settlement approach:

    \begin{itemize}
    \item For supervised methods, it is a straightforward CSC, where only claims settled in $[0,t_1]$ comprise the training set. If a training claim settles over 10 development periods, then it contributes 10 samples to the training set. Once the model is trained, it is fed claims development data spanning $[0,t_1]$ for claims that settle in $(t_1, M]$. This is repeated for different hyperparameter combinations, and the best-performing combination is selected.
    
    \item For RL, the difference lies in that we can extend the hold-one-out validation approach to use TS. The training set now contains all claim developments taking place during $[0, t_1]$ rather than only the claims that settle during this time, since the model is able to learn without needing a target. The ensuing steps remain the same.

    \end{itemize}

However, taking this approach sacrifices the averaging of folds that CV does, which may make the hyperparameter tuning less robust.

Note that the ``one" held out here refers to \emph{one fold} being held out and used for validation, not that one claim is being held out.

\textbf{Rolling Origin Validation (ROV)}

The rolling origin validation approach \citep{Tashman2000} respects the temporal ordering of time series data. Suppose the full calendar time of the dataset $[1, N]$ is split such that the training set comprises claims starting in $[1, t^*]$ ($t^*=M$ from our earlier notation is often called the valuation date, but in the context of rolling origin, $t^*$ is often used). The approach starts by defining an initial training ``window" (or fold 1), as $[1, t_0]$ with $t_0<t^*$, to be used to fit the model. Then $(t_0, t^*]$ is used to validate.

Then, for the expanding window approach, for fold $i$, train on $[1, t_0 + w(i-1)]$, where $w$ is the width of each window expansion. Correspondingly, validate on $(t_0+w(i-1), t^*]$. The average loss across all folds is then used to choose the best hyperparameter configuration.

There is also a sliding window approach, where the width of the training window is fixed, and slides along rather than expanding for each fold.

This approach is, however, only used with aggregate triangles in the literature \citep{Balona2020, Al-Mudafer2022}. Hence, ``origin" refers to the origin of \textit{payments} in the triangle (diagonals), rather than the origin of claims (rows). Each successive ``fold" includes new diagonal(s) compared to the previous fold in the training set, such that we effectively ``roll forward" in time. This makes sense in an aggregate triangles context, but for our micro-level reserving problem where we are working at a claims-level, ``origin" can only refer to either the accident period or the period of notification for the claim. In both cases, we cannot split based on the ``origin" of the claim, because it does not consider the future developments of the claim. This motivates development of a rolling \emph{settlement} approach.

\subsubsection{Rolling Settlement Validation (RSV)}

The idea of the ROV approach is suitable for our purposes, except that in the micro-level reserving context, it makes sense to split by the settlement time rather than the ``origin" (accident period or period of notification), so that we respect the development of the full claim instead of just when it originated. Hence, we term our approach \textit{Rolling Settlement Validation} (RSV) instead.

The idea of rolling settlement is essentially repeated applications of TS in the case of RL, and CSC for supervised methods. 

For the train-test-split, recall that we have decided to split the entire calendar time of the dataset $[0,N]$ into the training set $[0,M]$ and the test set $(M, N]$. For our dataset, $M=40$. The goal now is to describe the rolling settlement approach, which acts on the training set.

\begin{figure}[htb]
    \centering
    \includegraphics[width=.7\linewidth]{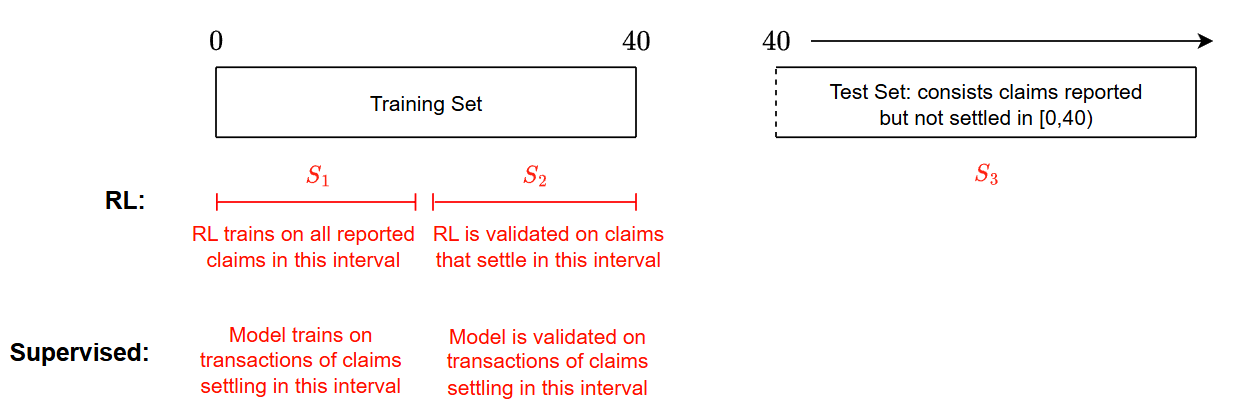}
    \caption{RL vs supervised methods illustration for the rolling settlement approach}
    \label{fig:rolling_settlement}
\end{figure}

We can partition the training time interval $\left[ 0, M\right]$ into $k$ equal length time intervals $S_1,..., S_{k}$. We need to know the actual OCL for claims to evaluate performance, so for RL, each interval should include only claims that settle during the next immediate interval; see Figure \ref{fig:rolling_settlement}, and supervised method folds should contain claims settling in that interval. Therefore, for each hyperparameter combination, we train on $S_1$ to predict the OCL for development periods of claims that settle in $S_2$, then train on $S_1+S_2$ and predict for $S_3$, and so on. The $k$ results are averaged to compare across all hyperparameter combinations.

This is best understood by referring to the simple diagram in Figure \ref{fig:rolling_settlement}, which summarises both splits for evaluation and tuning purposes. For simplicity and because we are working with a small dataset, we choose $k=3$ (Figure \ref{fig:rolling_settlement} depicts $k=2$).

\begin{enumerate}
    \item For RL, we are simply applying one more temporal split on the training set $[0,40]$. All claims that are reported (known) in $S_1$ are used to train, including still-open claims, and the performance is evaluated by looking at claims that were notified in $S_1$ and have settled in $S_2$.
    \item For supervised methods such as a feed-forward neural network, we take each successive sequence of transactions for claims settling in $S_1$ to train on, then evaluate on all sequences of claims that are settled in $S_2$ to train the model ($S_2$ is now the ``validation'' set in the sense of Remarks \ref{R_validation}--\ref{R_validation2}).
\end{enumerate}

There are two important considerations:
    \begin{itemize}
    
    \item There is a limit to how large $k$ can be depending on the size of the data, since we should have a sufficient number of claims settling in each interval $S_1,...,S_k$ for the validation results to be stable.
    
    \item If the data spans a long enough time frame such that there is possibly non-stationarity across calendar time, then $k$ needs to be smaller (or discard earlier times) so that the initial folds aren't completely outdated.
    \end{itemize}

\begin{remark} \label{R_validation2}
As already mentioned in Remark \ref{R_validation}, it is important, in many machine learning applications, to implement early stopping in both training and validation for variance reduction and to avoid overfitting across too many epochs (complete passes over the training data). However, this doesn't apply to RL since it only takes one pass through the training data. For supervised methods, this is done easily by an 80-20 split of the rolling-settlement fold, where the 20\% sub-fold can be used for early stopping. In our case of Figure \ref{fig:rolling_settlement}, we would train on the first 80\% of $S_1$ and use the remaining 20\% of $S_1$ for early stopping, then evaluate on claims that settle in $S_2$ for the validation results.
\end{remark}

\section{Downward Bias Adjustment} \label{Downward Bias Adjustment}

While individual-claim accuracy is important, actuarial reserving ultimately requires aggregate consistency at the portfolio level. In the absence of an explicit constraint or weighting mechanism, models optimised for average claim-level accuracy may systematically understate total OCL due to the rarity of large claims, much like class imbalance issues in classification tasks. We observed in our experiments that models suffer from a large and consistent downward bias, whereby they underestimate the aggregate OCL at the portfolio level. We further discuss the possible underlying reasons in detail in \ref{downward bias appendix}. 

We propose an \textbf{OCL importance weighting} idea to address the issue. This approach can be interpreted as aligning the learning objective with a materiality-weighted loss function, where errors on larger claims carry proportionally greater financial significance.
From this perspective, the weighting scheme is not an ex post correction, but an explicit articulation of the actuarial objective that portfolio-level estimates should be approximately unbiased in expectation.

A natural idea to address the rarity of large claims problem is to simply assign greater importance to larger claim developments by weighting claims proportionally to their true OCL. The next two sections implement this idea for supervised methods and RL, respectively.

It is worth emphasising that we are implementing the same fundamental OCL importance weighting idea for both RL and supervised learning methods, though there are some nuances that arise for RL when dealing with RBNS (open) claims.

\subsection{Bias correction for supervised methods}

For supervised methods, we can optimise a \textit{weighted} MSE objective
\begin{equation} \label{fnn_weighted_mse}
    \frac{\sum_{k=1}^K w_k(\hat{\text{OCL}}_k-\text{OCL}_k)^2}{\sum_{k=1}^K w_k},
\end{equation}
where $\text{OCL}_k$ is the true OCL and $\hat{\text{OCL}}_k$ is the predicted OCL for the $k$th \textbf{training} observation (where an observation is specified by a $(\text{claim}, \text{development period})$ tuple). All we have added here is the weight
\begin{equation}\label{E_wksup}
    w_k=\exp\left( \alpha\ln  \frac{\text{OCL}_k}{s} \right), \quad s=\bar{\text{OCL}},
\end{equation}
where $s$ is the scaling factor, and the idea is that larger-than-average OCL steps should be assigned more weight than 1. The (tuned) temperature $\alpha$ controls the convexity, and tapers the growth of the weight for increasing OCL if set to be less than one.

\subsection{Bias correction for RL}

For RL, the correction is achieved by simply increasing the size of the accuracy reward signal for claim developments with larger OCL. The multiplier $w_\tau$ acts on each development period of the claim, to ``inflate" $r_\text{acc}$ for claims with more high-OCL-periods. This idea of using the reward magnitude to control where the agent pays more attention to is well-founded in the RL literature to deal with class imbalance issues \citep[see, e.g.][and the references therein]{Lin2020}. Hence, \eqref{E_racc} becomes
\begin{equation} \label{r_acc_adj}
    r_{\text{acc}} = C\sum_{\tau=1}^{T_n-1} {\color{red} w_\tau} \cdot \frac{\gamma^{\tau-1}}{\sum_{\tau=1}^{T_n-1} \gamma^{\tau-1}}\cdot h\left(\text{OCL}_\tau, \hat{\text{OCL}}_{\tau}\right),
\end{equation}
where
\begin{equation} \label{w_tau}
    w_\tau = \begin{cases}
        \exp\left( \alpha \ln \frac{\text{OCL}_\tau}{s} \right), & \text{if the claim is settled in the training set};
        \\
        \left( \frac{\max(P_{\tau_\text{curr}}, \;_0\text{UL}) - P_\tau}{s} \right)^\alpha, & \text{if the claim is not settled in the training set};
    \end{cases}
\end{equation}
and where $\tau_\text{curr}<T_n$ denote the current periods since notification for the claim at the valuation (cutoff) date, $P_{\tau}$ denote the cumulative paid amount at the $j$th period since notification, and $_0\text{UL}$ is the initial ultimate loss value (see Section \ref{Initialisation}).

The multiplier $w_\tau$ is defined in the same way as for supervised methods for settled claims in \eqref{E_wksup}. The caveat here with RL is that for claim developments seen during training for claims that do not settle during the training interval, we cannot set $w_\tau$ in the same way since the true OCL is not yet known. 
Instead of naively setting $w_\tau=1$ for these claims, we can do something more informative as in the second part of \eqref{w_tau}. The idea is that the last known cumulative paid amount for the claim is $P_{\tau_\text{curr}}$, so we know that the ultimate claim size for the claim should be at least this much. By using this as the estimate of ultimate, we can get a lower bound on the true OCL for all previous developments $\tau<\tau_\text{curr}$ to use for assigning a more informative weight. The issue with just using $P_{\tau_\text{curr}} - P_\tau$ as the estimate of true OCL, however, is that for immature claims and/or long-tailed claims, $P_{\tau_\text{curr}}$ can be small, and hence underestimate the ultimate. This would mean the weight $w_\tau$ under-emphasises the importance of the claim. To reduce the impact of this shortcoming, we will instead use our initial estimate of claim size $\text{UL}_0$ in these cases.

\begin{remark}
    Note that we do not further normalise with $w_\tau$ in \eqref{r_acc_adj}, that is, the definition of $r_\text{acc}$ does not have $\gamma^{\tau-1}  w_\tau$ in the denominator because this would erase the effect of scaling between claims (small and large claims would have comparable $r_\text{acc}$). 

\end{remark}

\begin{remark}
    We also considered using a conditional value-at-risk (CVaR) objective \citep{Rockafellar2002}. Instead of assigning larger weights to larger OCLs, one then takes the perspective of controlling the tail risk of mis-estimation. This was the objective used by \citet{Dong2025} in their RL formulation. There are several reasons for choosing the OCL importance weighting idea over a CVaR objective. Firstly, importance weighting is a simpler method, and it also ostensibly produces better results in our tests. Additionally, CVaR controls for the tail risk, which is not aligned with our objective of targeting the central estimate.
\end{remark}

\section{Evaluation of Model Performance} \label{Evaluation}

We briefly discuss here benchmark models, as well as the evaluation metrics we use to evaluate and compare model performance.

\subsection{Benchmark models}

We will compare RL with a feed-forward neural network (FNN) as well as the chain ladder. Details for these models are discussed in \ref{benchmark models appendix}, but we mention here that the chain ladder approach must be stripped of IBNRs to be comparable to individual claims reserving models, for which we adopt the procedure of \citet[described in \ref{A_CL}]{Delong2020}.

\subsection{Evaluation Metrics} \label{eval_metrics}

We will use two metrics to compare the model performance of RL against FNN and chain ladder.
\begin{enumerate}
    \item \textbf{Relative OCL:} Focusing on aggregate performance, we define
    \begin{equation}\label{E_relOCL}
        \text{relative OCL} = \frac{\text{predicted OCL}}{\text{true OCL}}.
    \end{equation}
    It should be made clear that this metric is aggregated at some pre-specified level. For example, if the level specified is the entire test set, then the relative OCL is calculated by aggregating the predicted and true OCLs for all claims, and taking the ratio. If the level specified is per accident period, then the predicted and true OCLs are aggregated, and the quotient is taken for claims in each accident period. 

    \item \textbf{RMSE:} Focusing now on (average) performance at the individual claim level, we will consider the
    \begin{equation}
        RMSE = \sqrt{\frac{1}{N} \sum_{n=1}^N (\hat{\text{OCL}}_n - \text{OCL}_n)^2},
    \end{equation}
    where $n$ denotes the $n$-th claim, $\hat{\text{OCL}}_n$ denotes the predicted OCL for the claim, and $\text{OCL}_n$ denotes the true OCL. Of course, we want the RMSE to be as small as possible (if seeking the mean); a zero RMSE would be achieved by a ``saturated" model.
\end{enumerate}

It should be said that the primary metric of interest is the relative OCL, as we first and foremost want the predictions to be accurate at an aggregate level. The RMSE complements our analysis by giving insight into how the model fares at an individual claims level \textit{given} some level of aggregate performance.

\subsection{Hyperparameter tuning and subsequent training and testing}

Choosing an optimality criterion for the hyperparameters deserves some thought. A first idea would be to use the weighted MSE \eqref{fnn_weighted_mse}. However, it includes the hyperparameter $\alpha$ which we would like to tune. Even though one could use a fixed point type approach, we prefer avoiding such a manual procedure (which additionally has no guarantee of reaching an equilibrium).

Another option (which we decided to implement here) is to focus on the aggregate performance measured by the ratio \eqref{E_relOCL}; more specifically, to choose the set of hyperparameters that leads to a predicted (overall) OCL that is the closest to its actual value. This `aggregate' focus aligns nicely with the real problem at hand. Note that each set of hyperparameters leads to a single value of \eqref{E_relOCL}, but this ratio is the (scaled) sum of all our predictions, and hence is likely to be relatively stable (and symmetrical) due to the central limit theorem. Nevertheless, we did try to tune hyperparameters using \eqref{fnn_weighted_mse} and $\alpha=1$, and obtained very similar (in some cases such as SPLICE complexity 5, identical) results.

Following standard practice, after the best set of hyperparameters is obtained, we fix them and do one last training pass through the full training set to obtain the final model to use for testing. Finally, for testing purposes, we ``freeze'' the RL model so that it does not perform exploration; refer to Remark \ref{A comment on the Exploration vs Exploitation Trade-off} for a more detailed discussion on this.

\section{Empirical results on CAS and SPLICE synthetic data}

Models are first evaluated on 5 simulated property casualty insurance datasets provided online by the Casualty Actuarial Society \citep{CAS2025}. While we observe an extremely good performance by RL on these datasets, we cannot provide deep analysis, as we don't have any information on the underlying claim dynamics behind the data. Therefore, we proceeded to conduct a simulation study using data generated using the SynthETIC \citep{Avanzi2021,R-SynthETIC} and SPLICE \citep{Avanzi2023,R-SPLICE} generators. SynthETIC is a continuous-time individual claims experience simulator, that, for each claim, simulates a claim occurrence date, the total claim size, the notification and settlement dates, and partial payment amounts and distribution. SPLICE extends this to also include case estimates. We use simulated data for two reasons: firstly, there is no real publicly available transactional-level individual claims data, and secondly, it allows us to examine the performance of models across multiple datasets with the same underlying mechanisms. For details on the data structures for the CAS and SPLICE datasets, refer to \ref{Data Description}.

The CAS datasets are quite small, containing only 5 datasets spanning 16 periods (years), with each dataset containing around 1700 claims in total. The datasets contained unsettled claims and non-claims (claims with zero loss), which we removed. The CAS website states that the claim activity can be affected by a variety of changes that are common problems faced by members of CAS. The five datasets correspond to one or a combination of those challenges. 

For our simulation study using SPLICE, we simulated 30 small datasets with parameters of on average 2500 claims per period, and 40 accident periods with both complexity 1 and complexity 5 (maximum complexity) settings. Complexity 1 is the simplest type of data (no base or superimposed inflation), where chain-ladder is the perfect model. We use this to benchmark in particular against the chain ladder to demonstrate the feasibility of the RL approach. Complexity 5 is the most complex type of data that breaches assumptions underlying the chain-ladder, including a structural break at calendar period 20, and varying levels of inflation. In fact, it was designed to have a broad resemblance to the experience of a real Auto Bodily Injury portfolio. For more details on what the underlying mechanisms are for each level of data complexity, we refer to Table 4 in \citet{Avanzi2023}.

To evaluate the performance, we selected a valuation (cut-off) period of 15 and 40 for the CAS and SPLICE datasets, respectively. This means that all transactions prior to the end of year 15 and quarter 40 for CAS and SPLICE, respectively, are in the training set, while unsettled claims by the valuation date constitute test claims.

Below, we present several results based on the data complexity and model comparisons. We slice the data in several ways to classify results in terms of \textit{maturity} of each claim, where mature claims correspond to claims with earlier accident periods/later periods since notification, and immature claims correspond to claims with later accident periods/earlier periods since notification.

To this end, it is more important to first understand the share of the outstanding OCL dollars by accident period and periods since notification to understand where the bulk of the outstanding liability lies. In particular, accident periods up to and including 25 account for only roughly 20\% of all outstanding claims in the SPLICE data. As a result, the vast majority of the liability (roughly 80\%) is associated with the 15 last accident periods. Observations are similar in the CAS datasets, although they are lighter-tailed; see \ref{cum share OCL graphs appendix}. Furthermore, case estimates of immature claims are less likely to be accurate, which further stresses how important their estimation is. All in all, the prediction performance of immature claims is the most material, and we will focus on those in our analysis.

\begin{figure}[htb]
    \centering
    \includegraphics[width=1\linewidth]{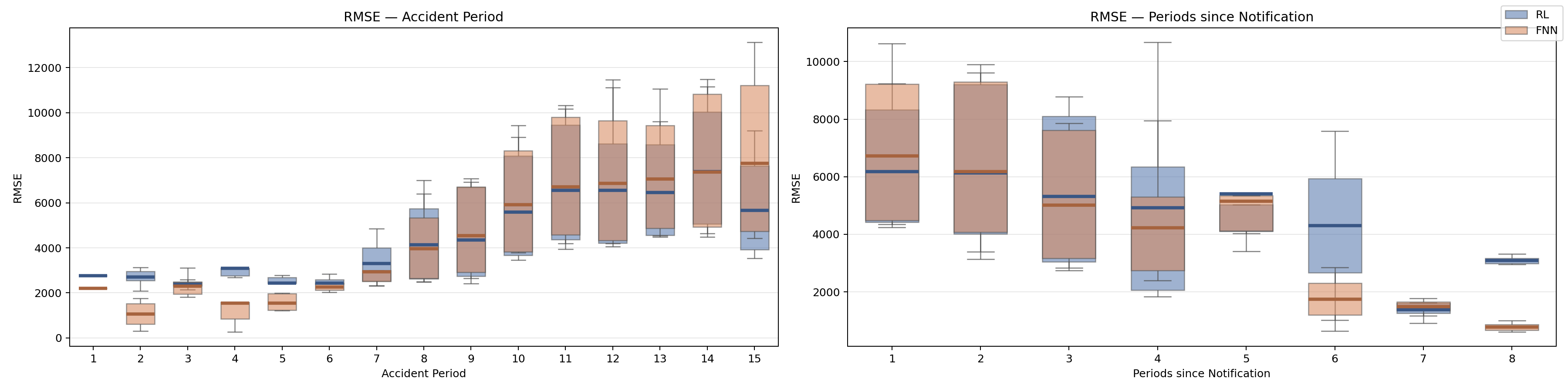}
    \caption{RMSE Performance of RL and FNN on CAS Test Data}
    \label{fig:rmse_cas}
\end{figure}

\begin{figure}[htb]
  \centering
  % first image
  \begin{subfigure}{1\linewidth}
    \centering
    \includegraphics[width=\linewidth]{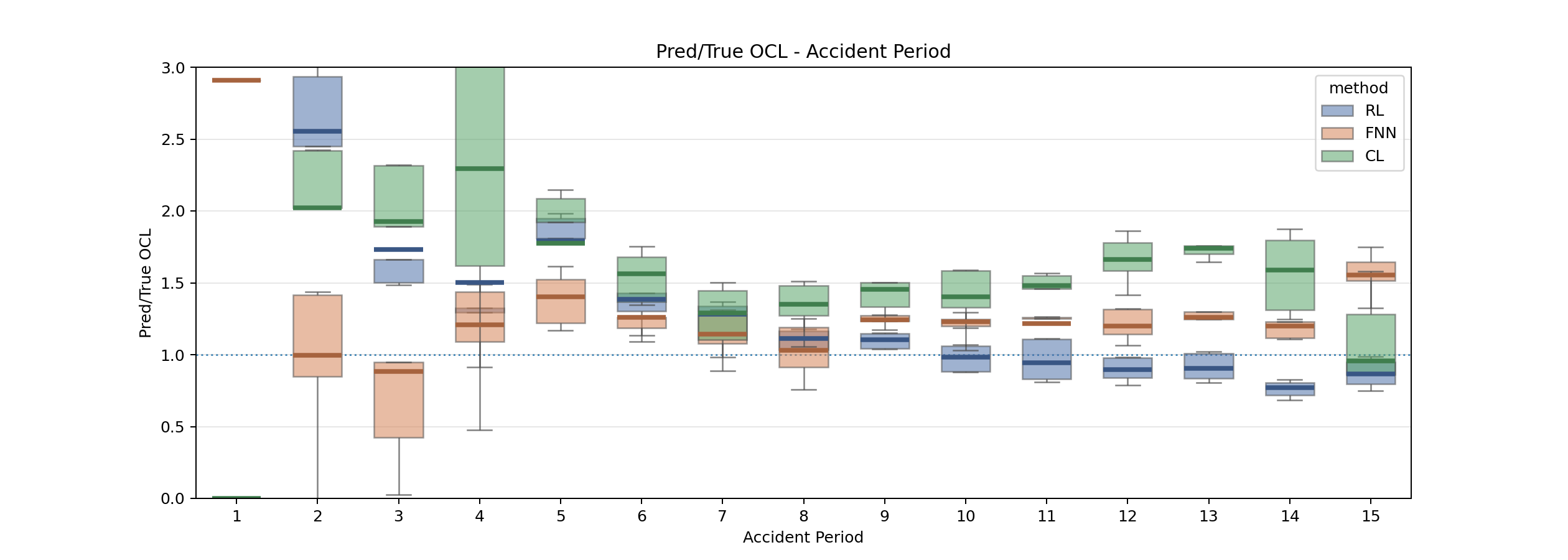}
    \caption{Boxplots of the relative OCL by accident period}
    \label{fig:Boxplots of the relative OCL by accident period CL CAS}
  \end{subfigure}\hfill
  % second image
  \begin{subfigure}{.64\linewidth}
    \centering
    \includegraphics[width=\linewidth]{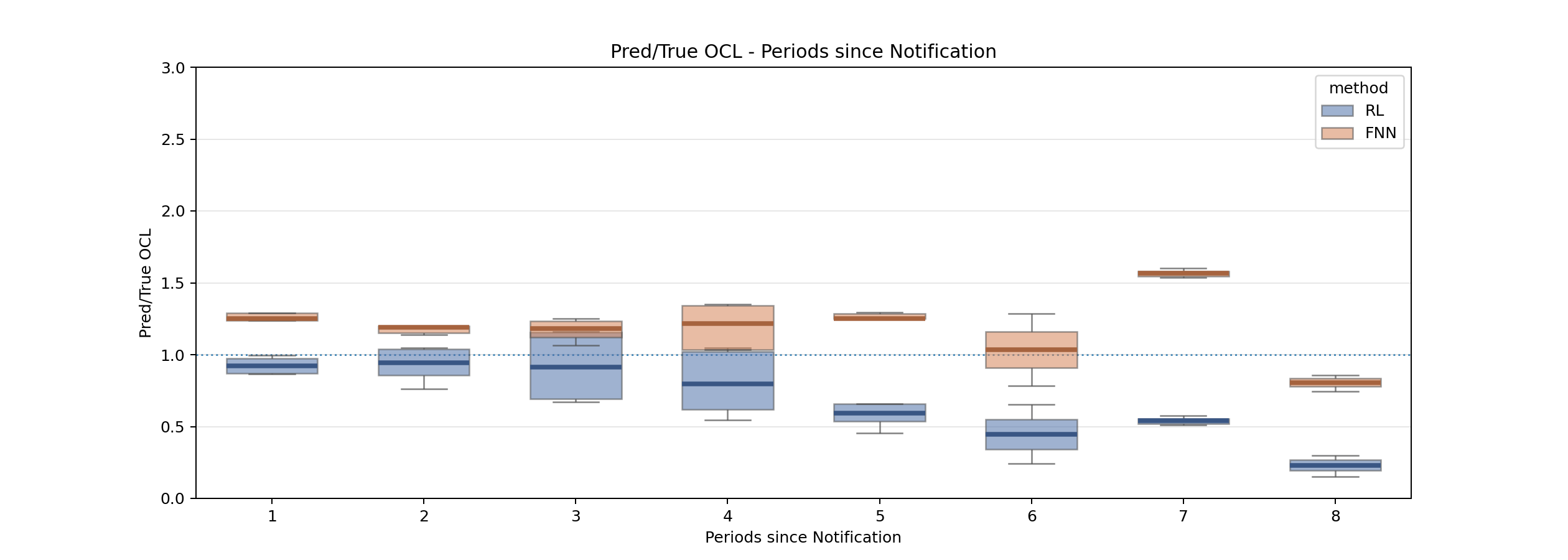}
    \caption{Boxplots of the relative OCL by periods since notification}
    \label{fig:Boxplots of the relative OCL by periods since notification CL CAS}
  \end{subfigure}
  % third image
  \begin{subfigure}{0.35\linewidth}
    \centering
    \includegraphics[width=\linewidth]{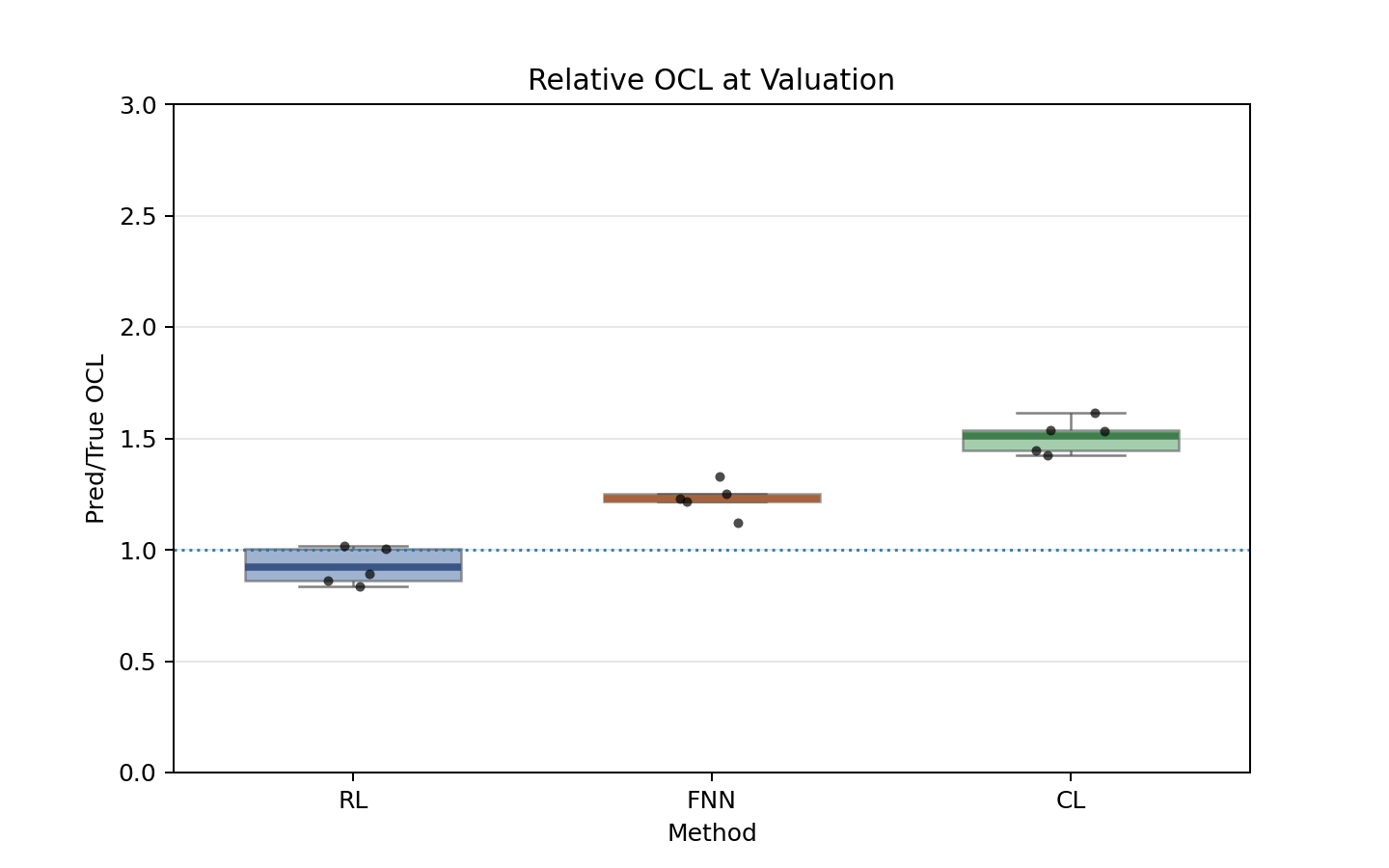}
    \caption{Boxplots of the aggregate relative OCL across all periods}
    \label{fig:Boxplots of the aggregate relative OCL across all periods CAS}
  \end{subfigure}

  \caption{Relative OCL Performance of RL, FNN, and CL on CAS Test Data}
  \label{fig:CAS CL}
\end{figure}

\subsection{Performance of RL and Benchmark Models on CAS Test Data} \label{Results_CAS}

We first briefly describe the graphical results. Figure \ref{fig:Boxplots of the relative OCL by accident period CL CAS} slices the data by the accident period (AP) of claim, and plots the relative OCL in aggregate for all claims with the same accident period. Figure \ref{fig:Boxplots of the relative OCL by periods since notification CL CAS} depicts the relative OCL by periods since notification (PSN) slices. Similarly, Figure \ref{fig:rmse_cas} depicts the RMSE by the AP and PSN slices. Finally, Figure \ref{fig:Boxplots of the aggregate relative OCL across all periods CAS} shows the aggregate relative OCL performance across all claims in the test set. Note that we have modified the boxplots to display the \textit{mean} rather than the median, as this corresponds to the central estimate required in most jurisdictions.

RL performs remarkably well on the CAS datasets, with an essentially perfect aggregate performance for two out of the five CAS datasets. We also observe the desirable property that performance is good for immature claims, with RL predicting quite accurately for the accident periods 8-15, which accounts for $\approx 90\%$ of the cumulative outstanding OCL. This is promising because, as we discussed earlier, immature claims are where predictive models can add a lot of value over case estimates. Slicing the data by periods since notification view in Figure \ref{fig:Boxplots of the relative OCL by periods since notification CL CAS} is similarly impressive, with performance deteriorating substantially only for latter four periods. 

\begin{table}[htb]
  \centering
  \begin{tabular}{lccc}
    \toprule
    Metric & RL & FNN & CL \\
    \midrule
    Average Relative OCL & 92.26\% & 123.00\% & 150.99\% \\
    Average RMSE         & 4698 & 4528 & \\
    \bottomrule
  \end{tabular}
  \caption{Model comparison at valuation: RL vs FNN vs CL on CAS Test Data}
  \label{tab:rl-fnn-cas}
\end{table}

We observe that the chain ladder (stripped of IBNRs) is failing on the CAS dataset, which indicates that the underlying dynamics of the CAS data do not adhere to chain ladder assumptions. The FNN posts a respectable performance in aggregate, but noticeably falls short for the important group of immature claims, particularly for accident period 15 and the first period since notification.

In terms of individual claim performance, RL and FNN have comparable RMSEs; refer to Table \ref{tab:rl-fnn-cas} and Figure \ref{fig:rmse_cas}. It stands that RL has a desirable performance in the sense that it is equal or better than FNN for the important immature claims, though worse for the mature claims.

As we don't know the underlying dynamics of the CAS dataset, it is difficult to derive further insights with these results. Therefore, we shift our attention to a simulation study using SPLICE-generated data in Section \ref{Results}.

\newpage

\begin{figure}[htb]
  \centering
  % first image
  \begin{subfigure}{\linewidth}
    \centering
    \includegraphics[width=\linewidth]{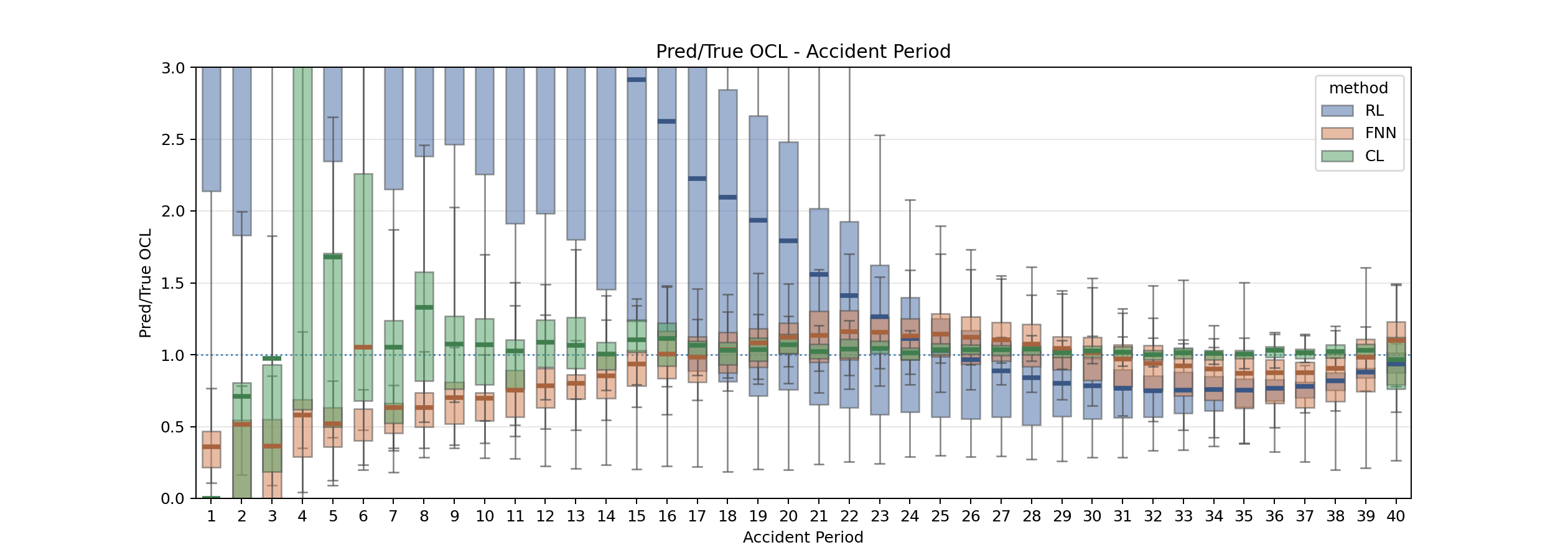}
    \caption{Boxplots of the relative OCL by accident period}
    \label{fig:Boxplots of the relative OCL by accident period CL}
  \end{subfigure}\hfill
  % second image
  \begin{subfigure}{\linewidth}
    \centering
    \includegraphics[width=\linewidth]{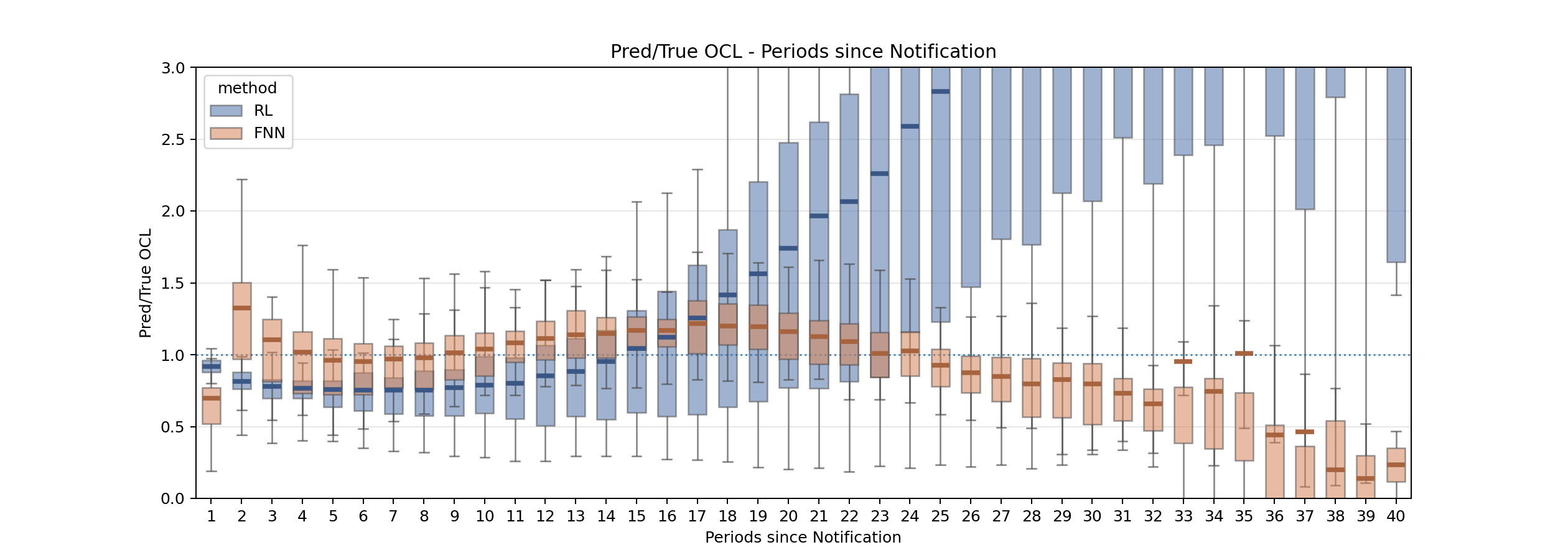}
    \caption{Boxplots of the relative OCL by periods since notification}
    \label{fig:Boxplots of the relative OCL by periods since notification CL}
  \end{subfigure}
  % third image
  \begin{subfigure}{0.6\linewidth}
    \centering
    \includegraphics[width=\linewidth]{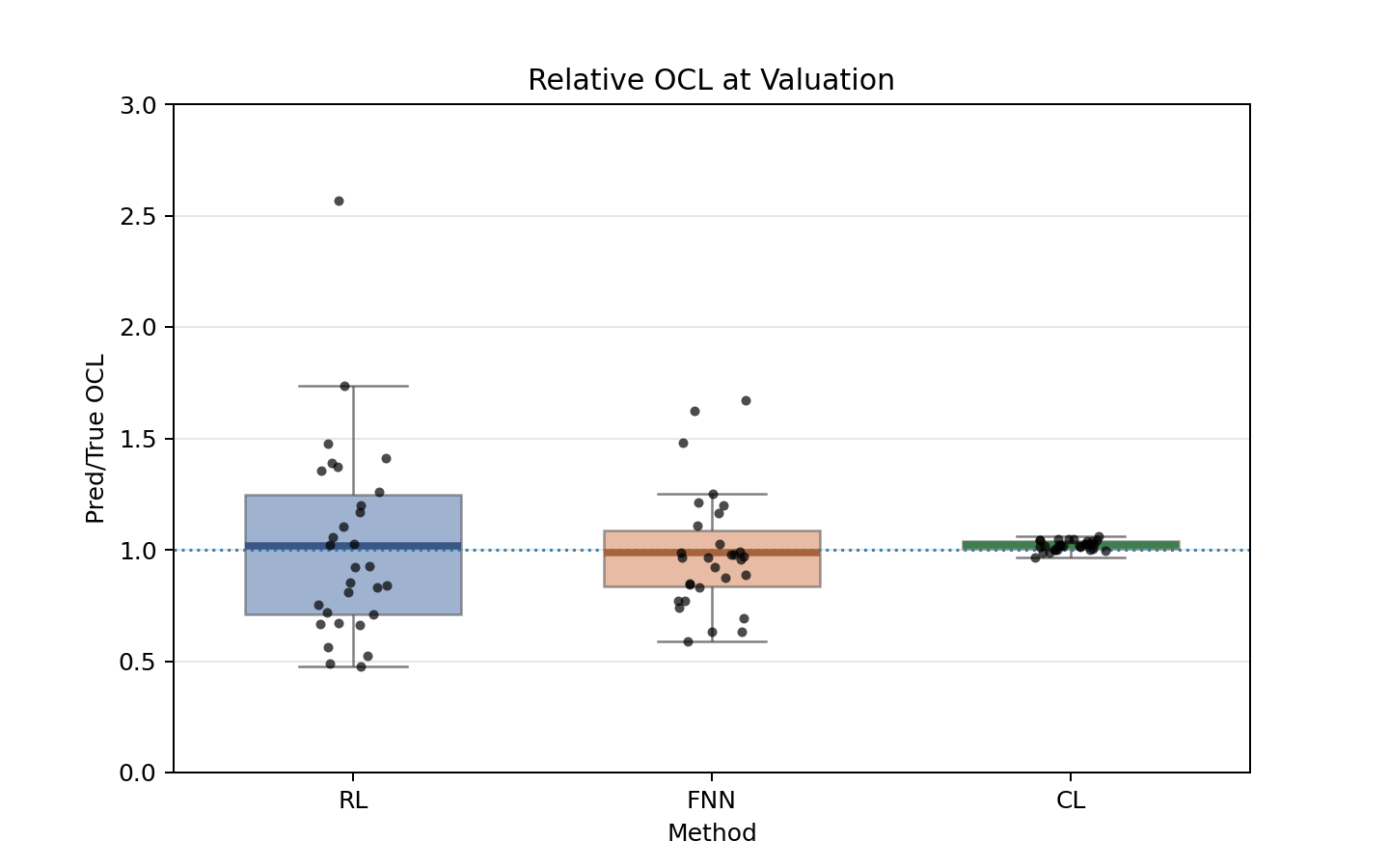}
    \caption{Boxplots of the aggregate relative OCL across all periods}
    \label{fig:Boxplots of the aggregate relative OCL across all periods}
  \end{subfigure}

  \caption{Relative OCL Performance of RL, FNN, and CL on Complexity 1 Test Data}
  \label{fig:Complexity 1 CL}
\end{figure}

\clearpage

\subsection{Simulation Study on the Test Set of SPLICE Generated Data} \label{Results}

\subsubsection{Complexity 1 - RL vs FNN vs CL}

\begin{table}[htb]
  \centering
  \begin{tabular}{lccc}
    \toprule
    Metric & RL & FNN & CL \\
    \midrule
    Average Relative OCL & 101.84\% & 98.54\% & 101.95\% \\
    Average RMSE         & 1182792 & 409841 & \\
    \bottomrule
  \end{tabular}
  \caption{Model comparison at valuation: RL vs FNN vs CL on complexity 1 Test Data}
  \label{tab:rl-fnn}
\end{table}

\begin{figure}[htb]
    \centering
    \includegraphics[width=1\linewidth]{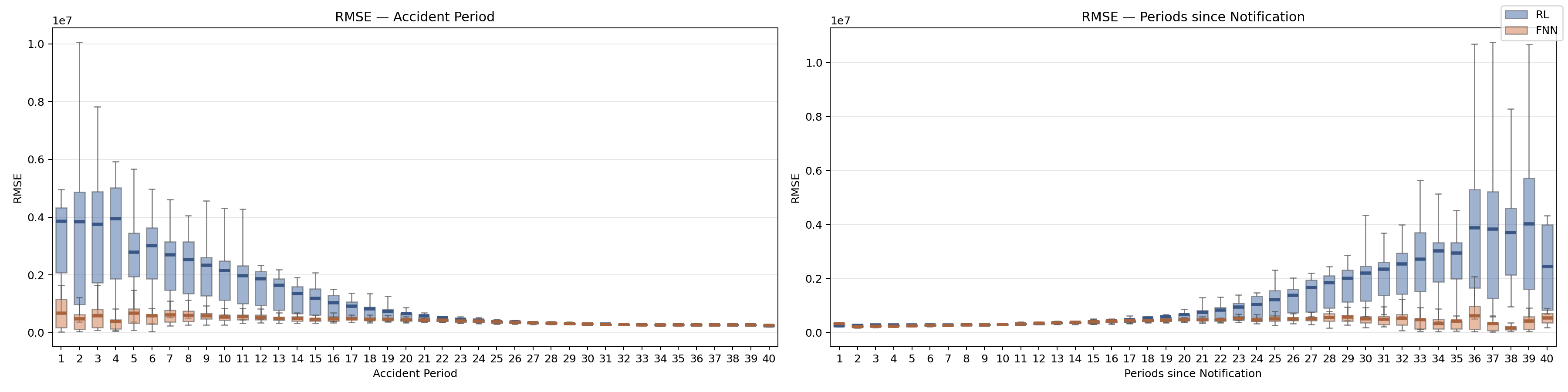}
    \caption{RMSE Performance of RL and FNN for Complexity 1 Test Data}
    \label{fig:Complexity 1 RMSE}
\end{figure}

We are first interested in benchmarking the performance of RL (and FNN) against the chain ladder model as a proof of concept and to ascertain that they are able to perform on a simple, well-behaved and well-known data structure where chain ladder is the perfect model. Before discussing RL and FNN, it is worthwhile to briefly examine chain-ladder's performance; in green in Figure \ref{fig:Complexity 1 CL}. Firstly, recall that the chain ladder (CL) here is stripped of IBNRs in an estimation procedure. This procedure may be unstable for earlier accident periods, where most claims have already settled. Chain ladder does not have a large systematic downward bias because it is a macro-level method, and so it doesn't suffer from our postulated rarity of large claims hypothesis (\ref{downward bias appendix}). Finally, we remark that CL is not present in Figure \ref{fig:Boxplots of the relative OCL by periods since notification CL} because periods since notification requires us to be working on an individual claims basis.

Now, let us turn our attention to the comparative performance of RL and FNN in Figure \ref{fig:Complexity 1 CL} in terms of relative OCL. It is hard to beat the CL results because they are essentially perfect, but it is promising that RL (and FNN) performs almost as well at the aggregate level (Figure \ref{fig:Boxplots of the aggregate relative OCL across all periods}) in terms of getting the mean correct across datasets. All three methods closely identify the correct mean relative OCL of one, with the superior performance of chain ladder showing through the low variance of its predictions. Aside from one outlier for RL, it has similar variation in predictions to the FNN. We believe that the reason behind the outlier, and indeed, the overall volatililty of RL and FNN can be largely explained by the models struggling to adjust sharply enough to final payments. We investigate this in \ref{final_payments}.

When we investigate the more granular AP and PSN breakdowns of predictions (Figures \ref{fig:Boxplots of the relative OCL by accident period CL} and \ref{fig:Boxplots of the relative OCL by periods since notification CL}, respectively), RL and FNN are comparable for the \textit{immature} claims, with RL underestimating slightly more so than FNN. The exception is the last AP and the first PSN, where RL is better than FNN. However, the good performance by RL here is more so attributable to the initialisation procedure being suitable for complexity 1 data.

Additionally, we can observe in Figures \ref{fig:Boxplots of the relative OCL by accident period CL} and \ref{fig:Boxplots of the relative OCL by periods since notification CL} that RL has significantly higher variance in predictions for \textit{mature} claims and tends to overestimate substantially. While this means that the RL model is not useful for these periods, it can also be viewed as a good thing as it shows that RL is able to ``recognise" the fact that estimates are volatile for mature claims. There are some periods where there are only a couple of claims being predicted in total, and this uncertainty is not being recognised by FNN. A reason why RL tends to overestimate may be that claims tend to have a sharp decrease in OCL when the final payment is made, and the constrained action space is unable to capture this abrupt decline in OCL. Furthermore, the OCL importance weighting adjustment could compound this effect as it decreases the importance of getting these smaller OCL predictions correct.

It is not surprising that RL has a higher RMSE (recall that this is the average \textit{per-claim} RMSE) than FNN. RL has large variations for mature claims, which evidently is the major source of the higher RMSE; see Figure \ref{fig:Complexity 1 RMSE}. However, the RMSE is approximately the same for the immature claims, which indicates that RL and FNN tend to make similarly-sized mistakes on a per-claim basis for the important immature claims.

\subsubsection{Complexity 5 - RL vs FNN}

\begin{figure}[htb]
    \centering
    \includegraphics[width=1\linewidth]{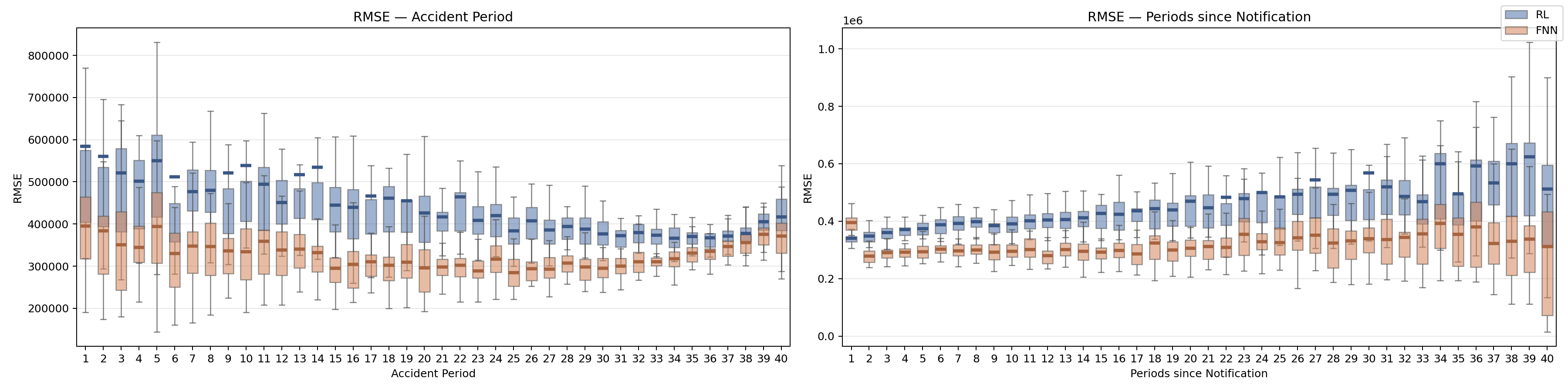}
    \caption{RMSE Performance of RL and FNN on Complexity 5 Test Data}
    \label{fig:Complexity 5 RMSE}
\end{figure}

\begin{figure}[htb]
  \centering
  % first image
  \begin{subfigure}{\linewidth}
    \centering
    \includegraphics[width=\linewidth]{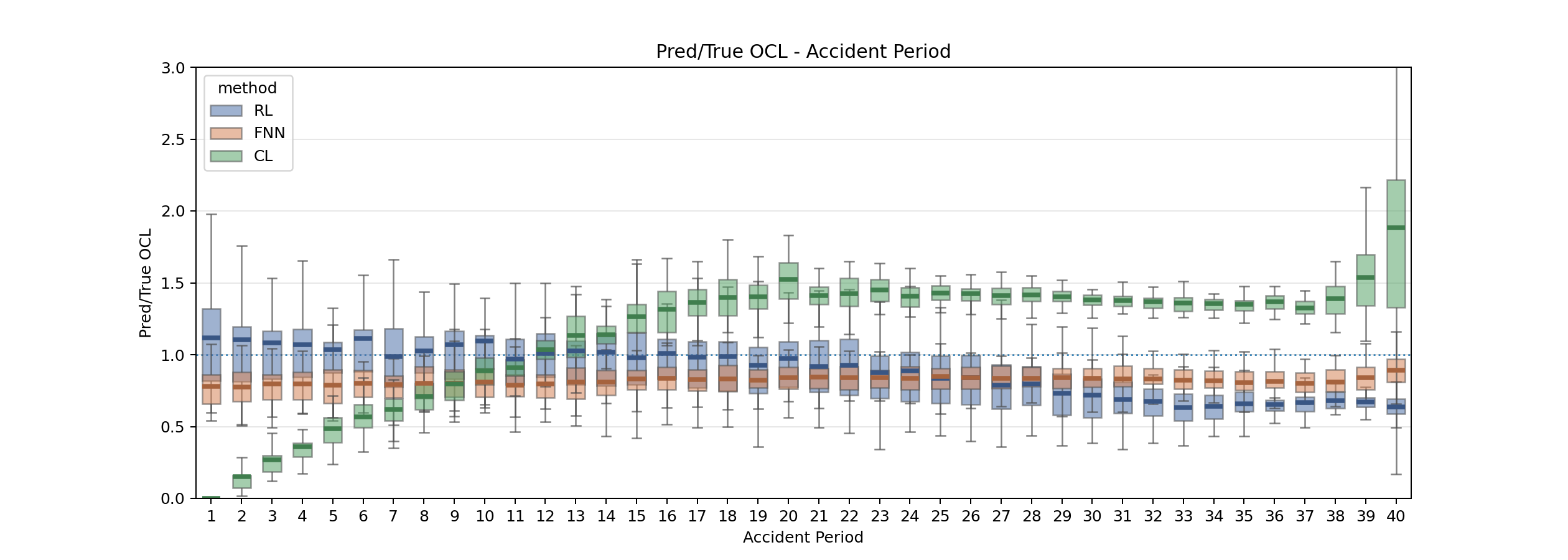}
    \caption{Boxplots of the relative OCL by accident period}
    \label{fig:Boxplots of the relative OCL by accident period C5}
  \end{subfigure}\hfill
  % second image
  \begin{subfigure}{\linewidth}
    \centering
    \includegraphics[width=\linewidth]{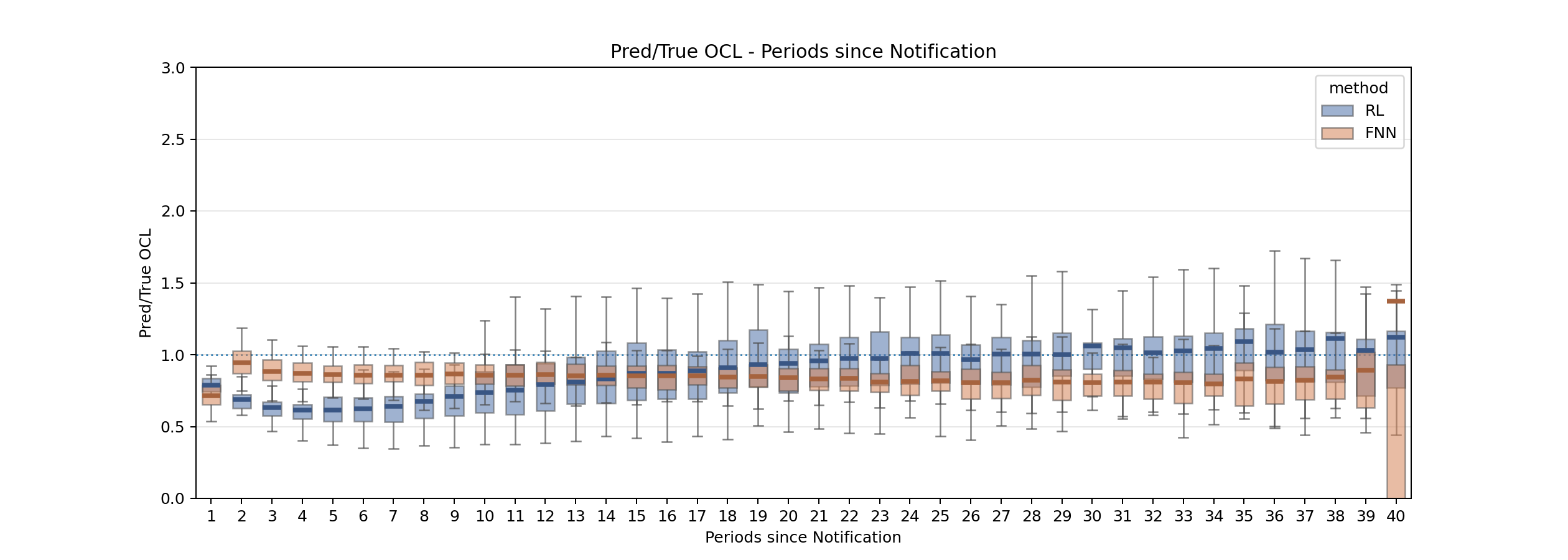}
    \caption{Boxplots of the relative OCL by periods since notification}
    \label{fig:Boxplots of the relative OCL by periods since notification C5}
  \end{subfigure}
  % third image
  \begin{subfigure}{.6\linewidth}
    \centering
    \includegraphics[width=0.75\linewidth]{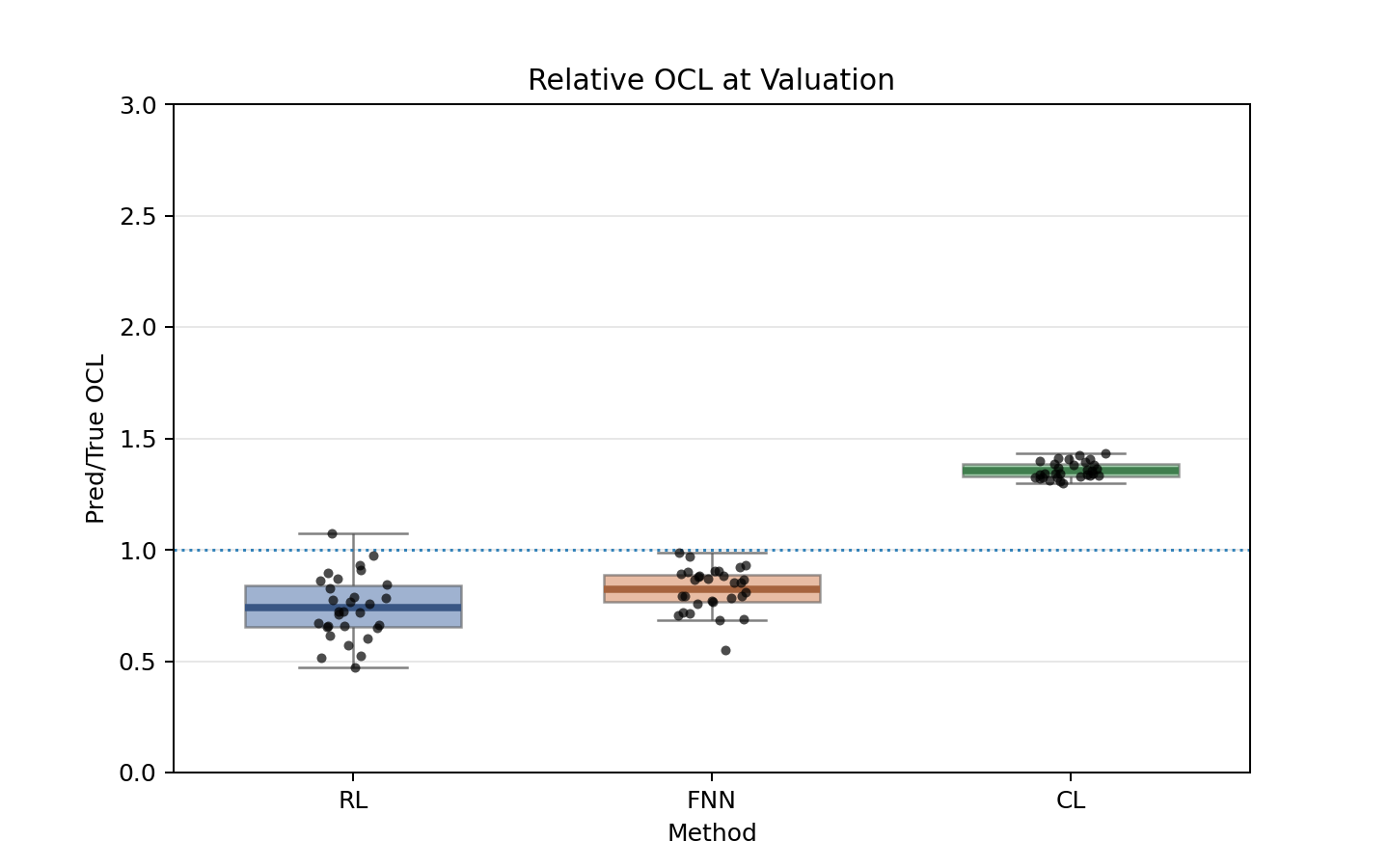}
    \caption{Boxplots of the aggregate relative OCL across all periods}
    \label{fig:Boxplots of the aggregate relative OCL and UL across all periods C5}
  \end{subfigure}

  \caption{Relative OCL Performance of RL, FNN, and CL on Complexity 5 Test Data}
  \label{fig:Complexity 5 CL}
\end{figure}

Much of our observations in complexity 1 remain true for complexity 5. As we can see in Figure \ref{fig:Complexity 5 CL}, RL and FNN are still comparable in terms of aggregate performance, with RL possessing higher variance than FNN for mature claims, and RMSE again indicates that both RL and FNN predict similarly on a per-claim basis for the immature claims (Figure \ref{fig:Complexity 5 RMSE}). 

There are, however, some interesting nuances to discuss. Firstly, RL appears to perform well for the mature claims up until around accident period 25 (albeit with a higher variance than FNN). This is very different to what we had observed for complexity 1. This is likely because for complexity 1, there are only around 300 RBNS claims from the first accident periods, whereas for complexity 5, there are around 1300. With the scarcity of claims, it is expected that there be a lot of volatility across datasets for complexity 1. What is very interesting about RL is that the decrease in performance (underestimation in later accident periods) correspond to the structural break at calendar period 20 on complexity 5 data. This indicates that our current formulation or algorithm choice are not perfectly equipped to deal with sudden changes to the underlying environment dynamics, which is to be expected. In comparison, FNN appears to systematically underestimate for all claims in all periods.

We can see that the chain ladder fails quite severely on the complex data, and is systematically overestimating. This overestimation is consistent with the results of Figure 5 in the original SPLICE paper \citep{Avanzi2023}.

\begin{table}[htb]
  \centering
  \begin{tabular}{lccc}
    \toprule
    Metric & RL & FNN & CL \\
    \midrule
    Average Relative OCL & 73.91\% & 82.30\% & 135.68\% \\
    Average RMSE         & 452121 & 323463 & \\
    \bottomrule
  \end{tabular}
  \caption{Model comparison at valuation: RL vs FNN vs CL on complexity 5 Test Data}
  \label{tab:rl-fnn}
\end{table}

\clearpage

\subsection{Discussion and Limitations} \label{Discussion}

In this paper, we have formulated a framework to apply RL to micro-level reserving, and implemented it with benchmarking against the chain-ladder model and a vanilla FNN model. Section \ref{Results_CAS} demonstrates superior RL performance on more short-tailed simulated data. Section \ref{Results} conducts a simulation study that demonstrates RL is a feasible approach to the micro-level reserving problem as it shows comparable performance to FNN and chain ladder on complexity 1 data in terms of the mean predictions across datasets. While RL is more volatile than FNN in the aggregate, this is purely due to high volatility in earlier accident periods rather than later. We discussed that this may, in fact, be a positive sign that RL is able to ``recognise" the uncertainty associated with these periods, with only a small number of claims remaining. That being said, it is also true that RL \textit{tends} to have higher variance in its predictions compared to their traditional supervised learning counterparts simply due to its underlying assumptions and structure. FNN assumes (incorrectly) i.i.d. observations, whereas RL uses a Markov model of dependence for claim developments. Part of the high variance could also be due to the fact that we have sparse reward signals for accuracy.

Additionally, we observed comparable performances of RL and FNN on complexity 5 data, with RL performing well on the mature claims. For the immature claims, we postulated that the current implementation is ill-equipped to deal with structural breaks where the underlying environment dynamics change abruptly. More complex RL algorithms and augmenting the MDP representation of the reserving problem to relax the stationarity assumptions could be possible ways to alleviate the problem.

Our experience on complexity 5 seems to indicate that inclusion of case estimates is comparatively more beneficial to RL's performance than FNN, as FNN is still systematically underestimating for all periods. It is plausible that additions of covariates to the model input could similarly be more advantageous for RL compared to supervised methods because of RL's capacity to learn from yet to settle claims. Currently, only the accuracy component of the reward is an informative signal, while the other components simply aid learning. One can imagine that RL could benefit significantly with the introduction of reward components that use case estimates and covariates to boost the density of informative reward signals. However, this would likely take some experimentation and expert knowledge. It is also possible that there will be less/no need for importance weighting to correct for the downward bias if there are discriminatory covariates available to the models.

\section{Conclusion} \label{S_Conclusion}

In this paper, we developed and implemented a reinforcement learning framework for claims reserving at the individual claim level, with a particular focus on modelling the sequential nature of reserve updates over claim development. By formulating outstanding claim liability (OCL) estimation as a Markov decision process, we move beyond static, one-shot prediction approaches and instead model reserving as a controlled, dynamic decision problem that evolves as new information becomes available.

A central advantage of this formulation is that it allows the learning algorithm to exploit information from \emph{all observed claim trajectories}, including claims that remain open at valuation. This contrasts with supervised learning approaches trained on ultimate outcomes, which necessarily restrict attention to settled claims and may therefore suffer from reduced sample sizes and selection effects, especially in long-tailed portfolios. By learning from partial claim histories, the proposed approach better reflects actuarial reserving practice and makes more efficient use of available data.

Several methodological choices were required to make reinforcement learning viable and meaningful in a reserving context. We proposed a continuous action space that updates OCL estimates multiplicatively, ensuring positivity and interpretability of reserves, and designed a reward function that balances accuracy at settlement with stability of reserve revisions over time. In particular, the reward structure penalises unnecessary reserve volatility in periods without new payment information, while allowing responsive adjustments when payments occur. Although inspired by potential-based reward shaping, the proposed reward design is tailored to the realities of claims development and reserving practice.

To support actuarial implementation, we also introduced a number of practical components that are often absent from generic RL applications. These include an explicit initialisation mechanism for new claims entering the portfolio, ensuring sensible starting reserves, and a temporally consistent validation and hyperparameter tuning strategy based on a rolling-settlement scheme. This evaluation framework respects the chronological structure of claims data and avoids information leakage, providing a more realistic assessment of out-of-sample reserving performance.

An important empirical finding of the paper is that optimising claim-level accuracy alone can lead to systematic underestimation of aggregate outstanding liabilities, driven by the rarity and materiality of large claims. This phenomenon is particularly relevant in a reinforcement learning setting, where the objective function directly shapes learning. To address this issue, we proposed an importance-weighting mechanism that increases the influence of large claims during training. Empirically, this adjustment substantially improves aggregate OCL estimates while preserving competitive claim-level performance, highlighting the importance of aligning learning objectives with portfolio-level reserving goals.

The proposed framework was evaluated on synthetic property--casualty datasets from the CAS and SPLICE libraries. Across these experiments, the reinforcement learning approach delivered strong performance at both the individual-claim and aggregate levels. In particular, the method performed well for immature claims and recent accident periods, which account for the largest share of outstanding liabilities and are typically the most challenging to reserve accurately. These results suggest that reinforcement learning is especially well suited to the segments of the portfolio that matter most for reserve adequacy.

While the results are encouraging, several avenues for future research remain. The current framework assumes a largely stationary claims environment, and further work is needed to address structural breaks and regime changes, for example through richer state representations or calendar-time effects. More informative intermediate rewards, potentially incorporating case estimates or expert judgments, may further improve learning efficiency and interpretability. Finally, extensions to real-world datasets and hybrid approaches combining reinforcement learning with traditional actuarial models represent promising directions for future investigation.

Overall, this paper demonstrates that reinforcement learning provides a flexible and powerful framework for individual claims reserving, capable of capturing the sequential nature of reserve updates while addressing practical actuarial concerns such as stability, aggregation, and validation. We hope that this work helps bridge the gap between modern reinforcement learning techniques and actuarial reserving practice, and encourages further exploration of sequential decision-making methods in insurance analytics.

\section*{Acknowledgments}
Benjamin Avanzi and Bernard Wong acknowledge support from the Australian Research Council’s Discovery Project funding scheme (DP200101859). The views expressed herein are those of the authors and do not necessarily reflect those of the supporting organisations.

\section*{References}

\bibliographystyle{elsarticle-harv}
\bibliography{references}

\begin{thebibliography}{74}
\expandafter\ifx\csname natexlab\endcsname\relax\def\natexlab#1{#1}\fi
\providecommand{\url}[1]{\texttt{#1}}
\providecommand{\href}[2]{#2}
\providecommand{\path}[1]{#1}
\providecommand{\DOIprefix}{doi:}
\providecommand{\ArXivprefix}{arXiv:}
\providecommand{\URLprefix}{URL: }
\providecommand{\Pubmedprefix}{pmid:}
\providecommand{\doi}[1]{\href{http://dx.doi.org/#1}{\path{#1}}}
\providecommand{\Pubmed}[1]{\href{pmid:#1}{\path{#1}}}
\providecommand{\bibinfo}[2]{#2}
\ifx\xfnm\relax \def\xfnm[#1]{\unskip,\space#1}\fi
%Type = Article
\bibitem[{Al-Mudafer et~al.(2022)Al-Mudafer, Avanzi, Taylor and
  Wong}]{Al-Mudafer2022}
\bibinfo{author}{Al-Mudafer, M.T.}, \bibinfo{author}{Avanzi, B.},
  \bibinfo{author}{Taylor, G.}, \bibinfo{author}{Wong, B.},
  \bibinfo{year}{2022}.
\newblock \bibinfo{title}{Stochastic loss reserving with mixture density neural
  networks}.
\newblock \bibinfo{journal}{Insurance: Mathematics and Economics}
  \bibinfo{volume}{105}, \bibinfo{pages}{144--174}.
\newblock \DOIprefix\doi{10.1016/j.insmatheco.2022.03.010}.
%Type = Techreport
\bibitem[{Antonio and Plat(2012)}]{Antonio2012}
\bibinfo{author}{Antonio, K.}, \bibinfo{author}{Plat, R.},
  \bibinfo{year}{2012}.
\newblock \bibinfo{title}{Micro-level stochastic loss reserving for general
  insurance}.
\newblock \bibinfo{type}{Technical Report}.
\newblock \URLprefix \url{https://ssrn.com/abstract=1620446}.
%Type = Article
\bibitem[{Arjas(1989)}]{Arjas1989}
\bibinfo{author}{Arjas, E.}, \bibinfo{year}{1989}.
\newblock \bibinfo{title}{The claims reserving problem in non-life insurance:
  Some structural ideas}.
\newblock \bibinfo{journal}{ASTIN Bulletin} \bibinfo{volume}{19},
  \bibinfo{pages}{139--152}.
\newblock \DOIprefix\doi{10.2143/ast.19.2.2014905}.
%Type = Techreport
\bibitem[{Asmussen and Taksar(1997)}]{Asmussen1997}
\bibinfo{author}{Asmussen, S.}, \bibinfo{author}{Taksar, M.},
  \bibinfo{year}{1997}.
\newblock \bibinfo{title}{Controlled diffusion models for optimal dividend
  pay-out}.
\newblock \bibinfo{type}{Technical Report}.
%Type = Unpublished
\bibitem[{Avanzi et~al.(2025)Avanzi, Lambrianidis, Taylor and
  Wong}]{avanzi2025}
\bibinfo{author}{Avanzi, B.}, \bibinfo{author}{Lambrianidis, M.},
  \bibinfo{author}{Taylor, G.}, \bibinfo{author}{Wong, B.},
  \bibinfo{year}{2025}.
\newblock \bibinfo{title}{On the use of case estimate and transactional payment
  data in neural networks for individual loss reserving}.
\newblock \URLprefix \url{https://arxiv.org/abs/2601.05274},
  \DOIprefix\doi{10.48550/arXiv.2601.05274}.
%Type = Misc
\bibitem[{Avanzi et~al.(2021a)Avanzi, Taylor and Wang}]{R-SPLICE}
\bibinfo{author}{Avanzi, B.}, \bibinfo{author}{Taylor, G.},
  \bibinfo{author}{Wang, M.}, \bibinfo{year}{2021}a.
\newblock \bibinfo{title}{\texttt{SPLICE}: Synthetic paid loss and incurred
  cost experience (splice) simulator}.
\newblock
  \bibinfo{howpublished}{\url{https://CRAN.R-project.org/package=SPLICE}}.
%Type = Article
\bibitem[{Avanzi et~al.(2023)Avanzi, Taylor and Wang}]{Avanzi2023}
\bibinfo{author}{Avanzi, B.}, \bibinfo{author}{Taylor, G.},
  \bibinfo{author}{Wang, M.}, \bibinfo{year}{2023}.
\newblock \bibinfo{title}{Splice: A synthetic paid loss and incurred cost
  experience simulator}.
\newblock \bibinfo{journal}{Annals of Actuarial Science} \bibinfo{volume}{17},
  \bibinfo{pages}{7--35}.
\newblock \DOIprefix\doi{10.1017/S1748499522000057}.
%Type = Article
\bibitem[{Avanzi et~al.(2021b)Avanzi, Taylor, Wang and Wong}]{Avanzi2021}
\bibinfo{author}{Avanzi, B.}, \bibinfo{author}{Taylor, G.},
  \bibinfo{author}{Wang, M.}, \bibinfo{author}{Wong, B.},
  \bibinfo{year}{2021}b.
\newblock \bibinfo{title}{Synthetic: An individual insurance claim simulator
  with feature control}.
\newblock \bibinfo{journal}{Insurance: Mathematics and Economics}
  \bibinfo{volume}{100}, \bibinfo{pages}{296--308}.
\newblock \DOIprefix\doi{10.1016/j.insmatheco.2021.06.004}.
%Type = Misc
\bibitem[{Avanzi et~al.(2021c)Avanzi, Taylor, Wang and Wong}]{R-SynthETIC}
\bibinfo{author}{Avanzi, B.}, \bibinfo{author}{Taylor, G.},
  \bibinfo{author}{Wang, M.}, \bibinfo{author}{Wong, B.},
  \bibinfo{year}{2021}c.
\newblock \bibinfo{title}{\texttt{SynthETIC}: Synthetic experience tracking
  insurance claims}.
\newblock
  \bibinfo{howpublished}{\url{https://CRAN.R-project.org/package=SynthETIC}}.
\newblock \bibinfo{note}{{R} package version 1.0.1}.
%Type = Article
\bibitem[{Avanzi et~al.(2015)Avanzi, Wong and Yang}]{Avanzi2015}
\bibinfo{author}{Avanzi, B.}, \bibinfo{author}{Wong, B.},
  \bibinfo{author}{Yang, X.}, \bibinfo{year}{2015}.
\newblock \bibinfo{title}{A micro-level claim count model with overdispersion
  and reporting delays}.
\newblock \bibinfo{journal}{SSRN Electronic Journal}
  \DOIprefix\doi{10.2139/ssrn.2705241}.
%Type = Techreport
\bibitem[{Balona and Richman(2020)}]{Balona2020}
\bibinfo{author}{Balona, C.}, \bibinfo{author}{Richman, R.},
  \bibinfo{year}{2020}.
\newblock \bibinfo{title}{The Actuary and IBNR Techniques: A Machine Learning
  Approach}.
\newblock \bibinfo{type}{Technical Report}.
\newblock \URLprefix \url{https://ssrn.com/abstract=3697256}.
%Type = Article
\bibitem[{Baudry and Robert(2019)}]{Baudry2019}
\bibinfo{author}{Baudry, M.}, \bibinfo{author}{Robert, C.Y.},
  \bibinfo{year}{2019}.
\newblock \bibinfo{title}{A machine learning approach for individual claims
  reserving in insurance}.
\newblock \bibinfo{journal}{Applied Stochastic Models in Business and Industry}
  \bibinfo{volume}{35}, \bibinfo{pages}{1127--1155}.
\newblock \DOIprefix\doi{10.1002/asmb.2455}.
%Type = Article
\bibitem[{Bornhuetter and Ferguson(1972)}]{Bornhuetter1972}
\bibinfo{author}{Bornhuetter, R.L.}, \bibinfo{author}{Ferguson, R.E.},
  \bibinfo{year}{1972}.
\newblock \bibinfo{title}{The actuary and ibnr}.
\newblock \bibinfo{journal}{Proceedings of the Casualty Actuarial Society}
  \bibinfo{volume}{LIX}, \bibinfo{pages}{181--195}.
%Type = Article
\bibitem[{Boute et~al.(2022)Boute, Gijsbrechts, van Jaarsveld and
  Vanvuchelen}]{Boute2022}
\bibinfo{author}{Boute, R.N.}, \bibinfo{author}{Gijsbrechts, J.},
  \bibinfo{author}{van Jaarsveld, W.}, \bibinfo{author}{Vanvuchelen, N.},
  \bibinfo{year}{2022}.
\newblock \bibinfo{title}{Deep reinforcement learning for inventory control: A
  roadmap}.
\newblock \bibinfo{journal}{European Journal of Operational Research}
  \bibinfo{volume}{298}, \bibinfo{pages}{401--412}.
\newblock \DOIprefix\doi{10.1016/j.ejor.2021.07.016}.
%Type = Article
\bibitem[{Buehler et~al.(2019)Buehler, Gonon, Teichmann and Wood}]{Buehler2019}
\bibinfo{author}{Buehler, H.}, \bibinfo{author}{Gonon, L.},
  \bibinfo{author}{Teichmann, J.}, \bibinfo{author}{Wood, B.},
  \bibinfo{year}{2019}.
\newblock \bibinfo{title}{Deep hedging}.
\newblock \bibinfo{journal}{Quantitative Finance} \bibinfo{volume}{19},
  \bibinfo{pages}{1271--1291}.
\newblock \DOIprefix\doi{10.1080/14697688.2019.1571683}.
%Type = Misc
\bibitem[{CAS(2025)}]{CAS2025}
\bibinfo{author}{CAS}, \bibinfo{year}{2025}.
\newblock \bibinfo{title}{Data sets now available for research rfp on
  forecasting future loss payments from policies sold in the past}.
\newblock \URLprefix
  \url{https://www.casact.org/article/data-sets-now-available-research-rfp-forecasting-future-loss-payments-policies-sold-past}.
%Type = Article
\bibitem[{Chaoubi et~al.(2023)Chaoubi, Besse, Cossette and
  Côté}]{Chaoubi2023}
\bibinfo{author}{Chaoubi, I.}, \bibinfo{author}{Besse, C.},
  \bibinfo{author}{Cossette, H.}, \bibinfo{author}{Côté, M.P.},
  \bibinfo{year}{2023}.
\newblock \bibinfo{title}{Micro-level reserving for general insurance claims
  using a long short-term memory network}.
\newblock \bibinfo{journal}{Applied Stochastic Models in Business and Industry}
  \bibinfo{volume}{39}, \bibinfo{pages}{382--407}.
\newblock \DOIprefix\doi{10.1002/asmb.2750}.
%Type = Article
\bibitem[{Chong et~al.(2023)Chong, Cui and Li}]{Chong2023}
\bibinfo{author}{Chong, W.F.}, \bibinfo{author}{Cui, H.}, \bibinfo{author}{Li,
  Y.}, \bibinfo{year}{2023}.
\newblock \bibinfo{title}{Pseudo-model-free hedging for variable annuities via
  deep reinforcement learning}.
\newblock \bibinfo{journal}{Annals of Actuarial Science} \bibinfo{volume}{18}.
\newblock \DOIprefix\doi{10.1017/S1748499523000027}.
%Type = Book
\bibitem[{Cover and Thomas(2006)}]{Cover2006}
\bibinfo{author}{Cover, T.M.}, \bibinfo{author}{Thomas, J.A.},
  \bibinfo{year}{2006}.
\newblock \bibinfo{title}{Elements of information theory}.
\newblock \bibinfo{publisher}{Wiley-Interscience}.
%Type = Article
\bibitem[{Delong et~al.(2022)Delong, Lindholm and Wüthrich}]{Delong2022}
\bibinfo{author}{Delong, L.}, \bibinfo{author}{Lindholm, M.},
  \bibinfo{author}{Wüthrich, M.V.}, \bibinfo{year}{2022}.
\newblock \bibinfo{title}{Collective reserving using individual claims data}.
\newblock \bibinfo{journal}{Scandinavian Actuarial Journal}
  \bibinfo{volume}{2022}, \bibinfo{pages}{1--28}.
\newblock \DOIprefix\doi{10.1080/03461238.2021.1921836}.
%Type = Article
\bibitem[{Delong and Wüthrich(2020)}]{Delong2020}
\bibinfo{author}{Delong, L.}, \bibinfo{author}{Wüthrich, M.V.},
  \bibinfo{year}{2020}.
\newblock \bibinfo{title}{Neural networks for the joint development of
  individual payments and claim incurred}.
\newblock \bibinfo{journal}{Risks} \bibinfo{volume}{8}.
\newblock \DOIprefix\doi{10.3390/risks8020033}.
%Type = Article
\bibitem[{Deng et~al.(2017)Deng, Bao, Kong, Ren and Dai}]{Deng2017}
\bibinfo{author}{Deng, Y.}, \bibinfo{author}{Bao, F.}, \bibinfo{author}{Kong,
  Y.}, \bibinfo{author}{Ren, Z.}, \bibinfo{author}{Dai, Q.},
  \bibinfo{year}{2017}.
\newblock \bibinfo{title}{Deep direct reinforcement learning for financial
  signal representation and trading}.
\newblock \bibinfo{journal}{IEEE Transactions on Neural Networks and Learning
  Systems} \bibinfo{volume}{28}, \bibinfo{pages}{653--664}.
\newblock \DOIprefix\doi{10.1109/TNNLS.2016.2522401}.
%Type = Article
\bibitem[{Denuit et~al.(2021)Denuit, Charpentier and Trufin}]{Denuit2021}
\bibinfo{author}{Denuit, M.}, \bibinfo{author}{Charpentier, A.},
  \bibinfo{author}{Trufin, J.}, \bibinfo{year}{2021}.
\newblock \bibinfo{title}{Autocalibration and tweedie-dominance for insurance
  pricing with machine learning}.
\newblock \bibinfo{journal}{Insurance: Mathematics and Economics}
  \bibinfo{volume}{101}, \bibinfo{pages}{485--497}.
\newblock \DOIprefix\doi{10.1016/j.insmatheco.2021.09.001}.
%Type = Article
\bibitem[{Dong and Finlay(2025)}]{Dong2025}
\bibinfo{author}{Dong, S.C.}, \bibinfo{author}{Finlay, J.R.},
  \bibinfo{year}{2025}.
\newblock \bibinfo{title}{Adaptive insurance reserving with cvar-constrained
  reinforcement learning under macroeconomic regimes} \URLprefix
  \url{http://arxiv.org/abs/2504.09396}.
%Type = Article
\bibitem[{Duval and Pigeon(2019)}]{Duval2019}
\bibinfo{author}{Duval, F.}, \bibinfo{author}{Pigeon, M.},
  \bibinfo{year}{2019}.
\newblock \bibinfo{title}{Individual loss reserving using a gradient
  boosting-based approach}.
\newblock \bibinfo{journal}{Risks} \bibinfo{volume}{7}.
\newblock \DOIprefix\doi{10.3390/risks7030079}.
%Type = Techreport
\bibitem[{England and Verrall(2002)}]{England2002}
\bibinfo{author}{England, P.D.}, \bibinfo{author}{Verrall, R.J.},
  \bibinfo{year}{2002}.
\newblock \bibinfo{title}{STOCHASTIC CLAIMS RESERVING IN GENERAL INSURANCE}.
\newblock \bibinfo{type}{Technical Report}.
%Type = Article
\bibitem[{French(1999)}]{French1999}
\bibinfo{author}{French, R.}, \bibinfo{year}{1999}.
\newblock \bibinfo{title}{Catastrophic forgetting in connectionist networks}.
\newblock \bibinfo{journal}{Trends in Cognitive Sciences} \bibinfo{volume}{3},
  \bibinfo{pages}{128--135}.
\newblock \DOIprefix\doi{10.1016/S1364-6613(99)01294-2}.
%Type = Article
\bibitem[{Gabrielli(2021)}]{Gabrielli2021}
\bibinfo{author}{Gabrielli, A.}, \bibinfo{year}{2021}.
\newblock \bibinfo{title}{An individual claims reserving model for reported
  claims}.
\newblock \bibinfo{journal}{European Actuarial Journal} \bibinfo{volume}{11},
  \bibinfo{pages}{541--577}.
\newblock \DOIprefix\doi{10.1007/s13385-021-00271-4}.
%Type = Techreport
\bibitem[{Gabrielli et~al.(2018)Gabrielli, Richman and
  Wüthrich}]{Gabrielli2018}
\bibinfo{author}{Gabrielli, A.}, \bibinfo{author}{Richman, R.},
  \bibinfo{author}{Wüthrich, M.V.}, \bibinfo{year}{2018}.
\newblock \bibinfo{title}{Neural Network Embedding of the Over-Dispersed
  Poisson Reserving Model}.
\newblock \bibinfo{type}{Technical Report}.
\newblock \URLprefix \url{https://ssrn.com/abstract=3288454}.
%Type = Book
\bibitem[{Gerber(1979)}]{Gerber1979}
\bibinfo{author}{Gerber, H.U.}, \bibinfo{year}{1979}.
\newblock \bibinfo{title}{An introduction to mathematical risk theory}.
\newblock \bibinfo{publisher}{S.S. Huebner Foundation for Insurance Education,
  Wharton School, University of Pennsylvania ; Distributed by R.D. Irwin}.
%Type = Book
\bibitem[{Goodfellow et~al.(2017)Goodfellow, Bengio and
  Courville}]{Goodfellow2017}
\bibinfo{author}{Goodfellow, I.}, \bibinfo{author}{Bengio, Y.},
  \bibinfo{author}{Courville, A.}, \bibinfo{year}{2017}.
\newblock \bibinfo{title}{Deep learning}.
\newblock \bibinfo{publisher}{The MIT Press}.
%Type = Article
\bibitem[{Haarnoja et~al.(2018)Haarnoja, Zhou, Abbeel and
  Levine}]{Haarnoja2018}
\bibinfo{author}{Haarnoja, T.}, \bibinfo{author}{Zhou, A.},
  \bibinfo{author}{Abbeel, P.}, \bibinfo{author}{Levine, S.},
  \bibinfo{year}{2018}.
\newblock \bibinfo{title}{Soft actor-critic: Off-policy maximum entropy deep
  reinforcement learning with a stochastic actor} \URLprefix
  \url{http://arxiv.org/abs/1801.01290}.
%Type = Article
\bibitem[{Hambly et~al.(2023)Hambly, Xu and Yang}]{Hambly2023}
\bibinfo{author}{Hambly, B.}, \bibinfo{author}{Xu, R.}, \bibinfo{author}{Yang,
  H.}, \bibinfo{year}{2023}.
\newblock \bibinfo{title}{Recent advances in reinforcement learning in
  finance}.
\newblock \bibinfo{journal}{Mathematical Finance} \bibinfo{volume}{33},
  \bibinfo{pages}{437--503}.
\newblock \DOIprefix\doi{10.1111/mafi.12382}.
%Type = Techreport
\bibitem[{Howard(1960)}]{Howard1960}
\bibinfo{author}{Howard, R.}, \bibinfo{year}{1960}.
\newblock \bibinfo{title}{Dynamic Programming and Markov Processes}.
\newblock \bibinfo{type}{Technical Report}.
%Type = Article
\bibitem[{Jaderberg et~al.(2017)Jaderberg, Dalibard, Osindero, Czarnecki,
  Donahue, Razavi, Vinyals, Green, Dunning, Simonyan, Fernando and
  Kavukcuoglu}]{Jaderberg2017}
\bibinfo{author}{Jaderberg, M.}, \bibinfo{author}{Dalibard, V.},
  \bibinfo{author}{Osindero, S.}, \bibinfo{author}{Czarnecki, W.M.},
  \bibinfo{author}{Donahue, J.}, \bibinfo{author}{Razavi, A.},
  \bibinfo{author}{Vinyals, O.}, \bibinfo{author}{Green, T.},
  \bibinfo{author}{Dunning, I.}, \bibinfo{author}{Simonyan, K.},
  \bibinfo{author}{Fernando, C.}, \bibinfo{author}{Kavukcuoglu, K.},
  \bibinfo{year}{2017}.
\newblock \bibinfo{title}{Population based training of neural networks}
  \URLprefix \url{http://arxiv.org/abs/1711.09846}.
%Type = Techreport
\bibitem[{Kohavi(1995)}]{Kohavi1995}
\bibinfo{author}{Kohavi, R.}, \bibinfo{year}{1995}.
\newblock \bibinfo{title}{A Study of Cross-Validation and Bootstrap for
  Accuracy Estimation and Model Selection}.
\newblock \bibinfo{type}{Technical Report}.
\newblock \URLprefix \url{http//roboticsStanfordedu/"ronnyk}.
%Type = Techreport
\bibitem[{Konda and Tsitsiklis(1999)}]{Konda1999}
\bibinfo{author}{Konda, V.R.}, \bibinfo{author}{Tsitsiklis, J.N.},
  \bibinfo{year}{1999}.
\newblock \bibinfo{title}{Actor-Critic Algorithms}.
\newblock \bibinfo{type}{Technical Report}.
%Type = Article
\bibitem[{Krasheninnikova et~al.(2019)Krasheninnikova, García, Maestre and
  Fernández}]{Krasheninnikova2019}
\bibinfo{author}{Krasheninnikova, E.}, \bibinfo{author}{García, J.},
  \bibinfo{author}{Maestre, R.}, \bibinfo{author}{Fernández, F.},
  \bibinfo{year}{2019}.
\newblock \bibinfo{title}{Reinforcement learning for pricing strategy
  optimization in the insurance industry}.
\newblock \bibinfo{journal}{Engineering Applications of Artificial
  Intelligence} \bibinfo{volume}{80}, \bibinfo{pages}{8--19}.
\newblock \DOIprefix\doi{10.1016/j.engappai.2019.01.010}.
%Type = Article
\bibitem[{Kuo(2019)}]{Kuo2019}
\bibinfo{author}{Kuo, K.}, \bibinfo{year}{2019}.
\newblock \bibinfo{title}{Deeptriangle: A deep learning approach to loss
  reserving}.
\newblock \bibinfo{journal}{Risks} \bibinfo{volume}{7}.
\newblock \DOIprefix\doi{10.3390/risks7030097}.
%Type = Techreport
\bibitem[{Kuo(2020)}]{Kuo2020}
\bibinfo{author}{Kuo, K.}, \bibinfo{year}{2020}.
\newblock \bibinfo{title}{Individual Claims Forecasting with Bayesian Mixture
  Density Networks}.
\newblock \bibinfo{type}{Technical Report}.
%Type = Article
\bibitem[{Lin et~al.(2020)Lin, Chen and Qi}]{Lin2020}
\bibinfo{author}{Lin, E.}, \bibinfo{author}{Chen, Q.}, \bibinfo{author}{Qi,
  X.}, \bibinfo{year}{2020}.
\newblock \bibinfo{title}{Deep reinforcement learning for imbalanced
  classification}.
\newblock \bibinfo{journal}{Applied Intelligence} \bibinfo{volume}{50},
  \bibinfo{pages}{2488--2502}.
\newblock \DOIprefix\doi{10.1007/s10489-020-01637-z}.
%Type = Techreport
\bibitem[{Lin(1992)}]{Lin1992}
\bibinfo{author}{Lin, L.J.}, \bibinfo{year}{1992}.
\newblock \bibinfo{title}{Self-Improving Reactive Agents Based On Reinforcement
  Learning, Planning and Teaching}.
\newblock \bibinfo{type}{Technical Report}.
%Type = Misc
\bibitem[{Lopez et~al.(2019)Lopez, Milhaud and Thérond}]{Lopez2019}
\bibinfo{author}{Lopez, O.}, \bibinfo{author}{Milhaud, X.},
  \bibinfo{author}{Thérond, P.E.}, \bibinfo{year}{2019}.
\newblock \bibinfo{title}{Erratum: A tree-based algorithm adapted to microlevel
  reserving and long development claims (astin bulletin (2019) (1-22) doi:
  10.1017/asb.2019.12)}.
\newblock \DOIprefix\doi{10.1017/asb.2019.21}.
%Type = Article
\bibitem[{Mack(1993)}]{Mack1993}
\bibinfo{author}{Mack, T.}, \bibinfo{year}{1993}.
\newblock \bibinfo{title}{Distribution-free calculation of the standard error
  of chain ladder reserve estimates}.
\newblock \bibinfo{journal}{ASTIN Bulletin} \bibinfo{volume}{23},
  \bibinfo{pages}{213--225}.
\newblock \DOIprefix\doi{10.2143/ast.23.2.2005092}.
%Type = Article
\bibitem[{Martin-Löf(1983)}]{Martin-Lf1983}
\bibinfo{author}{Martin-Löf, A.}, \bibinfo{year}{1983}.
\newblock \bibinfo{title}{Premium control in an insurance system, an approach
  using linear control theory}.
\newblock \bibinfo{journal}{Scandinavian Actuarial Journal}
  \bibinfo{volume}{1983}, \bibinfo{pages}{1--27}.
\newblock \DOIprefix\doi{10.1080/03461238.1983.10408686}.
%Type = Article
\bibitem[{Martin-Löf(1994)}]{Martin-Lf1994}
\bibinfo{author}{Martin-Löf, A.}, \bibinfo{year}{1994}.
\newblock \bibinfo{title}{Lectures on the use of control theory in insurance}.
\newblock \bibinfo{journal}{Scandinavian Actuarial Journal}
  \bibinfo{volume}{1994}, \bibinfo{pages}{1--25}.
\newblock \DOIprefix\doi{10.1080/03461238.1994.10413927}.
%Type = Inproceedings
\bibitem[{Meier et~al.(2012)Meier, LeBlanc and Tan}]{Meier2012}
\bibinfo{author}{Meier, E.}, \bibinfo{author}{LeBlanc, G.},
  \bibinfo{author}{Tan, Y.E.}, \bibinfo{year}{2012}.
\newblock \bibinfo{title}{Orbit correction studies using neural networks}, in:
  \bibinfo{booktitle}{Proc. IPAC’12 (3rd Int. Particle Accelerator Conf.)},
  \bibinfo{publisher}{JACoW Publishing, Geneva, Switzerland}.
\newblock \URLprefix
  \url{https://proceedings.jacow.org/IPAC2012/papers/weppp057.pdf}.
%Type = Article
\bibitem[{Mnih et~al.(2013)Mnih, Kavukcuoglu, Silver, Graves, Antonoglou,
  Wierstra and Riedmiller}]{Mnih2013}
\bibinfo{author}{Mnih, V.}, \bibinfo{author}{Kavukcuoglu, K.},
  \bibinfo{author}{Silver, D.}, \bibinfo{author}{Graves, A.},
  \bibinfo{author}{Antonoglou, I.}, \bibinfo{author}{Wierstra, D.},
  \bibinfo{author}{Riedmiller, M.}, \bibinfo{year}{2013}.
\newblock \bibinfo{title}{Playing atari with deep reinforcement learning}
  \URLprefix \url{http://arxiv.org/abs/1312.5602}.
%Type = Article
\bibitem[{Moody and Saffell(2001)}]{Moody2001}
\bibinfo{author}{Moody, J.}, \bibinfo{author}{Saffell, M.},
  \bibinfo{year}{2001}.
\newblock \bibinfo{title}{Learning to trade via direct reinforcement}.
\newblock \bibinfo{journal}{IEEE Transactions on Neural Networks}
  \bibinfo{volume}{12}, \bibinfo{pages}{875--889}.
\newblock \DOIprefix\doi{10.1109/72.935097}.
%Type = Article
\bibitem[{Moor et~al.(2022)Moor, Gijsbrechts and Boute}]{Moor2022}
\bibinfo{author}{Moor, B.J.D.}, \bibinfo{author}{Gijsbrechts, J.},
  \bibinfo{author}{Boute, R.N.}, \bibinfo{year}{2022}.
\newblock \bibinfo{title}{Reward shaping to improve the performance of deep
  reinforcement learning in perishable inventory management}.
\newblock \bibinfo{journal}{European Journal of Operational Research}
  \bibinfo{volume}{301}, \bibinfo{pages}{535--545}.
\newblock \DOIprefix\doi{10.1016/j.ejor.2021.10.045}.
%Type = Techreport
\bibitem[{Ng et~al.(1999)Ng, Harada and Russell}]{Ng1999}
\bibinfo{author}{Ng, A.Y.}, \bibinfo{author}{Harada, D.},
  \bibinfo{author}{Russell, S.}, \bibinfo{year}{1999}.
\newblock \bibinfo{title}{Policy invariance under reward transformations:
  Theory and application to reward shaping}.
\newblock \bibinfo{type}{Technical Report}.
%Type = Article
\bibitem[{Norberg(1993)}]{Norberg1993}
\bibinfo{author}{Norberg, R.}, \bibinfo{year}{1993}.
\newblock \bibinfo{title}{Prediction of outstanding liabilities in non-life
  insurance}.
\newblock \bibinfo{journal}{ASTIN Bulletin} \bibinfo{volume}{23},
  \bibinfo{pages}{95--115}.
\newblock \DOIprefix\doi{10.2143/ast.23.1.2005103}.
%Type = Article
\bibitem[{Norberg(1999)}]{Norberg1999}
\bibinfo{author}{Norberg, R.}, \bibinfo{year}{1999}.
\newblock \bibinfo{title}{Prediction of outstanding liabilities ii. model
  variations and extensions}.
\newblock \bibinfo{journal}{ASTIN Bulletin} \bibinfo{volume}{29},
  \bibinfo{pages}{5--25}.
\newblock \DOIprefix\doi{10.2143/ast.29.1.504603}.
%Type = Article
\bibitem[{Palmborg and Lindskog(2023)}]{Palmborg2023}
\bibinfo{author}{Palmborg, L.}, \bibinfo{author}{Lindskog, F.},
  \bibinfo{year}{2023}.
\newblock \bibinfo{title}{Premium control with reinforcement learning}.
\newblock \bibinfo{journal}{ASTIN Bulletin} \bibinfo{volume}{53},
  \bibinfo{pages}{233--257}.
\newblock \DOIprefix\doi{10.1017/asb.2023.13}.
%Type = Article
\bibitem[{Rockafellar and Uryasev(2002)}]{Rockafellar2002}
\bibinfo{author}{Rockafellar, R.}, \bibinfo{author}{Uryasev, S.},
  \bibinfo{year}{2002}.
\newblock \bibinfo{title}{Conditional value-at-risk for general loss
  distributions}.
\newblock \bibinfo{journal}{Journal of Banking \& Finance}
  \bibinfo{volume}{26}, \bibinfo{pages}{1443--1471}.
\newblock \DOIprefix\doi{10.1016/S0378-4266(02)00271-6}.
%Type = Article
\bibitem[{Schulman et~al.(2017a)Schulman, Levine, Moritz, Jordan and
  Abbeel}]{schulman2017a}
\bibinfo{author}{Schulman, J.}, \bibinfo{author}{Levine, S.},
  \bibinfo{author}{Moritz, P.}, \bibinfo{author}{Jordan, M.I.},
  \bibinfo{author}{Abbeel, P.}, \bibinfo{year}{2017}a.
\newblock \bibinfo{title}{Trust region policy optimization} \URLprefix
  \url{http://arxiv.org/abs/1502.05477}.
%Type = Article
\bibitem[{Schulman et~al.(2017b)Schulman, Wolski, Dhariwal, Radford and
  Klimov}]{schulman2017b}
\bibinfo{author}{Schulman, J.}, \bibinfo{author}{Wolski, F.},
  \bibinfo{author}{Dhariwal, P.}, \bibinfo{author}{Radford, A.},
  \bibinfo{author}{Klimov, O.}, \bibinfo{year}{2017}b.
\newblock \bibinfo{title}{Proximal policy optimization algorithms} \URLprefix
  \url{http://arxiv.org/abs/1707.06347}.
%Type = Article
\bibitem[{Schwab and Schneider(2024)}]{Schwab2024}
\bibinfo{author}{Schwab, B.}, \bibinfo{author}{Schneider, J.C.},
  \bibinfo{year}{2024}.
\newblock \bibinfo{title}{Advancing loss reserving: A hybrid neural network
  approach for individual claim development prediction}.
\newblock \bibinfo{journal}{SSRN Electronic Journal}
  \DOIprefix\doi{10.2139/ssrn.4769020}.
%Type = Misc
\bibitem[{Silver(2015)}]{Silver2015}
\bibinfo{author}{Silver, D.}, \bibinfo{year}{2015}.
\newblock \bibinfo{title}{Lectures on reinforcement learning}.
%Type = Article
\bibitem[{Silver et~al.(2017)Silver, Hubert, Schrittwieser, Antonoglou, Lai,
  Guez, Lanctot, Sifre, Kumaran, Graepel, Lillicrap, Simonyan and
  Hassabis}]{Silver2017}
\bibinfo{author}{Silver, D.}, \bibinfo{author}{Hubert, T.},
  \bibinfo{author}{Schrittwieser, J.}, \bibinfo{author}{Antonoglou, I.},
  \bibinfo{author}{Lai, M.}, \bibinfo{author}{Guez, A.},
  \bibinfo{author}{Lanctot, M.}, \bibinfo{author}{Sifre, L.},
  \bibinfo{author}{Kumaran, D.}, \bibinfo{author}{Graepel, T.},
  \bibinfo{author}{Lillicrap, T.}, \bibinfo{author}{Simonyan, K.},
  \bibinfo{author}{Hassabis, D.}, \bibinfo{year}{2017}.
\newblock \bibinfo{title}{Mastering chess and shogi by self-play with a general
  reinforcement learning algorithm} \URLprefix
  \url{http://arxiv.org/abs/1712.01815}.
%Type = Techreport
\bibitem[{Sutton(1988)}]{Sutton1988}
\bibinfo{author}{Sutton, R.S.}, \bibinfo{year}{1988}.
\newblock \bibinfo{title}{Learning to Predict by the Methods of Temporal
  Differences}.
\newblock \bibinfo{type}{Technical Report}.
%Type = Techreport
\bibitem[{Sutton and Barto(2018)}]{Sutton2018}
\bibinfo{author}{Sutton, R.S.}, \bibinfo{author}{Barto, A.G.},
  \bibinfo{year}{2018}.
\newblock \bibinfo{title}{Reinforcement Learning: An Introduction Second
  edition, in progress}.
\newblock \bibinfo{type}{Technical Report}.
%Type = Techreport
\bibitem[{Tashman(2000)}]{Tashman2000}
\bibinfo{author}{Tashman, L.J.}, \bibinfo{year}{2000}.
\newblock \bibinfo{title}{Out-of-sample tests of forecasting accuracy: an
  analysis and review}.
\newblock \bibinfo{type}{Technical Report}.
\newblock \URLprefix \url{www.elsevier.com/locate/ijforecast}.
%Type = Book
\bibitem[{Taylor(2000)}]{Taylor2000}
\bibinfo{author}{Taylor, G.}, \bibinfo{year}{2000}.
\newblock \bibinfo{title}{Loss Reserving}. volume~\bibinfo{volume}{21}.
\newblock \bibinfo{publisher}{Springer US}.
\newblock \DOIprefix\doi{10.1007/978-1-4615-4583-5}.
%Type = Book
\bibitem[{Taylor and McGuire(2016)}]{Taylor2016}
\bibinfo{author}{Taylor, G.}, \bibinfo{author}{McGuire, G.},
  \bibinfo{year}{2016}.
\newblock \bibinfo{title}{Stochastic Loss Reserving Using Generalized Linear
  Models}.
\newblock \bibinfo{publisher}{Casualty Actuarial Society}.
\newblock \URLprefix
  \url{https://www.casact.org/sites/default/files/2021-02/03-Taylor.pdf}.
%Type = Article
\bibitem[{Wekwete et~al.(2023)Wekwete, Kufakunesu and van Zyl}]{Wekwete2023}
\bibinfo{author}{Wekwete, T.A.}, \bibinfo{author}{Kufakunesu, R.},
  \bibinfo{author}{van Zyl, G.}, \bibinfo{year}{2023}.
\newblock \bibinfo{title}{Application of deep reinforcement learning in asset
  liability management}.
\newblock \bibinfo{journal}{Intelligent Systems with Applications}
  \bibinfo{volume}{20}.
\newblock \DOIprefix\doi{10.1016/j.iswa.2023.200286}.
%Type = Misc
\bibitem[{Wuthrich et~al.(2025)Wuthrich, Richman, Avanzi, Lindholm, Mayer,
  Schelldorfer and Scognamiglio}]{Wuthrich2025}
\bibinfo{author}{Wuthrich, M.V.}, \bibinfo{author}{Richman, R.},
  \bibinfo{author}{Avanzi, B.}, \bibinfo{author}{Lindholm, M.},
  \bibinfo{author}{Mayer, M.}, \bibinfo{author}{Schelldorfer, J.},
  \bibinfo{author}{Scognamiglio, S.}, \bibinfo{year}{2025}.
\newblock \bibinfo{title}{Ai tools for actuaries}.
\newblock \DOIprefix\doi{10.2139/ssrn.5162304}.
%Type = Book
\bibitem[{Wüthrich and Merz(2008)}]{Wuthrich2008}
\bibinfo{author}{Wüthrich, M.V.}, \bibinfo{author}{Merz, M.},
  \bibinfo{year}{2008}.
\newblock \bibinfo{title}{Stochastic claims reserving methods in insurance}.
\newblock \bibinfo{publisher}{John Wiley \& Sons}.
%Type = Article
\bibitem[{Wüthrich(2020)}]{Wthrich2020}
\bibinfo{author}{Wüthrich, M.V.}, \bibinfo{year}{2020}.
\newblock \bibinfo{title}{Bias regularization in neural network models for
  general insurance pricing}.
\newblock \bibinfo{journal}{European Actuarial Journal} \bibinfo{volume}{10},
  \bibinfo{pages}{179--202}.
\newblock \DOIprefix\doi{10.1007/s13385-019-00215-z}.
%Type = Article
\bibitem[{Wüthrich and Ziegel(2024)}]{Wthrich2024}
\bibinfo{author}{Wüthrich, M.V.}, \bibinfo{author}{Ziegel, J.},
  \bibinfo{year}{2024}.
\newblock \bibinfo{title}{Isotonic recalibration under a low signal-to-noise
  ratio}.
\newblock \bibinfo{journal}{Scandinavian Actuarial Journal}
  \bibinfo{volume}{2024}, \bibinfo{pages}{279--299}.
\newblock \DOIprefix\doi{10.1080/03461238.2023.2246743}.
%Type = Article
\bibitem[{Xu et~al.(2021)Xu, Wang and Liang}]{Xu2021}
\bibinfo{author}{Xu, T.}, \bibinfo{author}{Wang, Z.}, \bibinfo{author}{Liang,
  Y.}, \bibinfo{year}{2021}.
\newblock \bibinfo{title}{Improving sample complexity bounds for (natural)
  actor-critic algorithms} \URLprefix \url{http://arxiv.org/abs/2004.12956}.
%Type = Unpublished
\bibitem[{Xu et~al.(2019)Xu, Dai, Kemp and Metz}]{Xu2019}
\bibinfo{author}{Xu, Z.}, \bibinfo{author}{Dai, A.M.}, \bibinfo{author}{Kemp,
  J.}, \bibinfo{author}{Metz, L.}, \bibinfo{year}{2019}.
\newblock \bibinfo{title}{Learning an adaptive learning rate schedule}.
\newblock \URLprefix \url{https://arxiv.org/abs/1909.09712},
  \DOIprefix\doi{10.48550/arXiv.1909.09712}.
%Type = Article
\bibitem[{Xu et~al.(2018)Xu, van Hasselt and Silver}]{Xu2018}
\bibinfo{author}{Xu, Z.}, \bibinfo{author}{van Hasselt, H.},
  \bibinfo{author}{Silver, D.}, \bibinfo{year}{2018}.
\newblock \bibinfo{title}{Meta-gradient reinforcement learning} \URLprefix
  \url{http://arxiv.org/abs/1805.09801}.
%Type = Techreport
\bibitem[{Ziebart and Fox(2010)}]{Ziebart2010}
\bibinfo{author}{Ziebart, B.D.}, \bibinfo{author}{Fox, D.},
  \bibinfo{year}{2010}.
\newblock \bibinfo{title}{Modeling Purposeful Adaptive Behavior with the
  Principle of Maximum Causal Entropy}.
\newblock \bibinfo{type}{Technical Report}.

\end{thebibliography}

\newpage

\appendix

\section{Causes of Downward Bias} \label{downward bias appendix}

We observed in our experiments that models suffer from a large and consistent downward bias, whereby they underestimate the aggregate OCL at the portfolio level. We investigate this issue further through a series of hypotheses. To the best of our knowledge, this issue has not been reported anywhere else in the reserving literature, but has been discussed in the pricing literature \citep[e.g., ][]{Denuit2021, Wthrich2024, Wthrich2020}, which we will discuss below in the 5th hypothesis. 

Note that in our investigation of hypotheses, we need to show some preliminary results using the relative OCL, which is defined formally in section \ref{Evaluation}. It is simply the ratio of the aggregate predicted OCL to the true aggregate OCL. Hence, the aim to to be as close to a relative OCL of 1 as possible.

\begin{enumerate}
    \item \textbf{Rarity of Large Claims:} We postulate that a key contributing cause to the downward bias is the imbalance between small vs large claims, sort of like a class imbalance problem for classification tasks. In the absence of context that the \textit{aggregate} prediction should be unbiased, the model will treat every claim as having the same training ``power", so the most frequent claims (i.e., the small claims) take most of the model's attention.

    We find evidence that supports this hypothesis in Figure \ref{fig:ocl_by_size}, whereby we can see that without any adjustment, the larger claims are severely underestimated, which drives down the overall portfolio level estimate. In fact, FNN suffers substantially more so than RL for the small and medium claims.

    \begin{figure}[htb]
        \centering
        \includegraphics[width=1\linewidth]{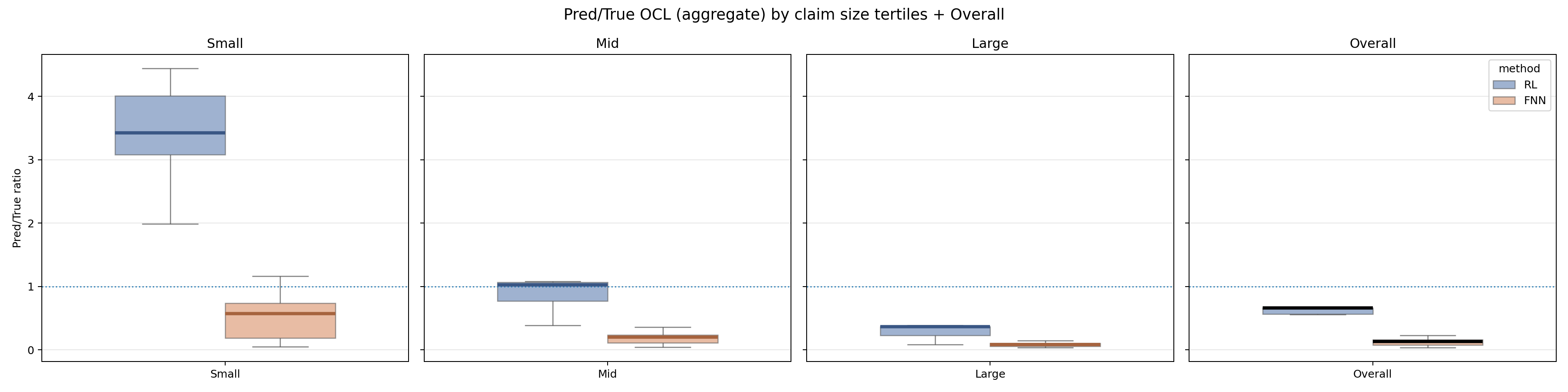}
        \caption{Performance of models on different-sized claims. Small, medium and large claims are simply claims split in thirds by their ultimate claim size. Note that this is using 10 datasets of complexity 1 test data (without bias correction from section \ref{Downward Bias Adjustment}), which is the simplest data where chain ladder assumptions hold; see section \ref{Results} for more detail.}
        \label{fig:ocl_by_size}
    \end{figure}

    \item \textbf{Lack of (Informative) Covariates:} Another likely contributing cause is the lack of covariates. With essentially just the accident period, development period, payment information, and case estimates, it is difficult to discern whether there is a little or a lot of OCL. This is in contrast to existing works \citep{Gabrielli2021, Kuo2020} using a different claims simulator and/or real data that includes potentially more discriminatory covariates like the injured body part. As we are more interested in the formulation of the RL model itself, we have decided to work without covariates, and leave this for future work.

    \item \textbf{Train-test Split Construction:} We noticed that a feature of SPLICE simulated data, when split temporally, is such that the claim sizes for claims in the test set are, on average, larger than those in the training set. An explanation for this is that by construction of our train-test split, the claims in the test set are those that have been reported in the training set time interval, but not yet settled. These would, on average, be claims with longer duration until settlement, and hence, tend to also be larger claims. Although realistic, this need not hold in practice, as these characteristics would depend on the line of business, but nonetheless, it appears to hold for the SPLICE simulated data. If we are to believe the rarity of large claims hypothesis, then this means that we would also underestimate the test set simply because it consists of larger claims than average.

    \item \textbf{Pricing Literature \& Auto-Calibration:} There is a strand of literature that discusses bias for pricing, and we discuss two things here. Firstly, there is a strand of literature pertaining to auto-calibration \citep{Denuit2021, Wthrich2024}. This is concerned primarily with addressing ``cross-financing", where some lines of business (LoBs) are overestimated and while others are underestimated. The objective of auto-calibration is then not to correct a large global bias, but instead to correct these biases between LoBs so that there is less/no cross-financing. This is not applicable in our case since we are concerned with an overall bias rather than, say, varying biases between accident periods. In a sort of adjacent paper, \citet{Wthrich2020} considers bias in pricing as well, but in an overall sense rather than just relating to cross-financing. Specifically, it is asserted that mechanisms like early-stopping cause neural networks to violate the balance property, and results in bias at the aggregate level. The proposed solution involves using an additional GLM step with the last hidden layer nodes as covariate inputs to the GLM. This may be worthwhile to pursue as an extension, but we do not consider it in this paper as it is not straightforward to adapt to the RL framework.

\end{enumerate}

\section{Benchmark Models} \label{benchmark models appendix}

\subsection{Chain-Ladder} \label{A_CL}

To compare the RL model's performance against Chain ladder, we use complexity 1 data generated from SPLICE, where chain ladder is expected to perform very well. Given that the RL model concerns RBNS claims only, we must strip out the IBNR claims from the chain ladder predictions for a fair comparison. For this, we follow the approach set out by \citet{Delong2020}.

\begin{enumerate}
    \item Aggregate the data into cumulative triangles for claim counts and cumulative paid amounts.

    \item Fit a simple Chain Ladder model.
    
    \item Project the ultimate paid and ultimate counts for each accident year $i$, then compute the average ultimate claim size $\mu_i$ for each accident year by taking the quotient.
    
    \item However, claims with longer reporting delays are often larger on average. So we fit a GAM to all the claims incurred (case estimates) available at the cutoff date $T^*$ that gives the scaling factor $s(d)$ standardised for $s(0)=1$. That is, given a reporting delay $d$, how much do we need to scale the average ultimate claims size $\mu_i$ by compared to if there was zero delay.
    
    \item For each accident period, the number of IBNR claims is the projected ultimate count less the currently already observed count. Multiply this by $\mu \times s(d)$, where $d$ is the reporting delay for the IBNR amount outstanding. I.e., the estimate of IBNR amount outstanding for accident period $i$ is:
    \begin{equation}
        \sum_j (\text{the incremental number of claims for each development period } j) \times \mu_i \times s(d)
    \end{equation}
    where $d=j-1$.

    \item The total RBNS portion of the UL is simply the total UL less the IBNR portion of the UL. However, this is the RBNS ultimate loss for \textit{all} claims, which includes those that have already settled. We need to remove the claims that settle in the training set when comparing with RL. 

\end{enumerate}

Once we have done this, we can compare the RBNS portion of the OCL (UL less cumulative paid) in chain ladder with the RL predictions by accident year aggregates.

\subsection{FNN}

We refer to \citet{Wuthrich2025} and \citet{Goodfellow2017} for details on feed-forward neural networks.

We fit a fully-connected FNN to the same claims dataset that the RL model sees. For a direct comparison, the train test split used for RL is also used for FNN. However, because the FNN targets the OCL, it can only train on claims that have already been fully settled in the training set (so that we have targets to work with). As such, we strip the training set of any claims that are not settled by the end of the training interval. We are training the FNN to minimise the importance weighted MSE (\ref{fnn_weighted_mse}). 

Each claim can contribute multiple observations to the training set. If a claim has $J$ development periods and its reporting delay is $d$ periods, then it contributes $J-d$ observations to the training set since we can use each known development of the claim to train. This is a practice done in the literature in various forms when working with individual claims data \citep{Schwab2024, Gabrielli2021, Delong2022, Chaoubi2023}, and is necessary for good performance. Of course, the shortcoming of this approach is that the FNN assumes all training observations are independent, when they clearly are not.

The input nodes for the FNN correspond to the state space of our RL model, except for the past predictions as it is difficult to embed within a vanilla FNN. This is yet another (minor) advantage of RL due to its sequential nature. The number of hidden layers and the number of nodes in a hidden layer are hyperparameters to be tuned. However, it is typical in practice to assume the same number of nodes in every hidden layer to reduce the complexity of hyperparameter tuning. Lastly, there is one output node corresponding to the model's prediction of the UL.

The FNN hyperparameters that we tune are: the number of hidden layers and how many nodes are in each hidden layer, the batch size, the dropout rate, and the learning rate. The validation procedure for tuning hyperparameters is described in section \ref{validation}.

\section{Data Description} \label{Data Description}

\subsection{SPLICE Data for the Simulated Study}

Simulated data was generated using the default \texttt{generate\_data()} function of SPLICE \citep{R-SPLICE,Avanzi2023}, which means we are using the default simulation assumptions outlined in the paper. Each row in the data corresponds to one transaction of a claim, with the following columns. We refer to the original papers by \citet{R-SPLICE,Avanzi2023} for technical details.

\begin{itemize}
    \item \texttt{claim\_no}: the claim number

    \item \texttt{claim\_size}: the ultimate loss (without inflation) of the claim
    
    \item \texttt{txn\_time}: the continuous transaction time at which the transaction occurred
    
    \item \texttt{txn\_type}: the transaction type can be one of

    \begin{itemize}
        \item \texttt{Mi}: minor revision to the case estimate
    
        \item \texttt{Ma}: major revision to the case estimate
        
        \item \texttt{P}: (partial) claim payment
        
        \item \texttt{PMi}: payment and minor revision
        
        \item \texttt{PMa}: payment and major revision
    \end{itemize}
    
    \item \texttt{incurred}: the case estimate of the ultimate claim size
    
    \item \texttt{OCL}: the case estimate of the outstanding claim liabilities
    
    \item \texttt{cumpaid}: the cumulative amount paid for the claim so far
    
    \item \texttt{accident\_period}: the accident period of the claim
\end{itemize}

\subsection{CAS Data}

The CAS datasets contain only a subset of the SPLICE features: \texttt{claim\_no}, \texttt{claim\_size}, \texttt{txn\_time}, \texttt{cumpaid}, and \texttt{accident\_period}.

Note that \texttt{txn\_time} is discrete in the CAS datasets (as opposed to continuous in the SPLICE datasets).

\newpage

\section{Cumulative Share of Outstanding OCL Graphs} \label{cum share OCL graphs appendix}

\begin{figure}[htb]
    \centering
    \includegraphics[width=1\linewidth]{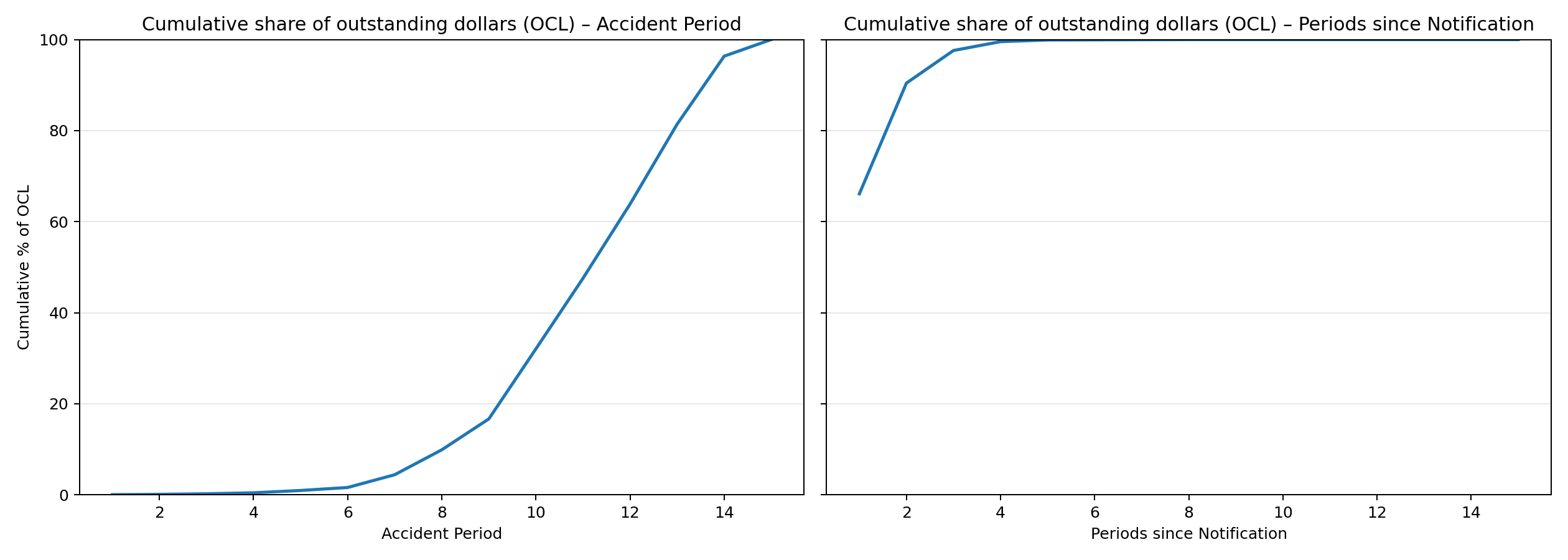}
    \caption{Cumulative share of outstanding OCL for CAS test data.}
    \label{fig:ocl_share_CAS}
\end{figure}

\begin{figure}[htb]
    \centering
    \includegraphics[width=1\linewidth]{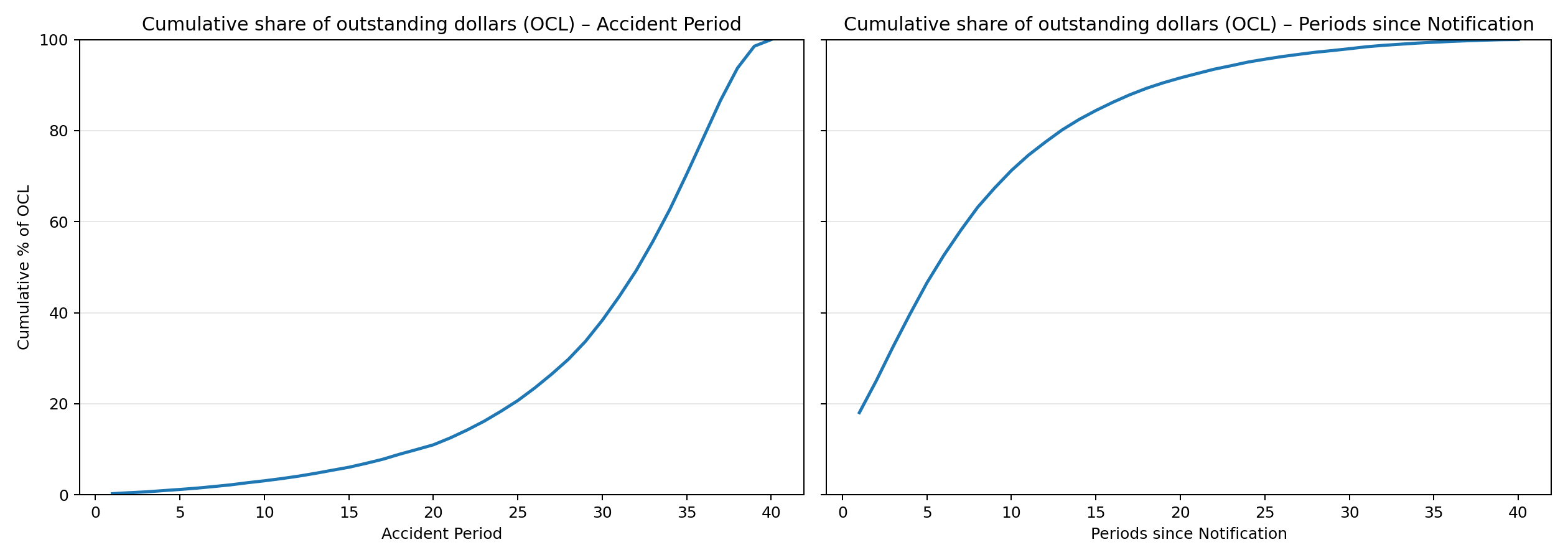}
    \caption{Cumulative share of outstanding OCL for complexity 5 test data (complexity 1 is essentially identical). Note: the periods since notification graph is cut off at 40 for visual purposes, but there are (rare) claims that have not settled by 40 periods since notification.}
    \label{fig:ocl_share_SPLICE}
\end{figure}

\newpage

\section{Effect of Final Payments} \label{final_payments}

\begin{figure}
    \centering
    \includegraphics[width=1\linewidth]{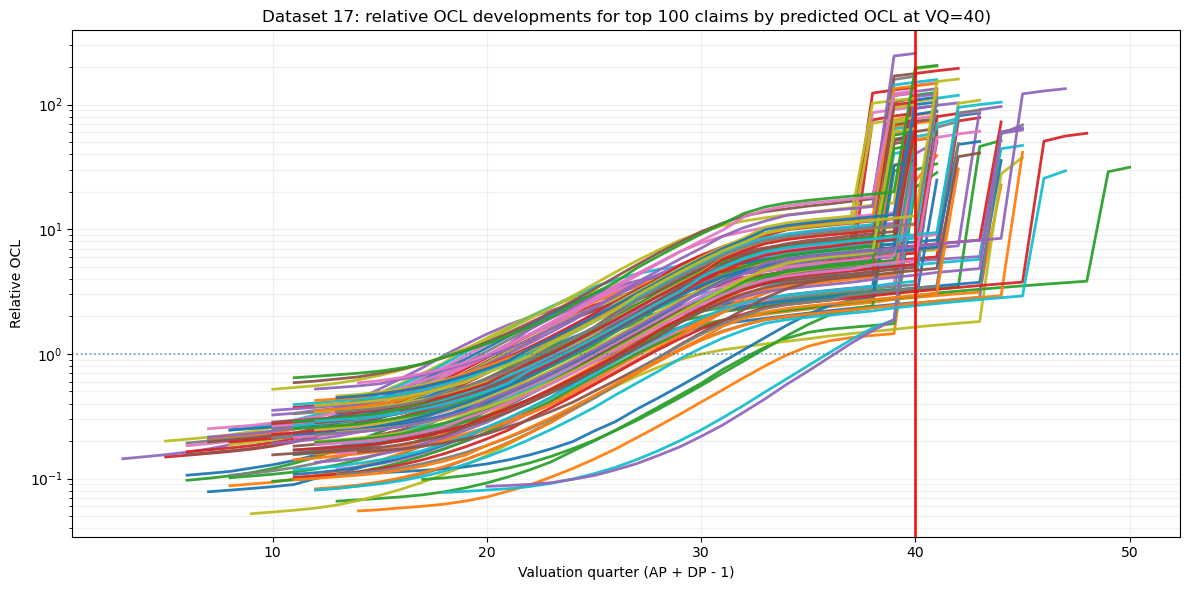}
    \caption{Relative OCL developments for the top 100 claims predicted by RL at valuation for Dataset 17}
    \label{fig:dataset17_rel_ocl}
\end{figure}

Figure \ref{fig:dataset17_rel_ocl} shows the relative OCLs for the top 100 claims by size of RL prediction at valuation for dataset 17 (the outlying dataset with an aggregate relative OCL of 2.56 in Figure \ref{fig:Boxplots of the aggregate relative OCL across all periods}). We compare this to dataset 15, which had an aggregate relative OCL of 1.02. These two graphs suggest that dataset 17 is so outlying because a large proportion of these top 100 predictions happen to overestimate significantly as they were unable to adjust to the final payment that occurred close to the valuation date. Only 36 of the 100 claims in dataset 17 had final payments after the valuation, while there were 56 from dataset 15. 

While we think this is a large reason behind dataset 17 being an outlier, in practice, this phenomenon would often be of little significance to insurers since the fact that these are final payments will likely be known to the insurer for most claims.

A related consideration is that we think a large portion of the volatility for RL and FNN is a result of this phenomenon. Without external signals, it is difficult for a method to adjust to a large and sudden change in OCL. This will be interesting to explore in future research.

\begin{figure}
    \centering
    \includegraphics[width=1\linewidth]{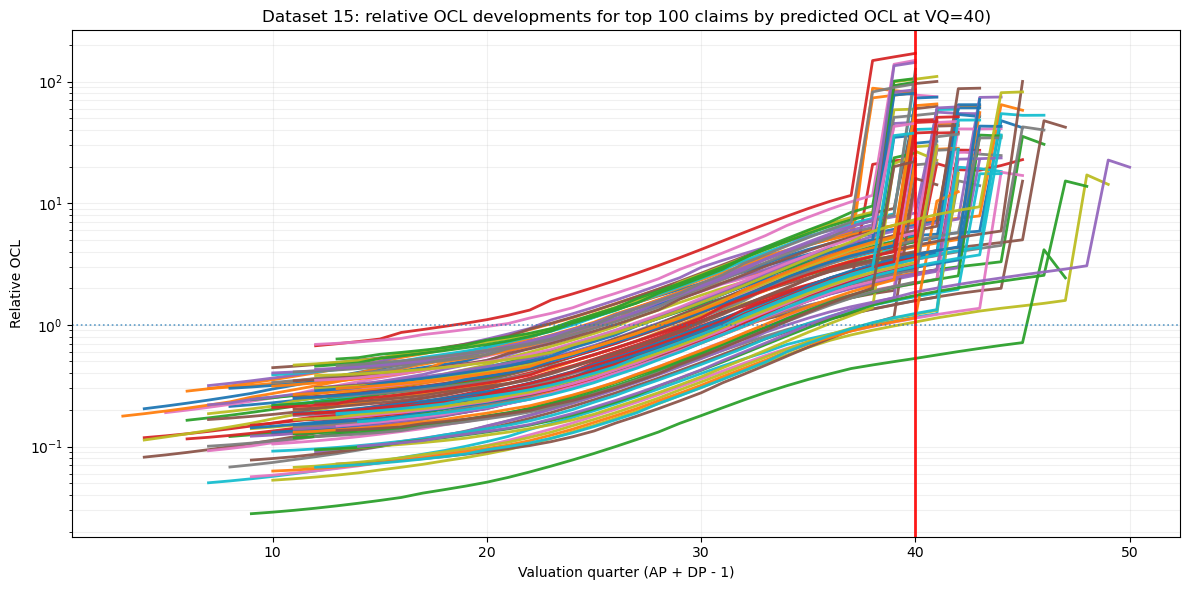}
    \caption{Relative OCL developments for the top 100 claims predicted by RL at valuation for Dataset 15}
    \label{fig:dataset15_rel_ocl}
\end{figure}

\section{Example of Iteration through State Space} \label{A_example}

See the Excel file ``Single Claim Example" (available upon request
) for an example of evolution through one claim from SPLICE complexity 5 data (claim $n$). We also include the table at the bottom of this section. Below, we take just the final development of the claim before its settlement to illustrate the calculations.

\begin{enumerate}
    \item Suppose the current state space that the agent observes is \\ 

    \begin{center} 
    $\begin{array}{ll} \text{State component} & \text{Current State Space at } j_n \\[6pt] 
    \hline AP=i_n & 34 \\[4pt] DP=j_n & 13 \\[4pt] \hat{\mathrm{OCL}}_{j_n-1} & \$675621.39 \\[4pt] 
    \mathrm{txn\_types}_{j_n} & [P] \\[4pt] 
    P_{j_n} & \$333857 \\[4pt] 
    \mathrm{repdel} & 1 \\[4pt] 
    n_{j_n}^{\mathrm{pay}} & 8 \\[4pt] 
    \mathrm{AQ} & 3 \\[4pt] 
    \mathrm{DQ}_{j_n} & 1 \\[6pt] 
    \mathrm{past\ predictions}_{j_n} & [332401,436967,448354,556969,675621] \\[4pt]
    \mathrm{case}_{j_n} & \$30615.11
    \end{array}$ 
    \end{center}

    and the agent predicts $\hat{\text{OCL}}_{j_n} = \$675411.78$.

    \item The agent takes an action $a_{j_n}$ based on the state space observed. This action is obtained from the underlying RL algorithm using what it has learnt so far. Because $\hat{\text{OCL}}_{j}=\hat{\text{OCL}}_{j-1}\exp(a_j)$ we can back out the action taken as:

    \begin{equation}
        a_{j_n} = \ln \left( \frac{\hat{\text{OCL}}_{j_n}}{\hat{\text{OCL}}_{j_n-1}} \right)=-0.0003
    \end{equation}
    
    \item The agent then receives a reward signal $r_{\tau=12}$.

    \begin{itemize}
        \item The claim has not yet settled (it will settle next period), hence, we need to calculate the stability and smoothing reward components. Now, as there is a payment, $r_{\text{stab}, 12} = r_{\text{smooth},12} = 0$. However, we show how they would be calculated \textit{if} there wasn't a payment made.

        \begin{itemize}
            \item We are at $\tau=T_n-1=12$ (one period before settlement), hence

            $$r_{\text{stab}, 12} = 0 - h\left( 675621+57577, 556968+48219 \right)$$

            Note that the inputs to $h(\cdot)$ need to be the ultimate losses; therefore, we are doing predicted OCL + cumulative paid to obtain the UL.

            \item The smoothing reward component is easy to calculate. Just note that more than the warm up number of predictions have been made, and the action space constraint is $K=2$ in this example. Hence

            $$r_{\text{smooth},12} = -1 \cdot \left( \frac{\lvert-0.0003\rvert}{\ln(2)} \right)^2$$
        \end{itemize}

        \item In the coding of the RL environment, as there are no predictions to be made at the settlement period of a claim, $r_\text{acc}$ is implemented to be given as at $\tau=T-1$ periods since notification instead of $\tau=T$ (settlement).

        Firstly, with $C=5$ in this example, we have the following accuracy reward \textit{without} OCL importance weighting.

        $$r_{\text{acc}} = 5 \cdot \frac{\sum_{\tau=1}^{12} \gamma^{\tau-1}\cdot h\left(\text{OCL}_\tau, \hat{\text{OCL}}_{\tau}\right)}{\sum_{\tau=1}^{12} \gamma^{\tau-1}}$$

        where $\text{OCL}_t$ is the true OCL remaining after $t$ periods since notification (only known upon settlement).

        However, we then need to apply the OCL importance weighting, so the $r_\text{acc}$ component is actually:

        $$r_{\text{acc}} = 5\cdot \sum_{\tau=1}^{12} {w_\tau} \cdot \frac{\gamma^{\tau-1}}{\sum_{\tau=1}^{12} \gamma^{\tau-1}}\cdot h\left(\text{OCL}_\tau, \hat{\text{OCL}}_\tau\right)=3.75$$

        Please refer back to Section \ref{Downward Bias Adjustment} for details and definitions.
    \end{itemize}
\end{enumerate}

\begin{sidewaystable}[p]
\centering
\caption{Example of how RL works with one claim (Note: AQ and DQ are arbitrarily chosen here for illustrative purposes)}
\label{tab:single-claim-example}

% slightly tighter than default
\setlength{\tabcolsep}{2.0pt}
\renewcommand{\arraystretch}{0.95}
\scriptsize

% scale to fit width, but do not force-fill page height
\begin{adjustbox}{max width=0.98\textheight,center}
\begin{tabular}{rr r l r r r r r p{6.2cm} r r r r r r}
\toprule
\makecell[c]{AP} &
\makecell[c]{DP} &
\makecell[c]{prev\\OCL} &
\makecell[c]{txn\\types} &
\makecell[c]{cum\\paid} &
\makecell[c]{rep\\delay} &
\makecell[c]{n\_pay} &
\makecell[c]{AQ} &
\makecell[c]{DQ} &
\makecell[c]{past preds} &
\makecell[c]{case\\OCL} &
\makecell[c]{RL\\pred} &
\makecell[c]{True\\OCL} &
\makecell[c]{R\\stab} &
\makecell[c]{R\\smooth} &
\makecell[c]{R\\acc} \\
\midrule
34 & 2  & 499175.5 & Ma  & 0.0      & 1 & 0 & 3 & 2 & \mbox{\texttt{[0,0,0,0,499175]}}                 & 289257.2 & 519377.1 & 364472.9 & -0.0033 & -0.0003 & 0.0000 \\
34 & 3  & 519377.1 & P   & 6768.9   & 1 & 1 & 3 & 2 & \mbox{\texttt{[0,0,0,499175,519377]}}           & 282488.3 & 484056.6 & 357704.0 &  0.0000 &  0.0000 & 0.0000 \\
34 & 4  & 484056.6 &     & 6768.9   & 1 & 1 & 3 & 4 & \mbox{\texttt{[0,0,499175,519377,484057]}}      & 282488.3 & 489150.9 & 357704.0 &  0.0363 & -0.0001 & 0.0000 \\
34 & 5  & 489150.9 & P   & 12678.5  & 1 & 2 & 3 & 4 & \mbox{\texttt{[0,499175,519377,484057,489151]}} & 276578.7 & 348606.7 & 351794.4 &  0.0000 &  0.0000 & 0.0000 \\
34 & 6  & 348606.7 & PMi & 19322.9  & 1 & 3 & 3 & 4 & \mbox{\texttt{[499175,519377,484057,489151,348607]}} & 320783.0 & 269355.0 & 345150.0 &  0.0000 &  0.0000 & 0.0000 \\
34 & 7  & 269355.0 &     & 19322.9  & 1 & 3 & 3 & 3 & \mbox{\texttt{[519377,484057,489151,348607,269355]}} & 320783.0 & 277343.3 & 345150.0 &  0.1864 & -0.0011 & 0.0000 \\
34 & 8  & 277343.3 & P   & 27497.3  & 1 & 4 & 3 & 3 & \mbox{\texttt{[484057,489151,348607,269355,277343]}} & 312608.5 & 332400.9 & 336975.5 &  0.0000 &  0.0000 & 0.0000 \\
34 & 9  & 332400.9 & PMi & 37795.6  & 1 & 5 & 3 & 3 & \mbox{\texttt{[489151,348607,269355,277343,332401]}} & 326677.3 & 436967.4 & 326677.3 &  0.0000 &  0.0000 & 0.0000 \\
34 & 10 & 436967.4 &     & 37795.6  & 1 & 5 & 3 & 2 & \mbox{\texttt{[348607,269355,277343,332401,436967]}} & 326677.3 & 448353.7 & 326677.3 &  0.2418 & -0.0012 & 0.0000 \\
34 & 11 & 448353.7 & P   & 48219.6  & 1 & 6 & 3 & 2 & \mbox{\texttt{[269355,277343,332401,436967,448354]}} & 316253.3 & 556969.0 & 316253.3 &  0.0000 &  0.0000 & 0.0000 \\
34 & 12 & 556969.0 & P   & 57578.0  & 1 & 7 & 3 & 1 & \mbox{\texttt{[277343,332401,436967,448354,556969]}} & 306894.9 & 675621.4 & 306894.9 &  0.0000 &  0.0000 & 0.0000 \\
34 & 13 & 675621.4 & P   & 333857.8 & 1 & 8 & 3 & 1 & \mbox{\texttt{[332401,436967,448354,556969,675621]}} & 30615.1  & 675411.8 & 30615.1  &  0.0000 &  0.0000 & 3.7502 \\
\bottomrule
\end{tabular}
\end{adjustbox}
\end{sidewaystable}

\section{Histograms of actions taken during testing of the RL model}

\begin{figure}[htb]
    \centering
    \includegraphics[width=0.75\linewidth]{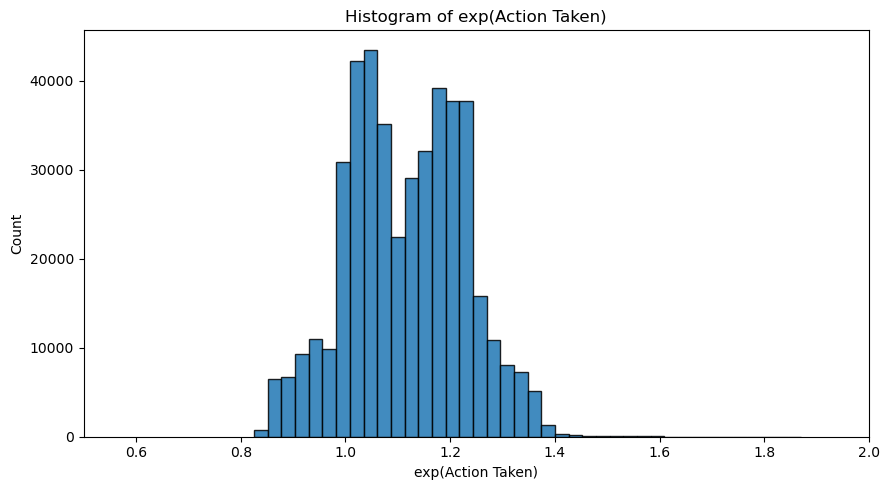}
    \caption{$\exp(a_j)\in[0.5, 2]$ undertaken during training \\We obtained this graph by plotting the exponential of the actions taken when RL predicted the remaining periods of the complexity 1 \textit{test} claims}
    \label{fig:actions_hist}
\end{figure}

\begin{figure}[htb]
    \centering
    \includegraphics[width=1\linewidth]{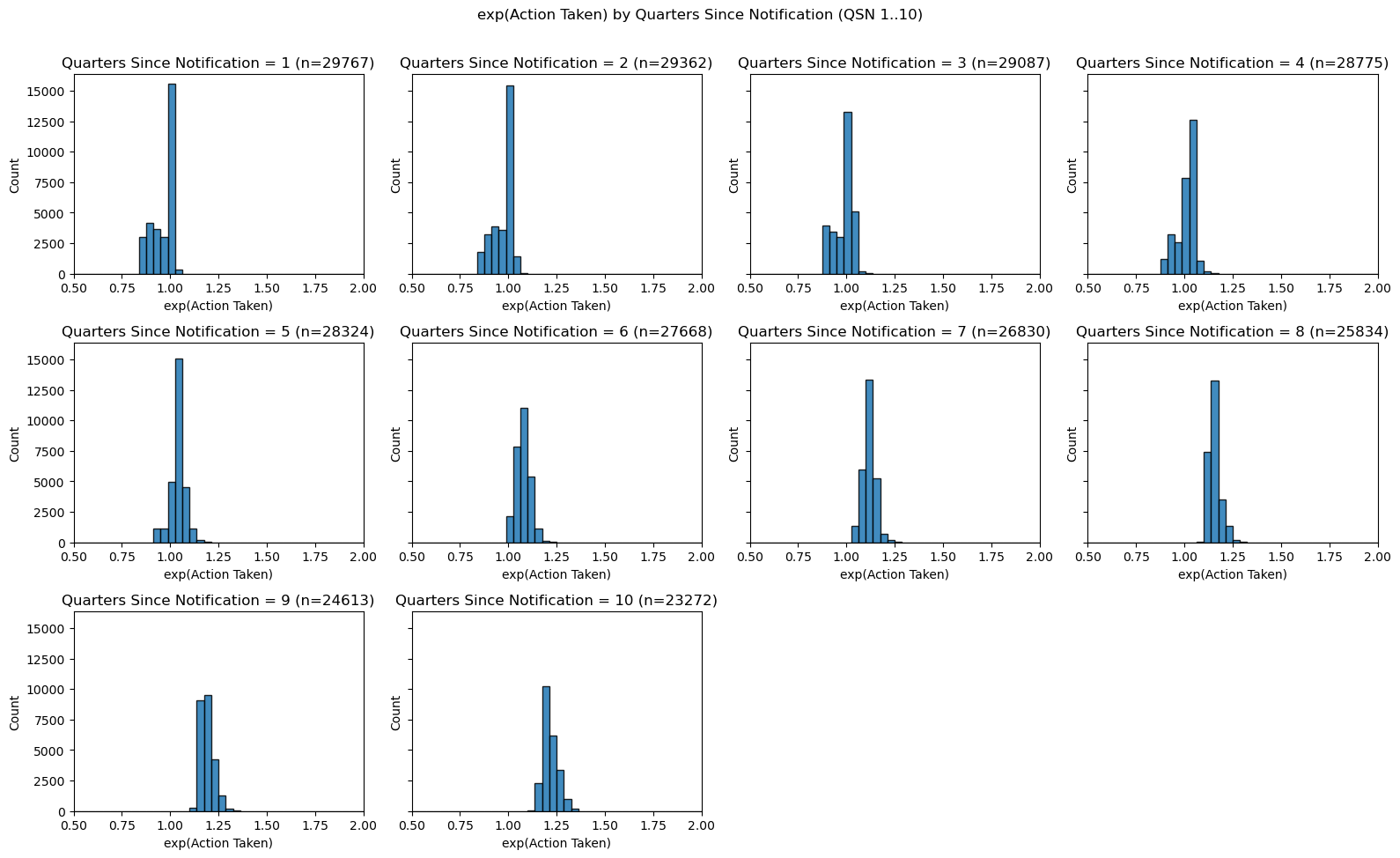}
    \caption{$\exp(a_j)\in[0.5, 2]$ undertaken during training for the first 10 quarters since notification for the complexity 1 test claims}
    \label{fig:actions_hist_by_qns}
\end{figure}

Figures \ref{fig:actions_hist}--\ref{fig:actions_hist_by_qns} were obtained from a complexity 1 SPLICE dataset.

Similarly, for complexity 5 of the SPLICE dataset:

\begin{figure}
    \centering
    \includegraphics[width=0.75\linewidth]{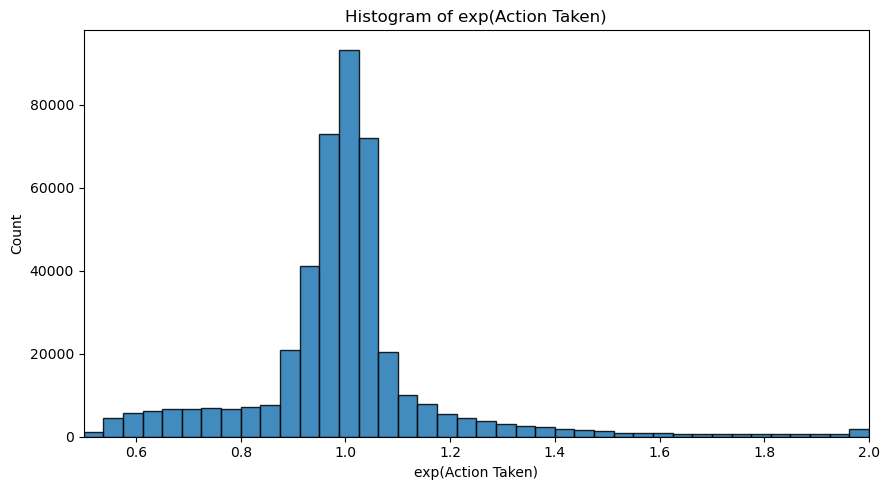}
    \caption{$\exp(a_j)\in[0.5, 2]$ undertaken during training \\We obtained this graph by plotting the exponential of the actions taken when RL predicted the remaining periods of the complexity 5 \textit{test} claims}
    \label{fig:actions_hist_C5}
\end{figure}

\begin{figure}
    \centering
    \includegraphics[width=1\linewidth]{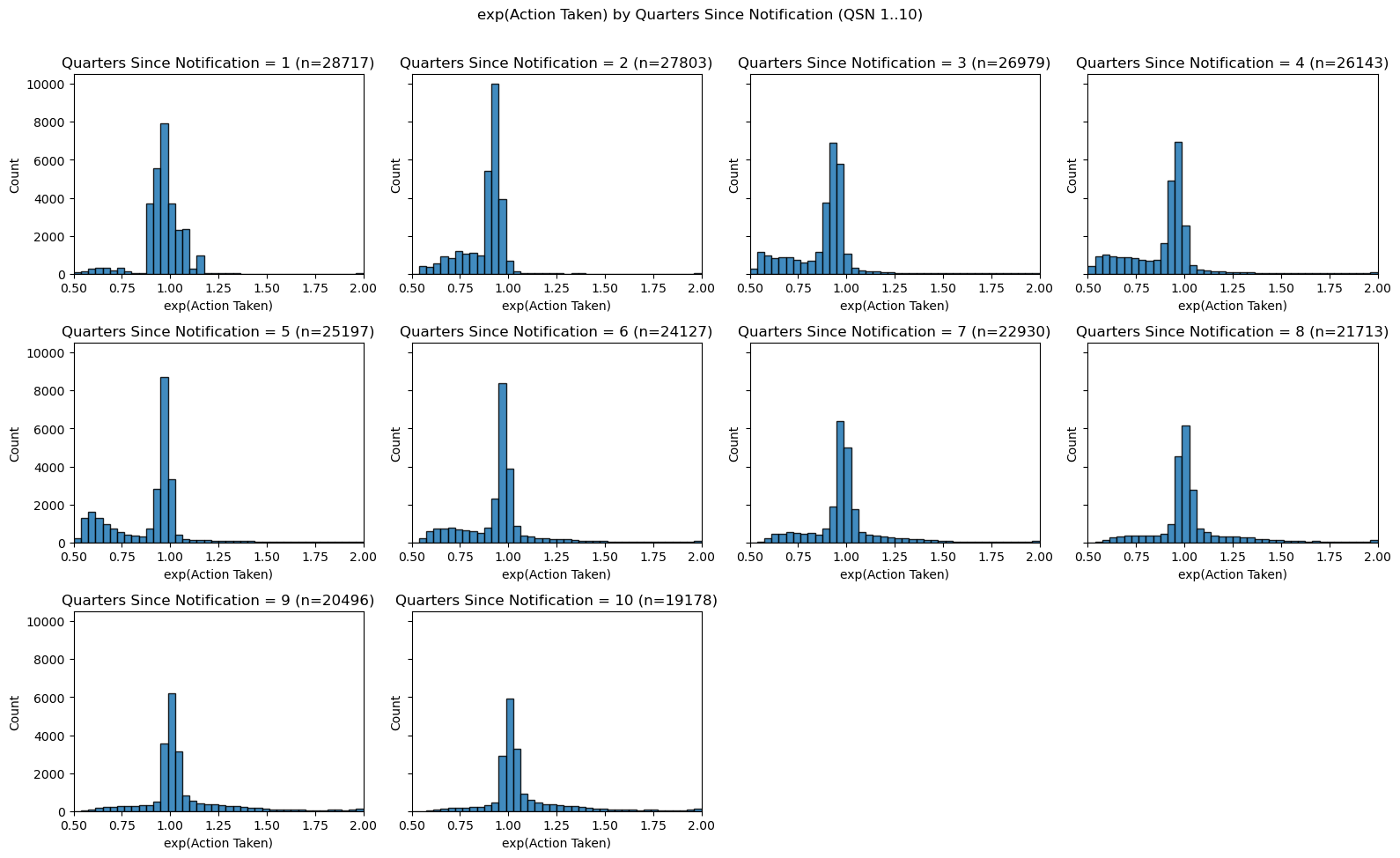}
    \caption{$\exp(a_j)\in[0.5, 2]$ undertaken during training for the first 10 quarters since notification for the complexity 5 test claims}
    \label{fig:action_hist_qsn_C5}
\end{figure}

\end{document}